\documentclass[12pt]{article}

\usepackage[T1]{fontenc}
\usepackage{authblk}
\usepackage{setspace}

\title{Matrix Decomposition Latent Growth Model Tree}

\author[1]{Naoya Todo}
\author[2]{Naoto Yamashita}
\author[3]{Satoshi Usami}

\affil[1]{Tokyo Metropolitan University}
\affil[2]{Kansai University}
\affil[3]{The University of Tokyo}

\date{}

\begin{document}
\def\spacingset#1{\renewcommand{\baselinestretch}{#1}\small\normalsize}
\spacingset{1.75}

\maketitle

\vspace{0.5cm}

\noindent\textbf{Data availability.}
The datasets generated and analyzed during the current study are available from the corresponding author on reasonable request.

\vspace{0.5cm}

\noindent\textbf{Funding.}
This research was supported by Japan Society for the Promotion of Science KAKENHI Grant Number JP23K02861.

\vspace{0.5cm}

\noindent\textbf{Competing interests.}
The authors have no competing interests to declare that are relevant to the content of this article.

\vspace{0.5cm}

\noindent \textbf{Corresponding author:}\\
Naoya Todo\\
Faculty of Humanities and Social Sciences, Tokyo Metropolitan University\\
1-1 Minami-Osawa, Hachioji-shi, Tokyo 192-0397, Japan\\
E-mail: \texttt{ntodo@tmu.ac.jp}

\thispagestyle{empty}

\end{document}


\def\spacingset#1{\renewcommand{\baselinestretch}{#1}\small\normalsize}
\spacingset{1.75}

\null
\vfill
\begin{center}
{\Huge\bfseries Supplementary Materials \par}
\end{center}
\vfill
\newpage

\section*{Supplementary materials}
\subsection{Results of the simulation used to determine $cp$ values}
As mentioned in the main text, to determine the appropriate $cp$ value, we conducted a small preliminary simulation. Specifically, we ran this simulation using basically the same settings as those in the main text. However, unlike the simulation described in the main text, we used only the MDLGM Tree1 and SEM Tree1 as the estimation method and calculated only the proportion in which the true number of nodes (= 4) was correctly estimated. Furthermore, we generated five data sets under each condition except for the conditions in which the true model was a quadratic LGM whereas the specified model was a linear LGM and manipulated $cp$ value as an additional factor. We set $cp=0.01$, $0.005$, $0.003$, and $0.001$ as the levels of $cp$ factor, and calculated, for each $cp$ value, the proportion in which the true number of nodes was correctly estimated across the conditions.

Table~\ref{tab:cp} shows the result. This table implied that for both methods, that is, splits based on the Mahalanobis distance and for splits based on the log-likelihood, the highest estimation accuracy would be obtained when $cp=0.003$. Therefore, for the simulations in the main text, $cp$ was set to $0.003$ for all six estimation methods.

\begin{table}[h]
    \centering
    \begin{tabular}{ccc} \hline
        $cp$ & MDLGM Tree1 & SEM Tree1 \\ \hline
        0.01 & 0.188 & 0.144 \\
        0.005 & 0.713 & 0.413 \\
        0.003 & 0.919 & 0.738 \\
        0.001 & 0.756 & 0.600 \\ \hline
    \end{tabular}
    \caption{\\Proportion in which the true number of nodes (= 4) was correctly estimated across the conditions under each $cp$ value. MDLGM Tree1 denotes splitting based on the Mahalanobis distance and SEM Tree1 denotes ML estimation respectively. $cp$ denotes the $cp$ value which was used in building trees.}
    \label{tab:cp}
\end{table} \clearpage

\subsection{Proportion in which improper solutions occur}
\begin{figure}[h] \centering \includegraphics[width=0.9\linewidth]{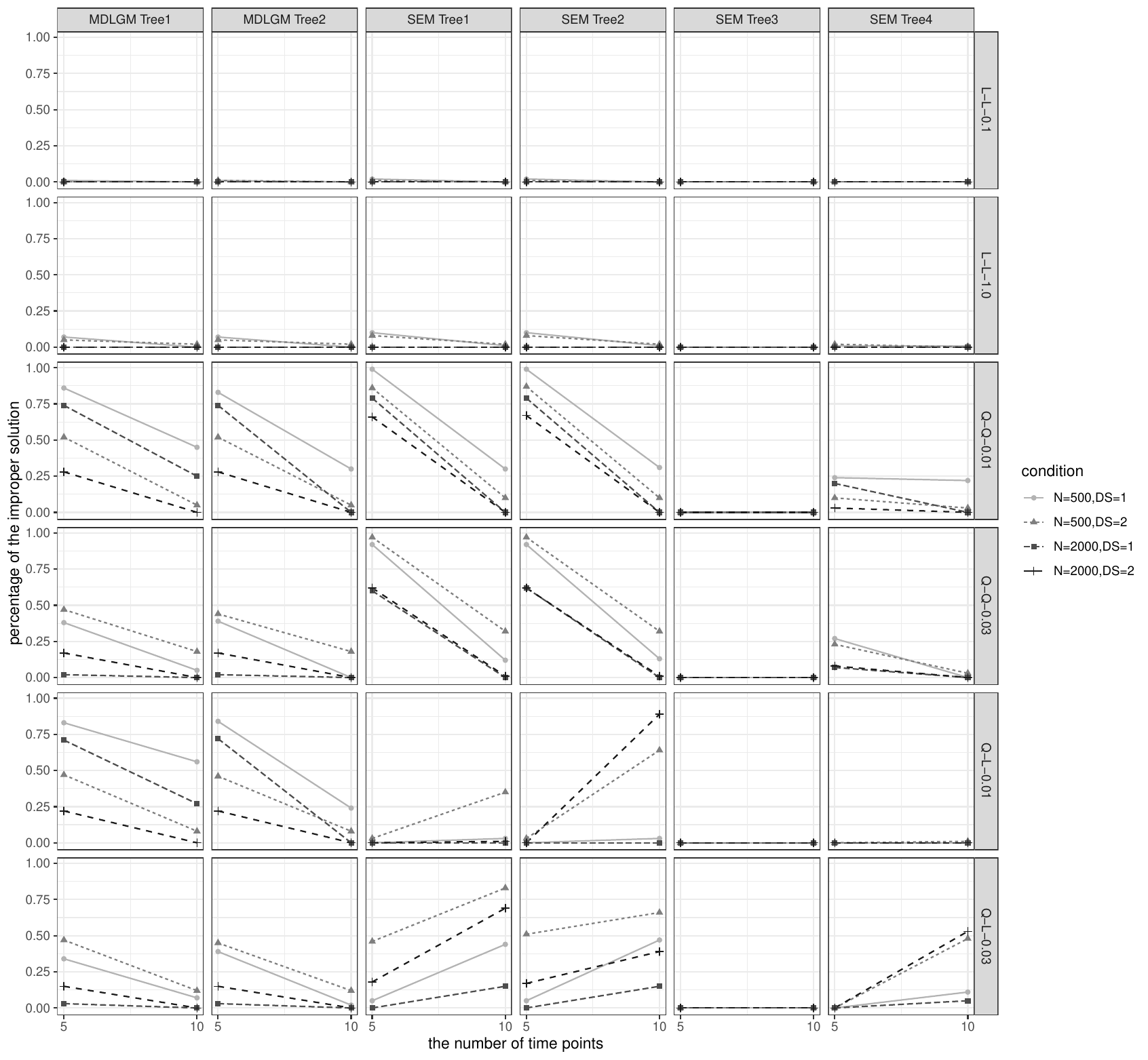} \caption{\\Proportion in which improper solutions occurred, using $10^{-2}$ as the threshold for variance parameters estimates. SEM Tree1 denotes ML estimation, SEM Tree2 denotes constrained ML (CML) estimation, SEM Tree3 denotes Bayesian estimation, and SEM Tree4 denotes ML estimation with an algorithm that avoids node splitting based on any estimated model that produces a warning. MDLGM Tree1 and MDLGM Tree2 denote splitting based on the Mahalanobis distance and deviance, respectively. $DS$ denotes the degree of separation, and the three-element labels on the vertical axis (e.g., Q-L-0.01) indicate, in order, the true model (linear or quadratic), the template model (linear or quadratic), and the specified value of the slope factor variance.} \label{fig:improper_soluion_2} \end{figure}

\begin{figure}[h] \centering \includegraphics[width=0.9\linewidth]{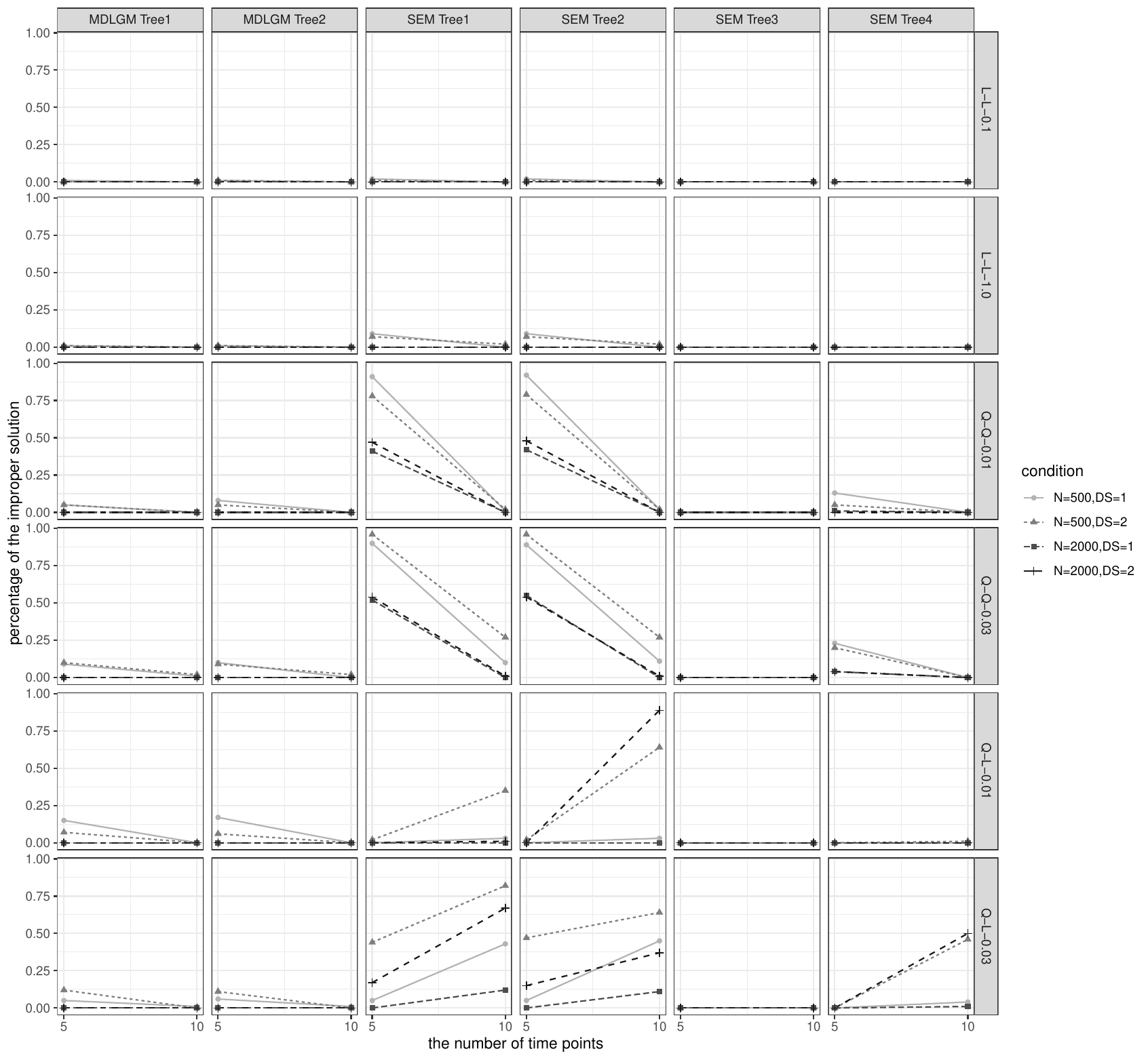} \caption{\\Proportion in which improper solutions occurred, using $10^{-3}$ as the threshold for variance parameters estimates. SEM Tree1 denotes ML estimation, SEM Tree2 denotes constrained ML (CML) estimation, SEM Tree3 denotes Bayesian estimation, and SEM Tree4 denotes ML estimation with an algorithm that avoids node splitting based on any estimated model that produces a warning. MDLGM Tree1 and MDLGM Tree2 denote splitting based on the Mahalanobis distance and deviance, respectively. $DS$ denotes the degree of separation, and the three-element labels on the vertical axis (e.g., Q-L-0.01) indicate, in order, the true model (linear or quadratic), the template model (linear or quadratic), and the specified value of the slope factor variance.} \label{fig:improper_soluion_3} \end{figure}

\begin{figure}[h] \centering \includegraphics[width=0.9\linewidth]{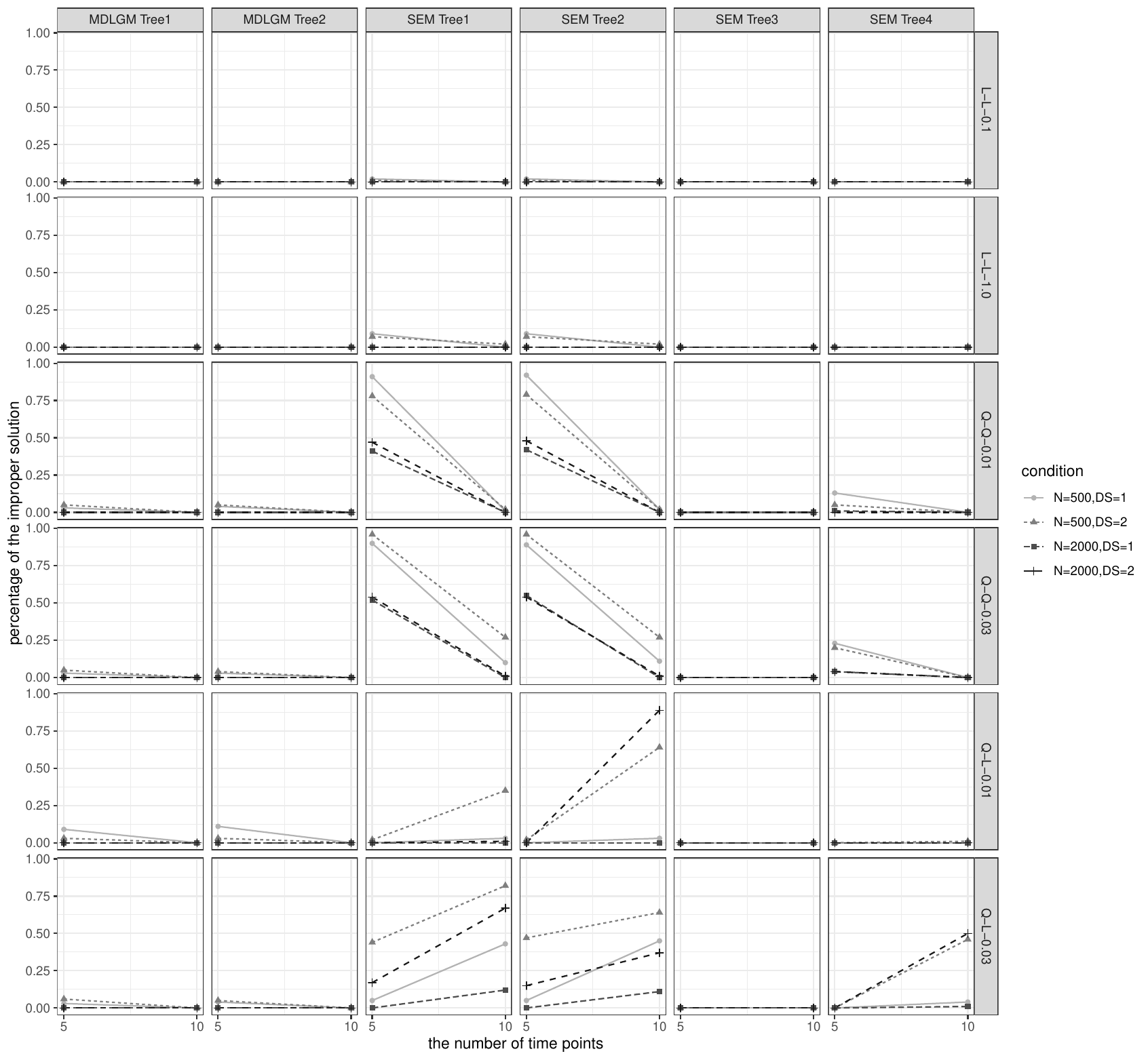} \caption{\\Proportion in which improper solutions occurred, using $10^{-4}$ as the threshold for variance parameters estimates. SEM Tree1 denotes ML estimation, SEM Tree2 denotes constrained ML (CML) estimation, SEM Tree3 denotes Bayesian estimation, and SEM Tree4 denotes ML estimation with an algorithm that avoids node splitting based on any estimated model that produces a warning. MDLGM Tree1 and MDLGM Tree2 denote splitting based on the Mahalanobis distance and deviance, respectively. $DS$ denotes the degree of separation, and the three-element labels on the vertical axis (e.g., Q-L-0.01) indicate, in order, the true model (linear or quadratic), the template model (linear or quadratic), and the specified value of the slope factor variance.} \label{fig:improper_soluion_4} \end{figure}

Figure~\ref{fig:improper_soluion_2}-\ref{fig:improper_soluion_4} show the proportion for each method and condition in which improper solutions occurred, using $10^{-2}$, $10^{-3}$, and $10^{-4}$ as the threshold for variance parameters estimates respectively. As mentioned in the main text, the results regarding differences among the estimation methods were basically similar even when different thresholds were applied to the variance estimates. 

 \clearpage 

 \clearpage

 \clearpage 

\subsection{Proportion of node splitting based on $z_3$ in the SEM Tree4} 
To examine whether the SEM Tree4 avoided node splitting based on $z_1$ and $z_2$ because of the occurrence of improper solutions while node splitting based on $z_3$ was still performed, we conducted an additional simulation. In the simulation, we investigated how frequently node splitting based on $z_3$ occurs in the SEM Tree4 under each condition. Specifically, we ran this simulation using exactly the same settings as as those in the main text. However, we used only the SEM Tree4 as the estimation method and calculated the number of times, out of 100, that $z_3$-based node splitting occurred. If the node splitting by $z_3$ occurred more frequently under the condition in which the average proportion of individuals whose node membership was correctly classified in the SEM Tree4 became lower, this would suggest that our prediction was correct.

\begin{figure}[h] \centering \includegraphics[width=0.5\linewidth]{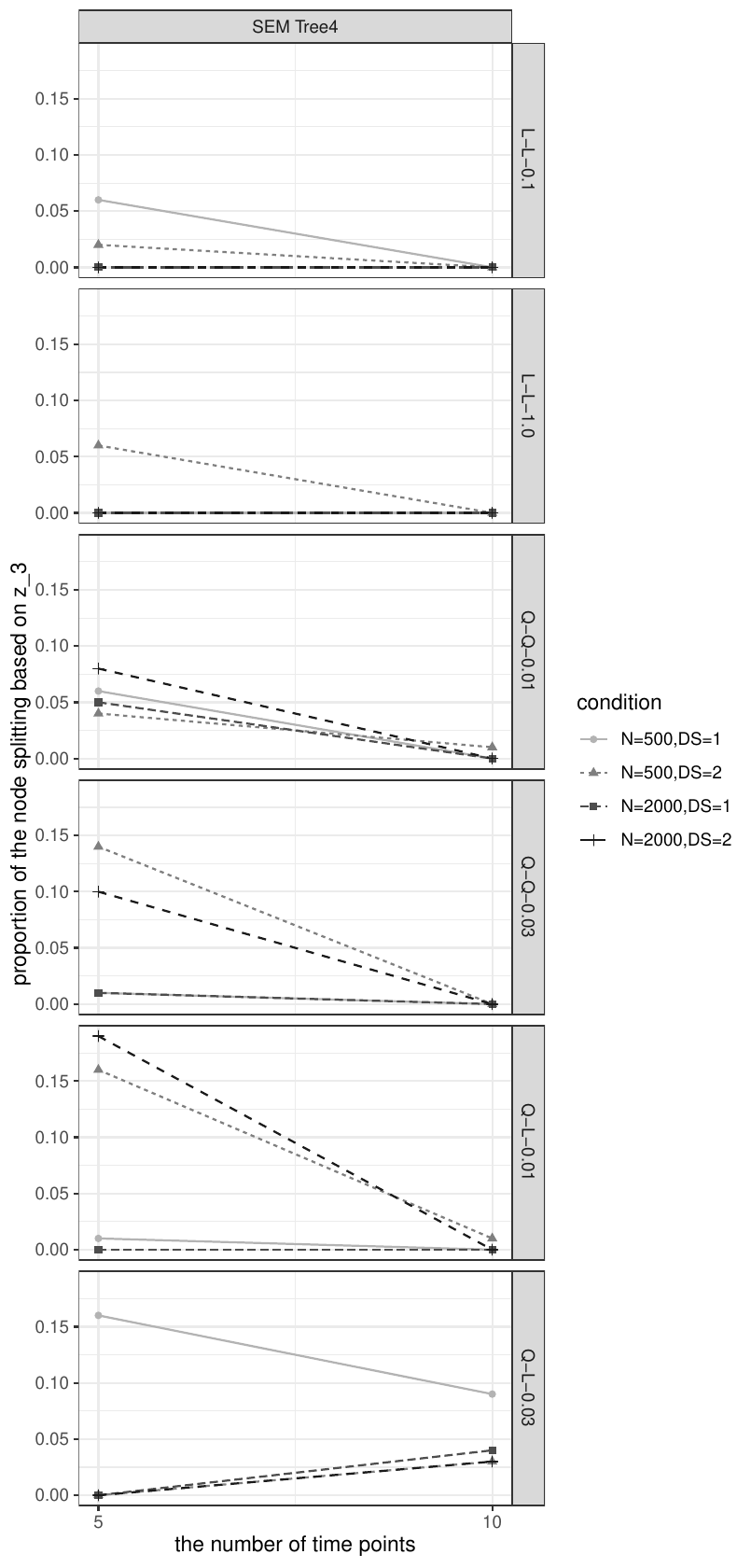} \caption{\\Proportion of node splitting based on $z_3$ in the SEM Tree4. SEM Tree4 denotes ML estimation with an algorithm that avoids node splitting based on any estimated model that produces a warning. $DS$ denotes the degree of separation, and the three-element labels on the vertical axis (e.g., Q-L-0.01) indicate, in order, the true model (linear or quadratic), the template model (linear or quadratic), and the specified value of the slope factor variance.} \label{fig:node_split_z3} \end{figure}

Figure~\ref{fig:node_split_z3} shows the result. Under the condition that the true and template models were linear LGMs, the node splitting by $z_3$ occurred only a very small number of times. In contrast, under the condition that the true models were quadratic LGMs and $T = 5$, the node splitting by $z_3$ was observed more frequently. As expected, under the condition in which the average proportion of individuals whose node membership was correctly classified in the SEM Tree4 became lower, the node splitting based on $z_3$ occurred more frequently. 

 \clearpage 

\newpage

\subsection{Bias and RMSE of parameters}
\subsubsection*{Bias and RMSE of means of the growth factors}
\begin{figure}[h]
    \centering
    \includegraphics[width=0.9\linewidth]{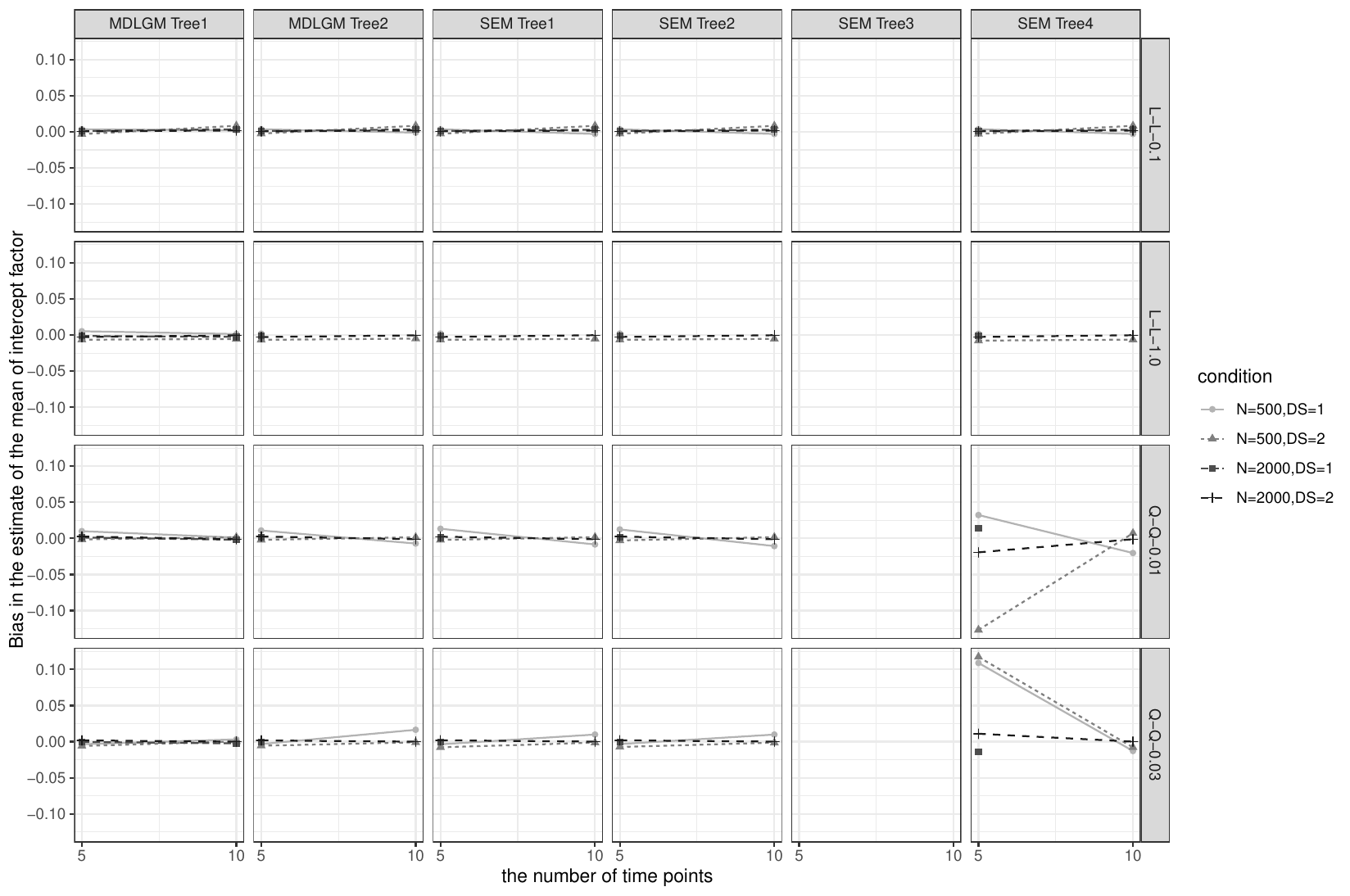}
    \caption{\\Bias in $\mu_I$ calculated using the data when the number of terminal nodes was correctly estimated under conditions in which the true model and the template model were identical. SEM Tree1 denotes ML estimation, SEM Tree2 denotes constrained ML (CML) estimation, SEM Tree3 denotes Bayesian estimation, and SEM Tree4 denotes ML estimation with an algorithm that avoids node splitting based on any estimated model that produces a warning. MDLGM Tree1 and MDLGM Tree2 denote splitting based on the Mahalanobis distance and deviance, respectively. $DS$ denotes the degree of separation, and the three-element labels on the vertical axis (e.g., Q-L-0.01) indicate, in order, the true model (linear or quadratic), the template model (linear or quadratic), and the specified value of the slope factor variance. The bias of the SEM Tree3 could not be calculated. }
    \label{fig:bias_mu_i}
\end{figure}

\begin{figure}[h]
    \centering
    \includegraphics[width=0.9\linewidth]{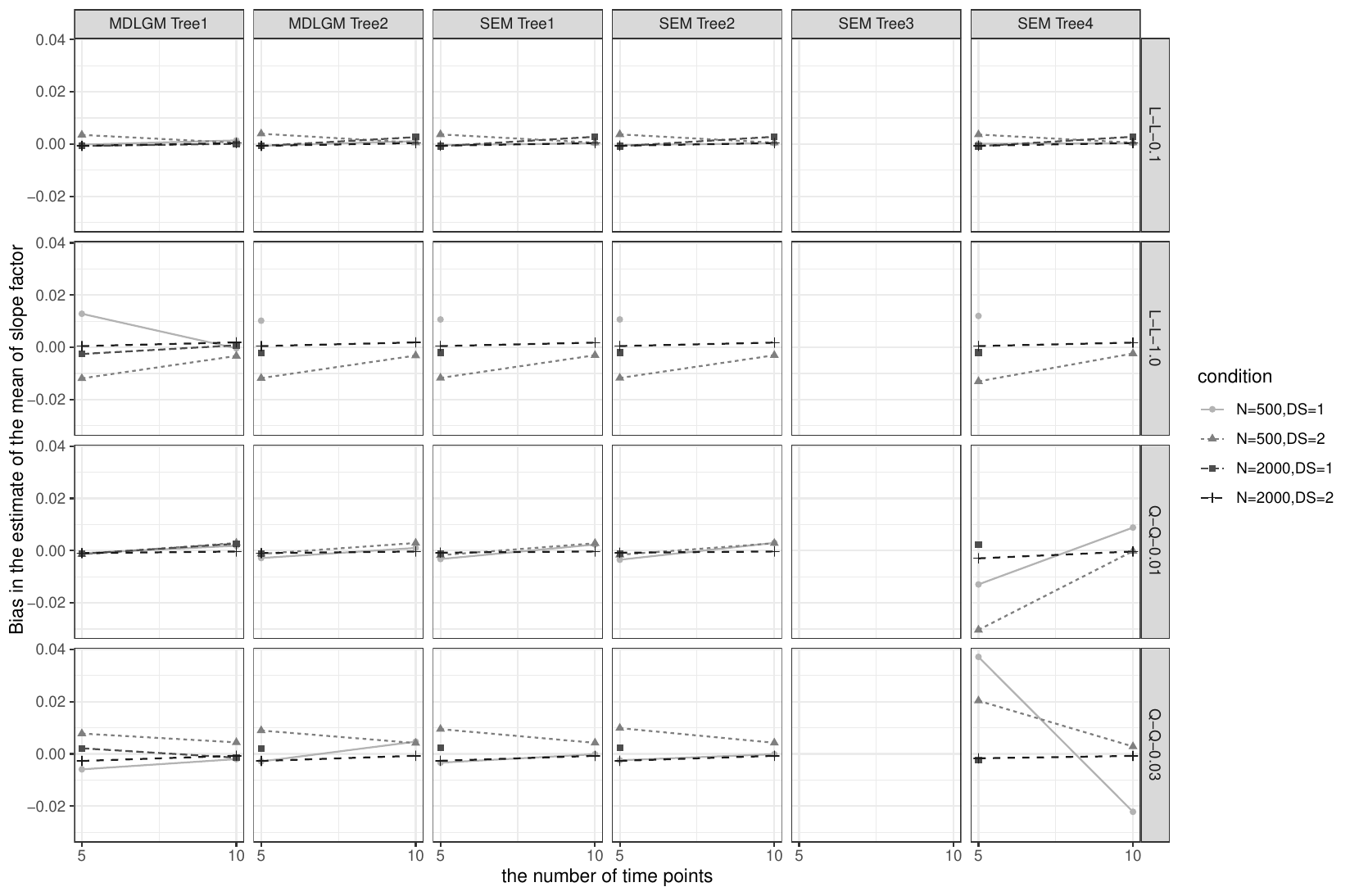}
    \caption{\\Bias in $\mu_S$ calculated using the data when the number of terminal nodes was correctly estimated under conditions in which the true model and the template model were identical. SEM Tree1 denotes ML estimation, SEM Tree2 denotes constrained ML (CML) estimation, SEM Tree3 denotes Bayesian estimation, and SEM Tree4 denotes ML estimation with an algorithm that avoids node splitting based on any estimated model that produces a warning. MDLGM Tree1 and MDLGM Tree2 denote splitting based on the Mahalanobis distance and deviance, respectively. $DS$ denotes the degree of separation, and the three-element labels on the vertical axis (e.g., Q-L-0.01) indicate, in order, the true model (linear or quadratic), the template model (linear or quadratic), and the specified value of the slope factor variance. The bias of the SEM Tree3 could not be calculated. }
    \label{fig:bias_mu_s}
\end{figure}

\begin{figure}[h]
    \centering
    \includegraphics[width=0.9\linewidth]{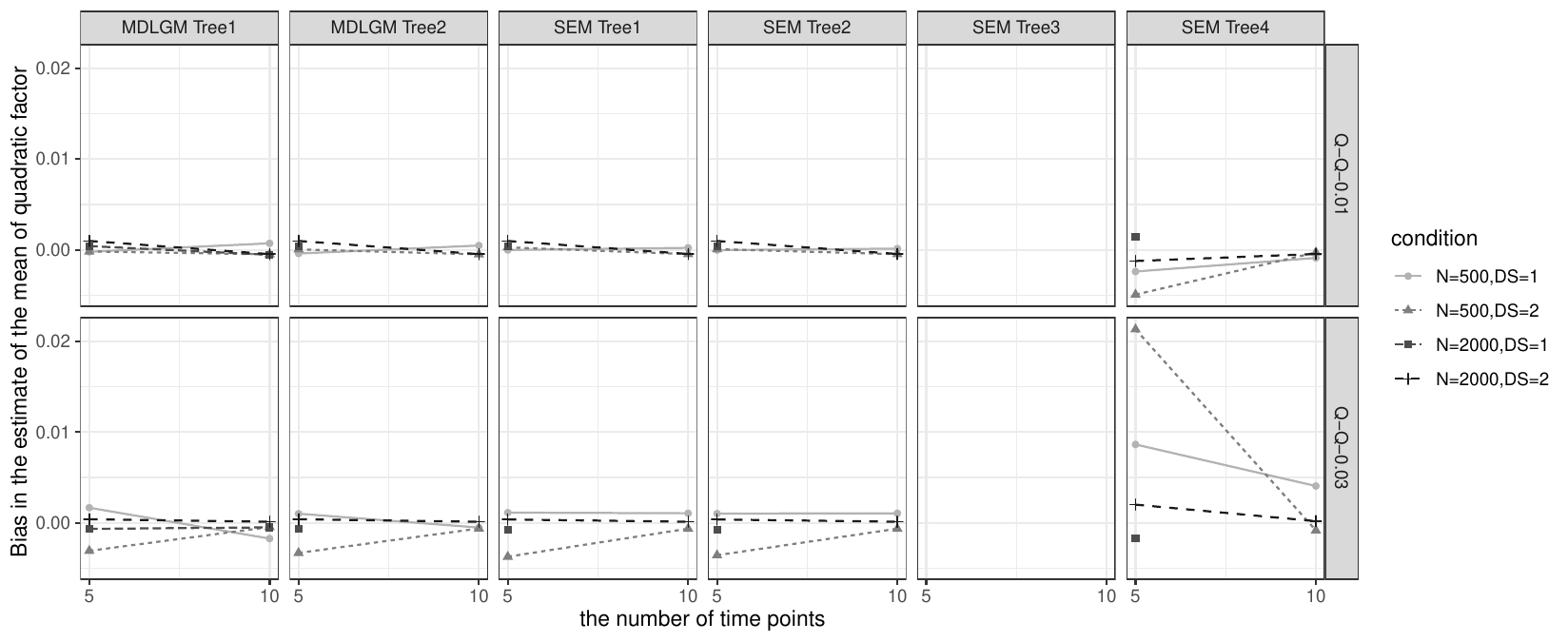}
    \caption{\\Bias in $\mu_Q$ calculated using the data when the number of terminal nodes was correctly estimated under conditions in which the true model and the template model were identical. SEM Tree1 denotes ML estimation, SEM Tree2 denotes constrained ML (CML) estimation, SEM Tree3 denotes Bayesian estimation, and SEM Tree4 denotes ML estimation with an algorithm that avoids node splitting based on any estimated model that produces a warning. MDLGM Tree1 and MDLGM Tree2 denote splitting based on the Mahalanobis distance and deviance, respectively. $DS$ denotes the degree of separation, and the three-element labels on the vertical axis (e.g., Q-L-0.01) indicate, in order, the true model (linear or quadratic), the template model (linear or quadratic), and the specified value of the slope factor variance. The bias of the SEM Tree3 could not be calculated. }
    \label{fig:bias_mu_q}
\end{figure}

\begin{figure}[h]
    \centering
    \includegraphics[width=0.9\linewidth]{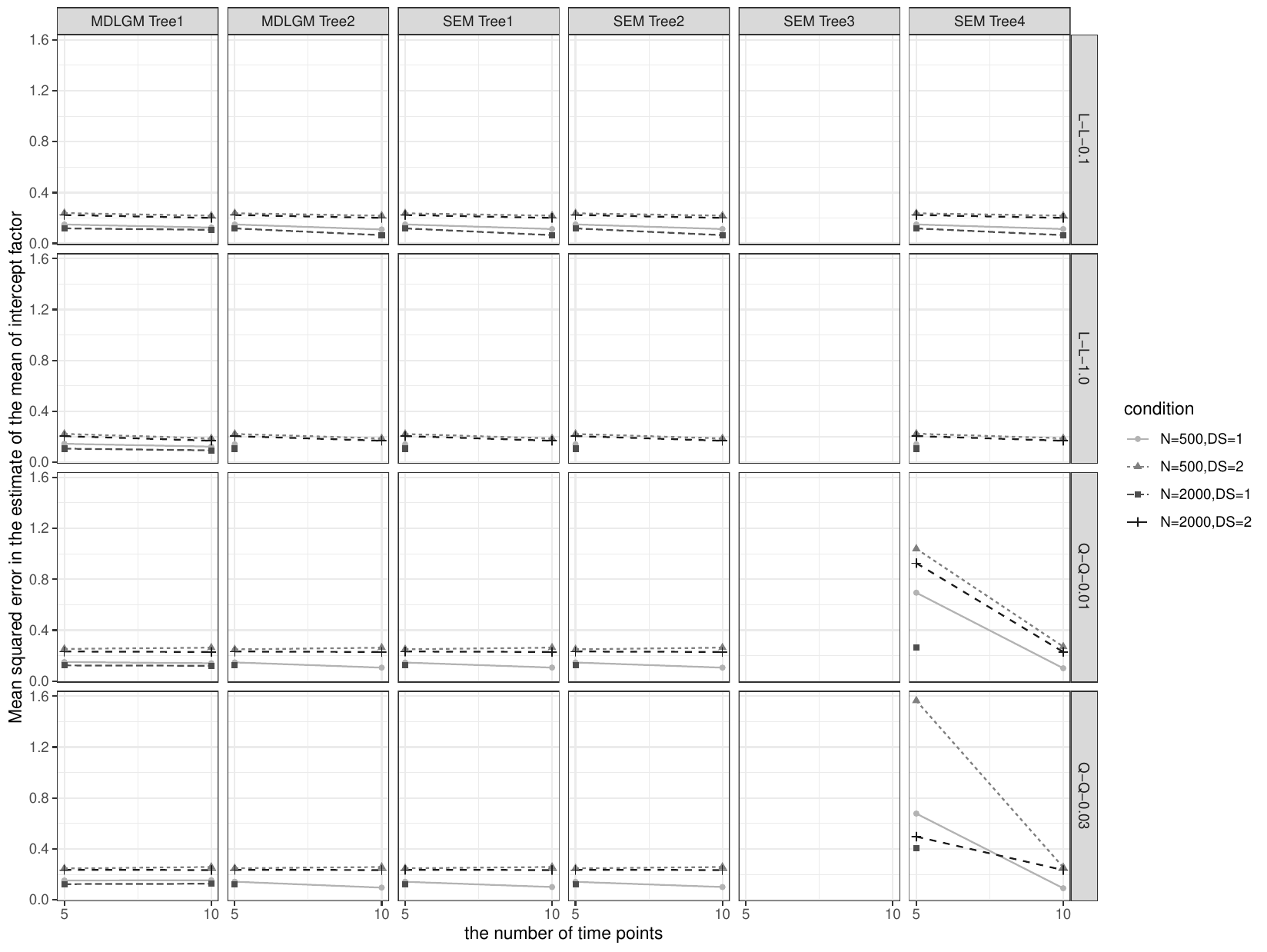}
    \caption{\\RMSE in $\mu_I$ calculated using the data when the number of terminal nodes was correctly estimated under conditions in which the true model and the template model were identical. SEM Tree1 denotes ML estimation, SEM Tree2 denotes constrained ML (CML) estimation, SEM Tree3 denotes Bayesian estimation, and SEM Tree4 denotes ML estimation with an algorithm that avoids node splitting based on any estimated model that produces a warning. MDLGM Tree1 and MDLGM Tree2 denote splitting based on the Mahalanobis distance and deviance, respectively. $DS$ denotes the degree of separation, and the three-element labels on the vertical axis (e.g., Q-L-0.01) indicate, in order, the true model (linear or quadratic), the template model (linear or quadratic), and the specified value of the slope factor variance. The RMSE of the SEM Tree3 could not be calculated. }
    \label{fig:rmse_mu_i}
\end{figure}

\begin{figure}[h]
    \centering
    \includegraphics[width=0.9\linewidth]{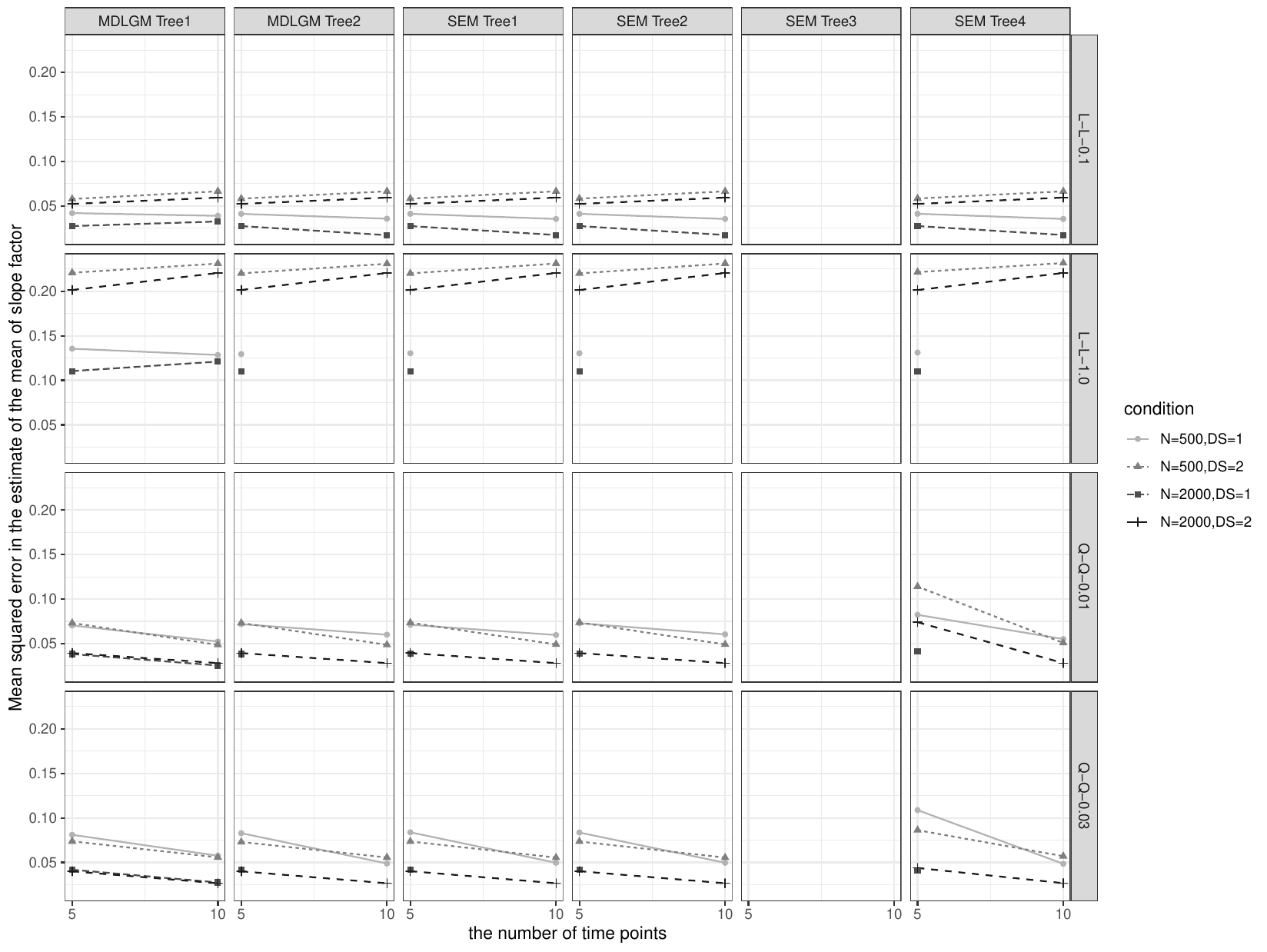}
    \caption{\\RMSE in $\mu_S$ calculated using the data when the number of terminal nodes was correctly estimated under conditions in which the true model and the template model were identical. SEM Tree1 denotes ML estimation, SEM Tree2 denotes constrained ML (CML) estimation, SEM Tree3 denotes Bayesian estimation, and SEM Tree4 denotes ML estimation with an algorithm that avoids node splitting based on any estimated model that produces a warning. MDLGM Tree1 and MDLGM Tree2 denote splitting based on the Mahalanobis distance and deviance, respectively. $DS$ denotes the degree of separation, and the three-element labels on the vertical axis (e.g., Q-L-0.01) indicate, in order, the true model (linear or quadratic), the template model (linear or quadratic), and the specified value of the slope factor variance. The RMSE of the SEM Tree3 could not be calculated. }
    \label{fig:rmse_mu_s}
\end{figure}

\begin{figure}[h]
    \centering
    \includegraphics[width=0.9\linewidth]{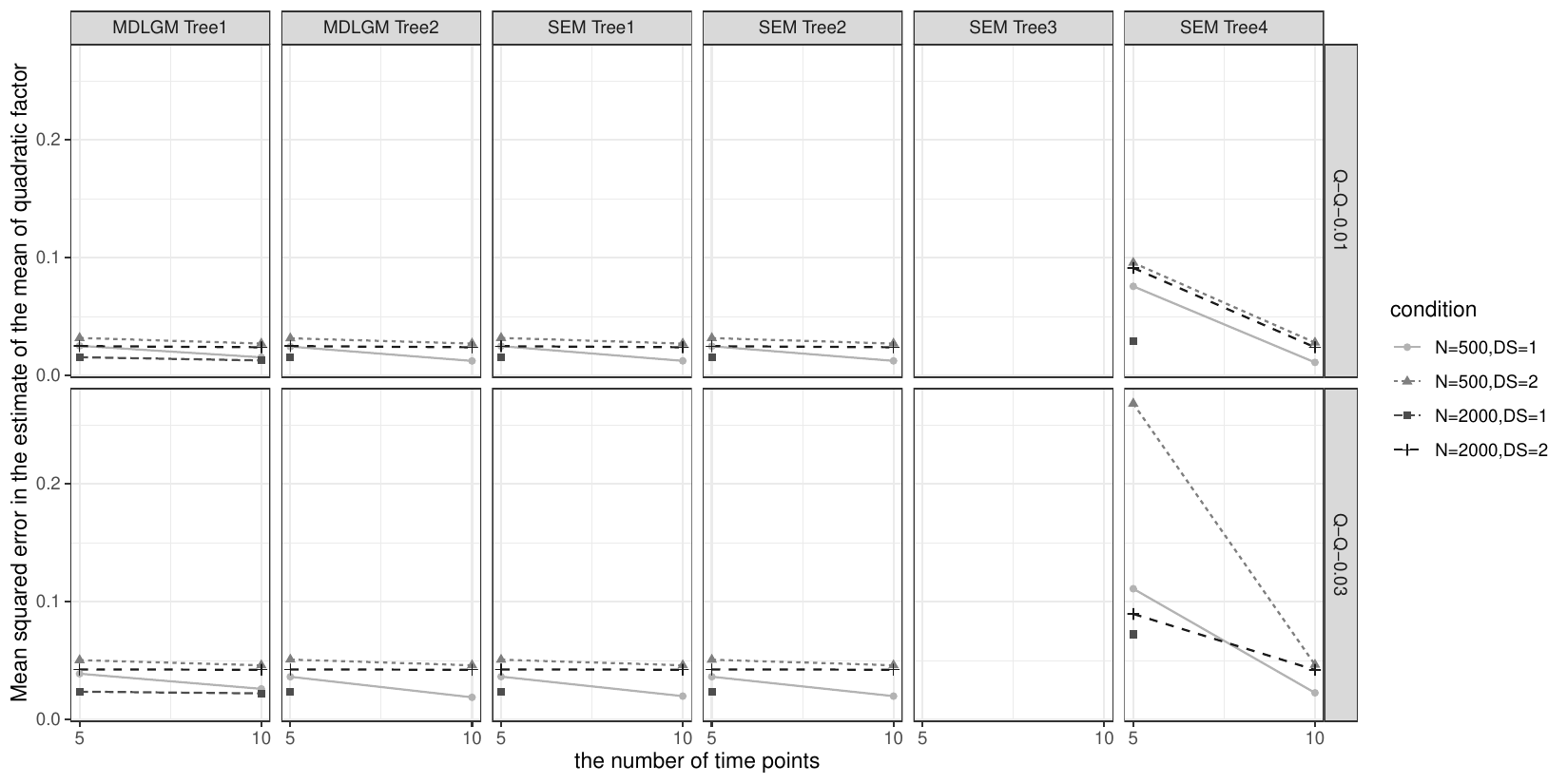}
    \caption{\\RMSE in $\mu_Q$ calculated using the data when the number of terminal nodes was correctly estimated under conditions in which the true model and the template model were identical. SEM Tree1 denotes ML estimation, SEM Tree2 denotes constrained ML (CML) estimation, SEM Tree3 denotes Bayesian estimation, and SEM Tree4 denotes ML estimation with an algorithm that avoids node splitting based on any estimated model that produces a warning. MDLGM Tree1 and MDLGM Tree2 denote splitting based on the Mahalanobis distance and deviance, respectively. $DS$ denotes the degree of separation, and the three-element labels on the vertical axis (e.g., Q-L-0.01) indicate, in order, the true model (linear or quadratic), the template model (linear or quadratic), and the specified value of the slope factor variance. The RMSE of the SEM Tree3 could not be calculated. }
    \label{fig:rmse_mu_q}
\end{figure}

Figure~\ref{fig:bias_mu_i}-\ref{fig:bias_mu_q} show the biases of $\mu_I$, $\mu_S$, and $\mu_Q$ respectively and Figure~\ref{fig:rmse_mu_i}-\ref{fig:rmse_mu_q} show the RMSEs of those in each method and condition. As mentioned in the main text, these figures show that all methods except the SEM Tree4 exhibited similar patterns. 

 \clearpage 

 \clearpage

 \clearpage

 \clearpage

 \clearpage

 \clearpage

\subsubsection*{Bias and RMSE of variances and covariances of the growth factors}
\begin{figure}[h]
    \centering
    \includegraphics[width=0.9\linewidth]{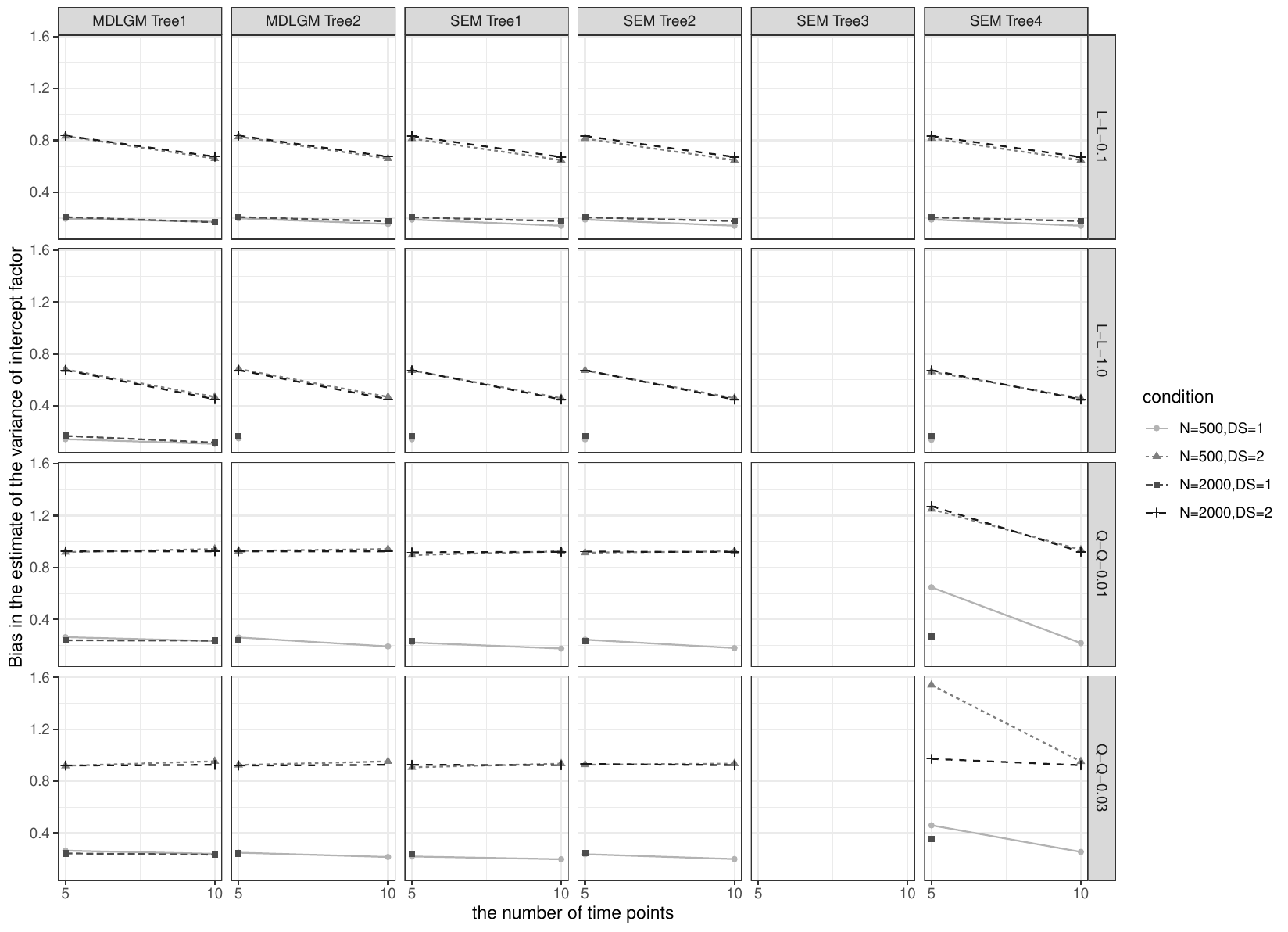}
    \caption{\\Bias in $\phi^2_I$ calculated using the data when the number of terminal nodes was correctly estimated under conditions in which the true model and the template model were identical. SEM Tree1 denotes ML estimation, SEM Tree2 denotes constrained ML (CML) estimation, SEM Tree3 denotes Bayesian estimation, and SEM Tree4 denotes ML estimation with an algorithm that avoids node splitting based on any estimated model that produces a warning. MDLGM Tree1 and MDLGM Tree2 denote splitting based on the Mahalanobis distance and deviance, respectively. $DS$ denotes the degree of separation, and the three-element labels on the vertical axis (e.g., Q-L-0.01) indicate, in order, the true model (linear or quadratic), the template model (linear or quadratic), and the specified value of the slope factor variance. The bias of the SEM Tree3 could not be calculated. }
    \label{fig:bias_phi_i}
\end{figure}

\begin{figure}[h]
    \centering
    \includegraphics[width=0.9\linewidth]{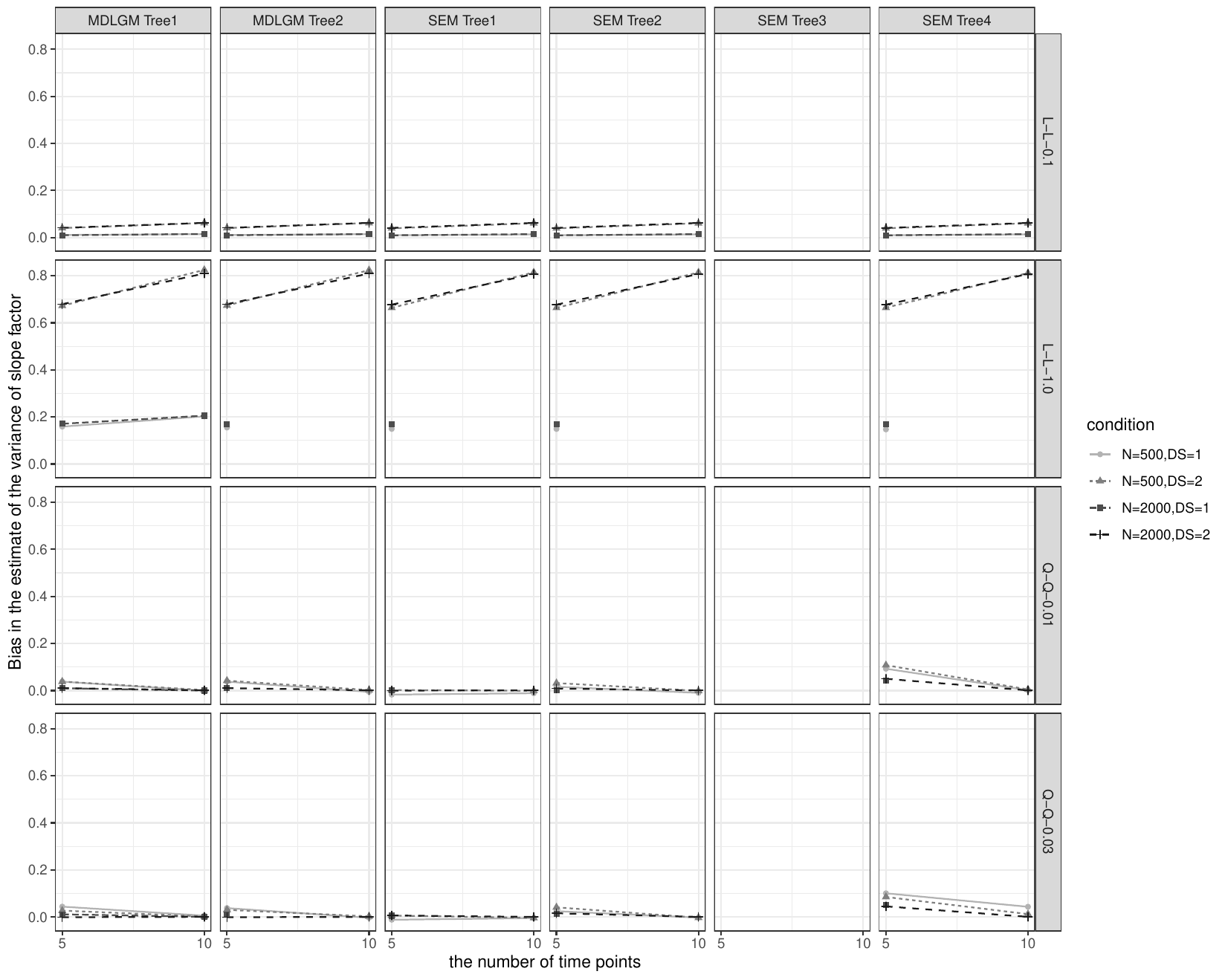}
    \caption{\\Bias in $\phi^2_S$ calculated using the data when the number of terminal nodes was correctly estimated under conditions in which the true model and the template model were identical. SSEM Tree1 denotes ML estimation, SEM Tree2 denotes constrained ML (CML) estimation, SEM Tree3 denotes Bayesian estimation, and SEM Tree4 denotes ML estimation with an algorithm that avoids node splitting based on any estimated model that produces a warning. MDLGM Tree1 and MDLGM Tree2 denote splitting based on the Mahalanobis distance and deviance, respectively. $DS$ denotes the degree of separation, and the three-element labels on the vertical axis (e.g., Q-L-0.01) indicate, in order, the true model (linear or quadratic), the template model (linear or quadratic), and the specified value of the slope factor variance. The bias of the SEM Tree3 could not be calculated. }
    \label{fig:bias_phi_s}
\end{figure}

\begin{figure}[h]
    \centering
    \includegraphics[width=0.9\linewidth]{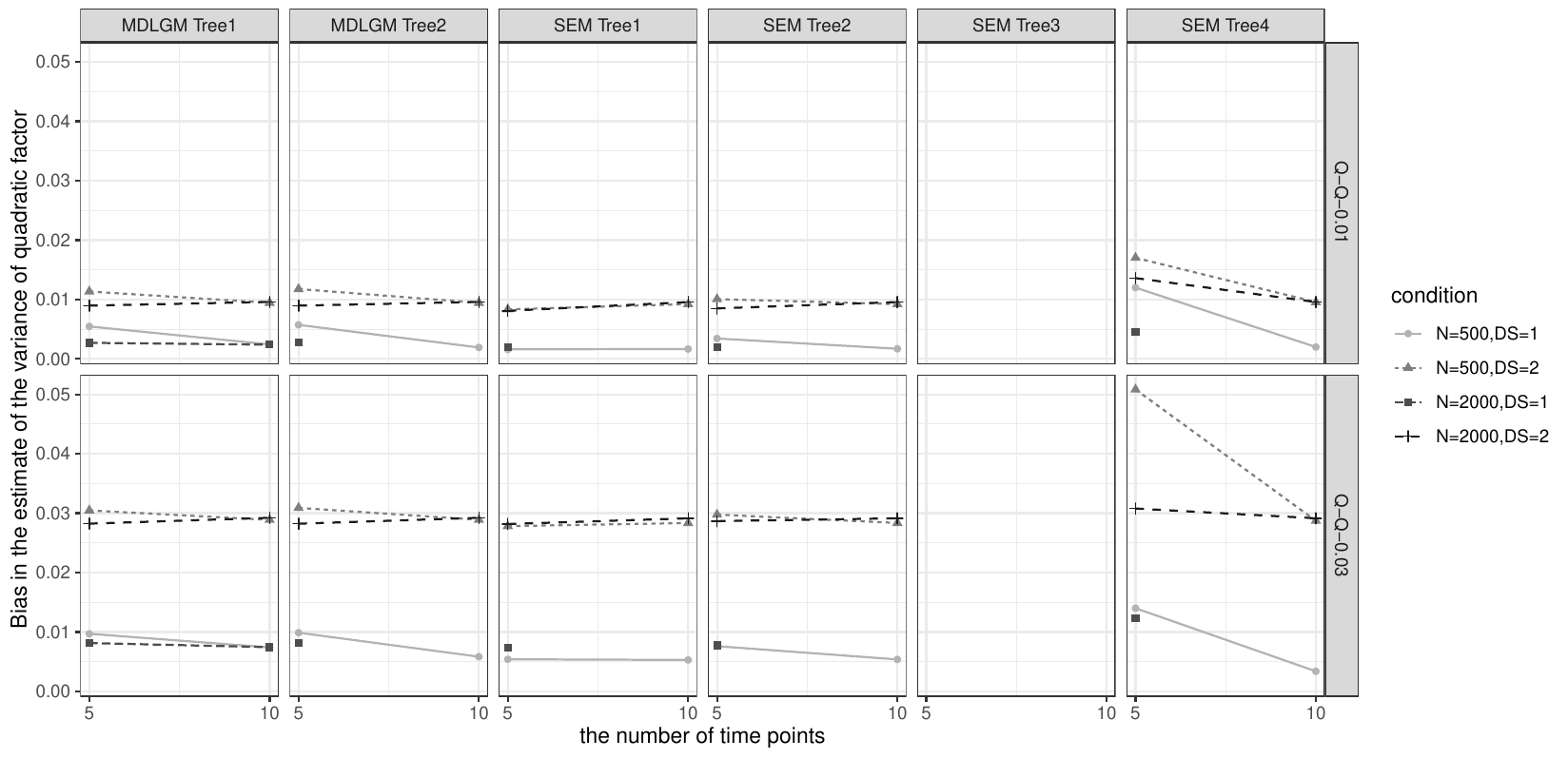}
    \caption{\\Bias in $\phi^2_Q$ calculated using the data when the number of terminal nodes was correctly estimated under conditions in which the true model and the template model were identical. SEM Tree1 denotes ML estimation, SEM Tree2 denotes constrained ML (CML) estimation, SEM Tree3 denotes Bayesian estimation, and SEM Tree4 denotes ML estimation with an algorithm that avoids node splitting based on any estimated model that produces a warning. MDLGM Tree1 and MDLGM Tree2 denote splitting based on the Mahalanobis distance and deviance, respectively. $DS$ denotes the degree of separation, and the three-element labels on the vertical axis (e.g., Q-L-0.01) indicate, in order, the true model (linear or quadratic), the template model (linear or quadratic), and the specified value of the slope factor variance. The bias of the SEM Tree3 could not be calculated. }
    \label{fig:bias_phi_q}
\end{figure}

\begin{figure}[h]
    \centering
    \includegraphics[width=0.9\linewidth]{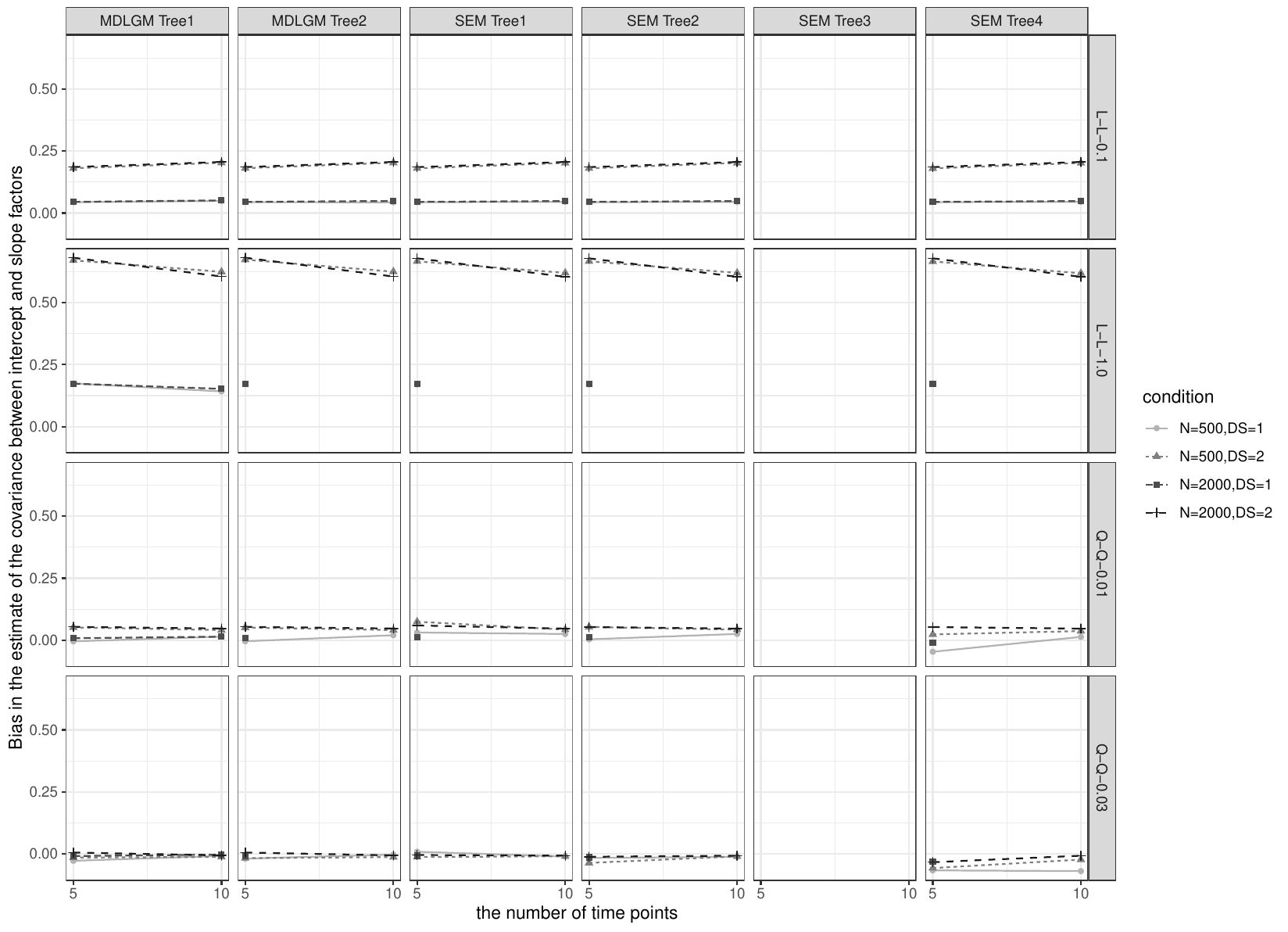}
    \caption{\\Bias in $\phi_{IS}$ calculated using the data when the number of terminal nodes was correctly estimated under conditions in which the true model and the template model were identical. SEM Tree1 denotes ML estimation, SEM Tree2 denotes constrained ML (CML) estimation, SEM Tree3 denotes Bayesian estimation, and SEM Tree4 denotes ML estimation with an algorithm that avoids node splitting based on any estimated model that produces a warning. MDLGM Tree1 and MDLGM Tree2 denote splitting based on the Mahalanobis distance and deviance, respectively. $DS$ denotes the degree of separation, and the three-element labels on the vertical axis (e.g., Q-L-0.01) indicate, in order, the true model (linear or quadratic), the template model (linear or quadratic), and the specified value of the slope factor variance. The bias of the SEM Tree3 could not be calculated. }
    \label{fig:bias_phi_is}
\end{figure}

\begin{figure}[h]
    \centering
    \includegraphics[width=0.9\linewidth]{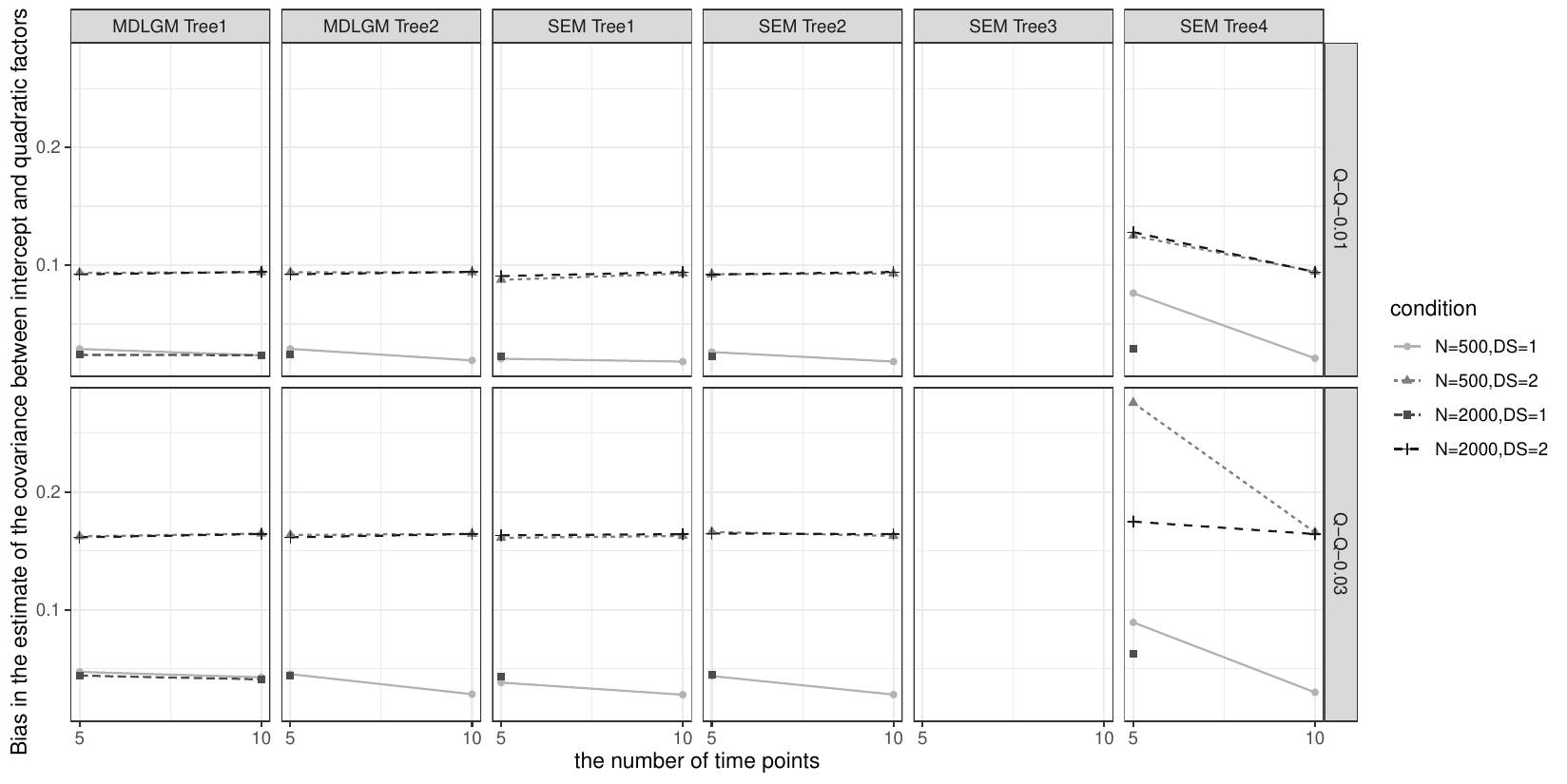}
    \caption{\\Bias in $\phi_{IQ}$ calculated using the data when the number of terminal nodes was correctly estimated under conditions in which the true model and the template model were identical. SEM Tree1 denotes ML estimation, SEM Tree2 denotes constrained ML (CML) estimation, SEM Tree3 denotes Bayesian estimation, and SEM Tree4 denotes ML estimation with an algorithm that avoids node splitting based on any estimated model that produces a warning. MDLGM Tree1 and MDLGM Tree2 denote splitting based on the Mahalanobis distance and deviance, respectively. $DS$ denotes the degree of separation, and the three-element labels on the vertical axis (e.g., Q-L-0.01) indicate, in order, the true model (linear or quadratic), the template model (linear or quadratic), and the specified value of the slope factor variance. The bias of the SEM Tree3 could not be calculated. }
    \label{fig:bias_phi_iq}
\end{figure}

\begin{figure}[h]
    \centering
    \includegraphics[width=0.9\linewidth]{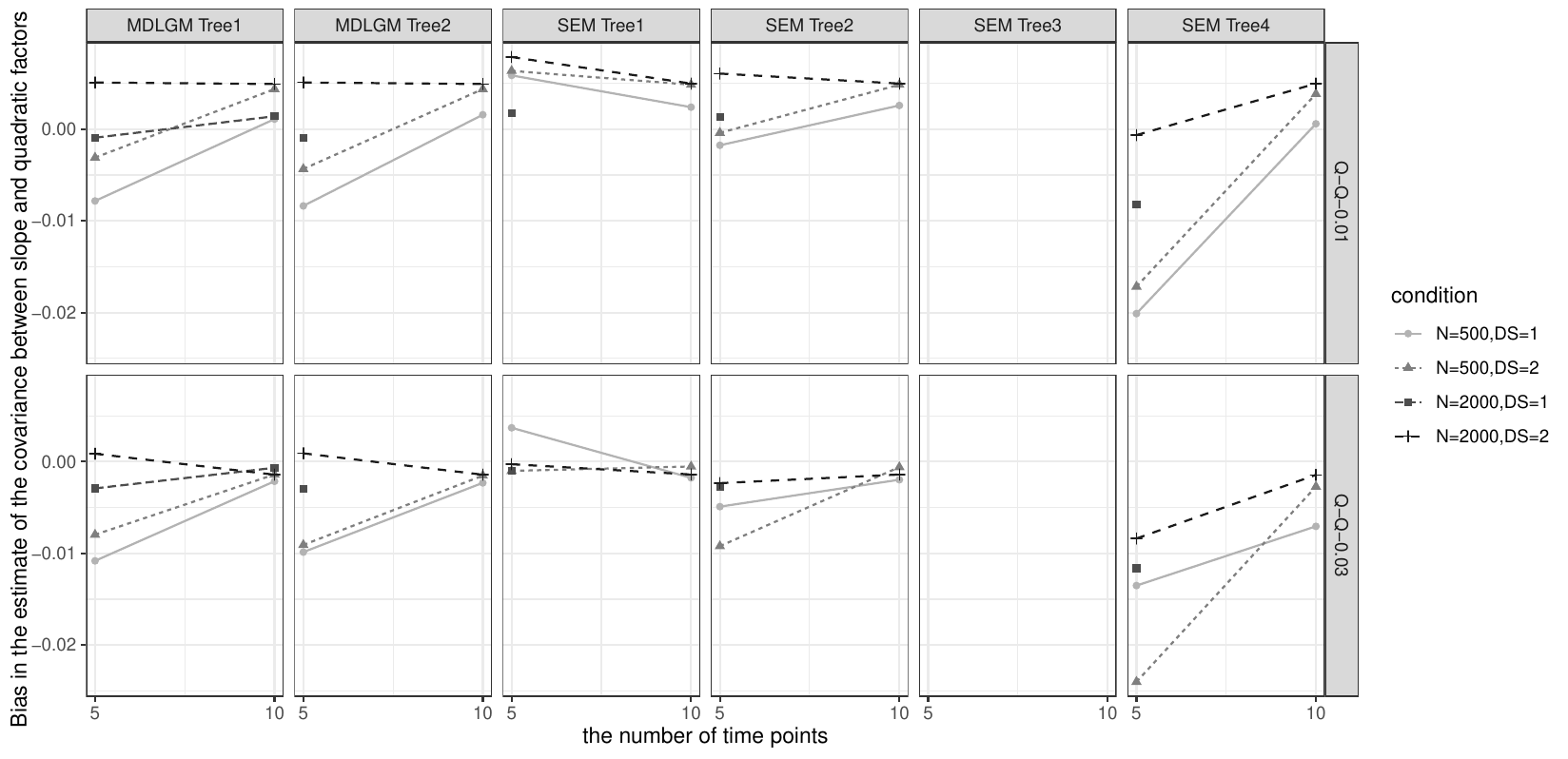}
    \caption{\\Bias in $\phi_{SQ}$ calculated using the data when the number of terminal nodes was correctly estimated under conditions in which the true model and the template model were identical. SEM Tree1 denotes ML estimation, SEM Tree2 denotes constrained ML (CML) estimation, SEM Tree3 denotes Bayesian estimation, and SEM Tree4 denotes ML estimation with an algorithm that avoids node splitting based on any estimated model that produces a warning. MDLGM Tree1 and MDLGM Tree2 denote splitting based on the Mahalanobis distance and deviance, respectively. $DS$ denotes the degree of separation, and the three-element labels on the vertical axis (e.g., Q-L-0.01) indicate, in order, the true model (linear or quadratic), the template model (linear or quadratic), and the specified value of the slope factor variance. The bias of the SEM Tree3 could not be calculated. }
    \label{fig:bias_phi_sq}
\end{figure}

\begin{figure}[h]
    \centering
    \includegraphics[width=0.9\linewidth]{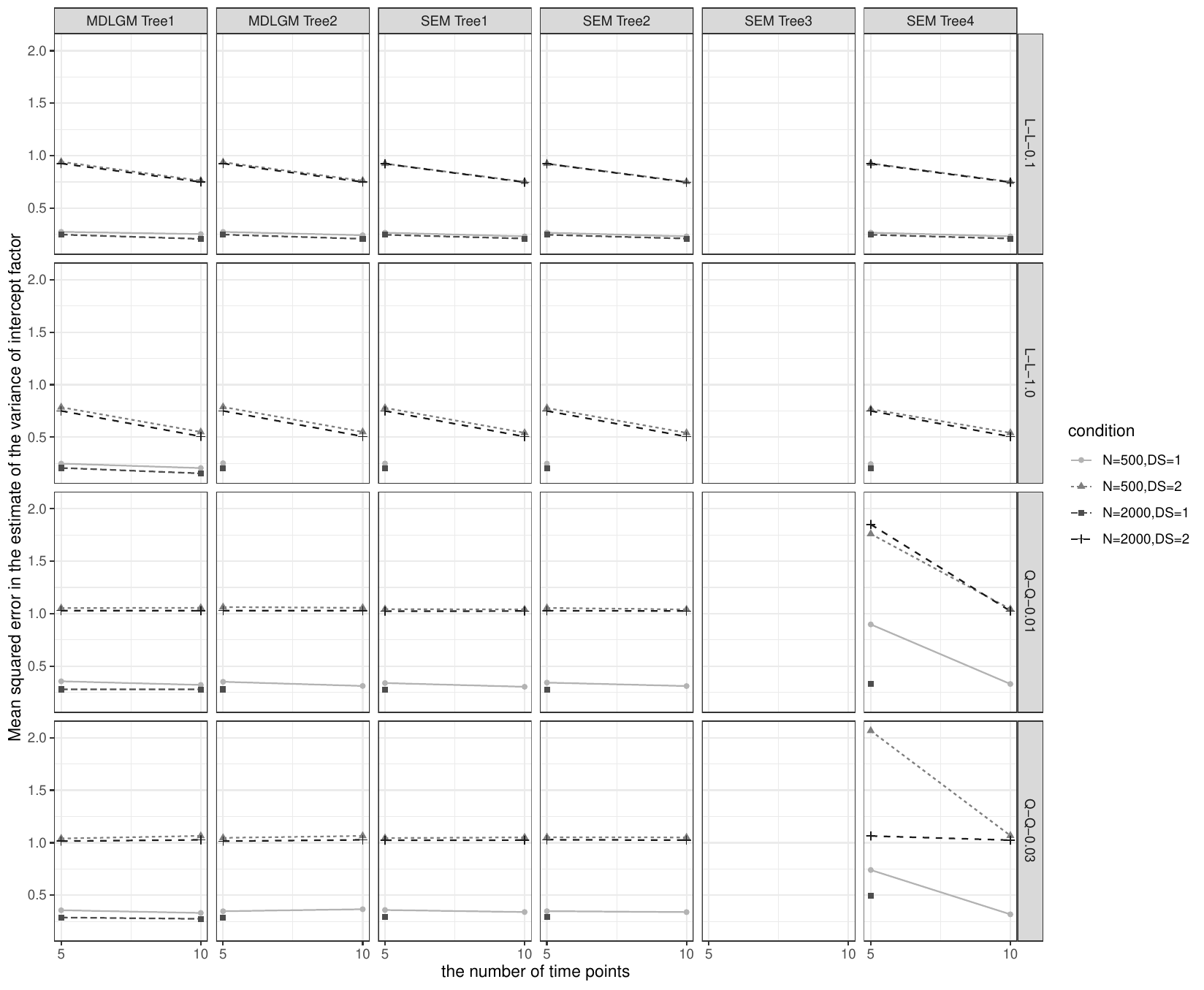}
    \caption{\\RMSE in $\phi^2_{I}$ calculated using the data when the number of terminal nodes was correctly estimated under conditions in which the true model and the template model were identical. SEM Tree1 denotes ML estimation, SEM Tree2 denotes constrained ML (CML) estimation, SEM Tree3 denotes Bayesian estimation, and SEM Tree4 denotes ML estimation with an algorithm that avoids node splitting based on any estimated model that produces a warning. MDLGM Tree1 and MDLGM Tree2 denote splitting based on the Mahalanobis distance and deviance, respectively. $DS$ denotes the degree of separation, and the three-element labels on the vertical axis (e.g., Q-L-0.01) indicate, in order, the true model (linear or quadratic), the template model (linear or quadratic), and the specified value of the slope factor variance. The RMSE of the SEM Tree3 could not be calculated. }
    \label{fig:rmse_phi_i}
\end{figure}

\begin{figure}[h]
    \centering
    \includegraphics[width=0.9\linewidth]{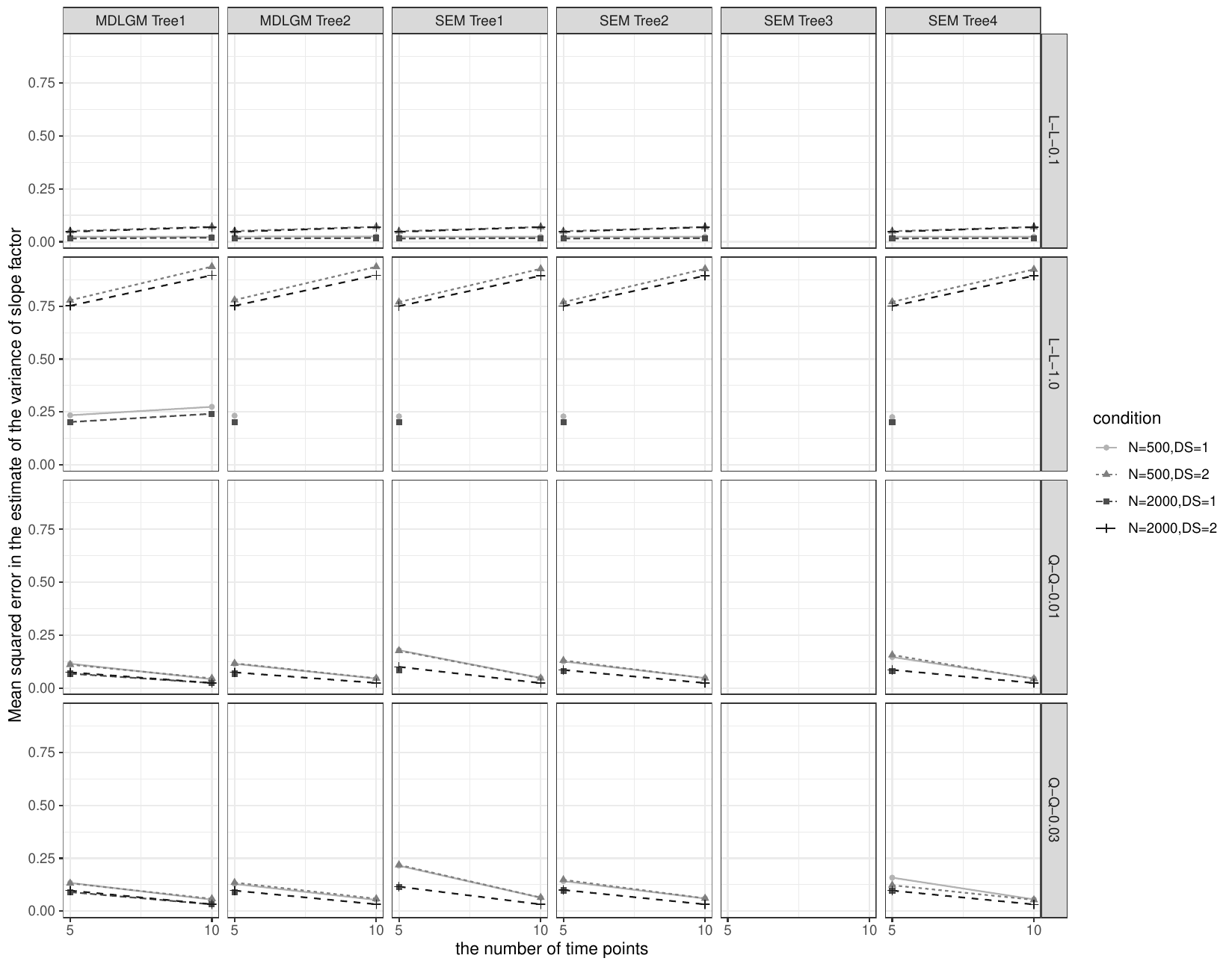}
    \caption{\\RMSE in $\phi^2_{S}$ calculated using the data when the number of terminal nodes was correctly estimated under conditions in which the true model and the template model were identical. SEM Tree1 denotes ML estimation, SEM Tree2 denotes constrained ML (CML) estimation, SEM Tree3 denotes Bayesian estimation, and SEM Tree4 denotes ML estimation with an algorithm that avoids node splitting based on any estimated model that produces a warning. MDLGM Tree1 and MDLGM Tree2 denote splitting based on the Mahalanobis distance and deviance, respectively. $DS$ denotes the degree of separation, and the three-element labels on the vertical axis (e.g., Q-L-0.01) indicate, in order, the true model (linear or quadratic), the template model (linear or quadratic), and the specified value of the slope factor variance. The RMSE of the SEM Tree3 could not be calculated. }
    \label{fig:rmse_phi_s}
\end{figure}

\begin{figure}[h]
    \centering
    \includegraphics[width=0.9\linewidth]{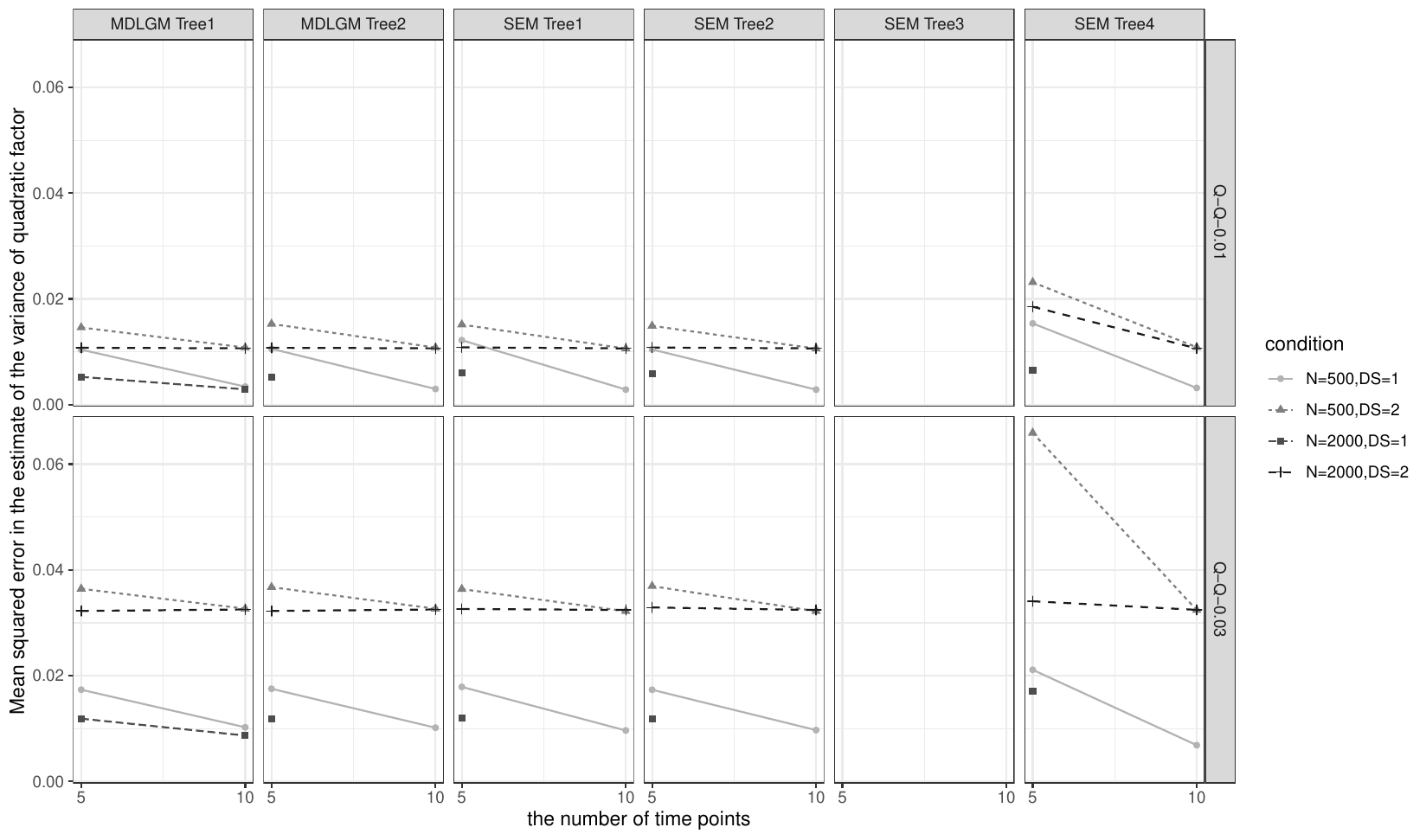}
    \caption{\\RMSE in $\phi^2_{Q}$ calculated using the data when the number of terminal nodes was correctly estimated under conditions in which the true model and the template model were identical. SEM Tree1 denotes ML estimation, SEM Tree2 denotes constrained ML (CML) estimation, SEM Tree3 denotes Bayesian estimation, and SEM Tree4 denotes ML estimation with an algorithm that avoids node splitting based on any estimated model that produces a warning. MDLGM Tree1 and MDLGM Tree2 denote splitting based on the Mahalanobis distance and deviance, respectively. $DS$ denotes the degree of separation, and the three-element labels on the vertical axis (e.g., Q-L-0.01) indicate, in order, the true model (linear or quadratic), the template model (linear or quadratic), and the specified value of the slope factor variance. The RMSE of the SEM Tree3 could not be calculated. }
    \label{fig:rmse_phi_q}
\end{figure}

\begin{figure}[h]
    \centering
    \includegraphics[width=0.9\linewidth]{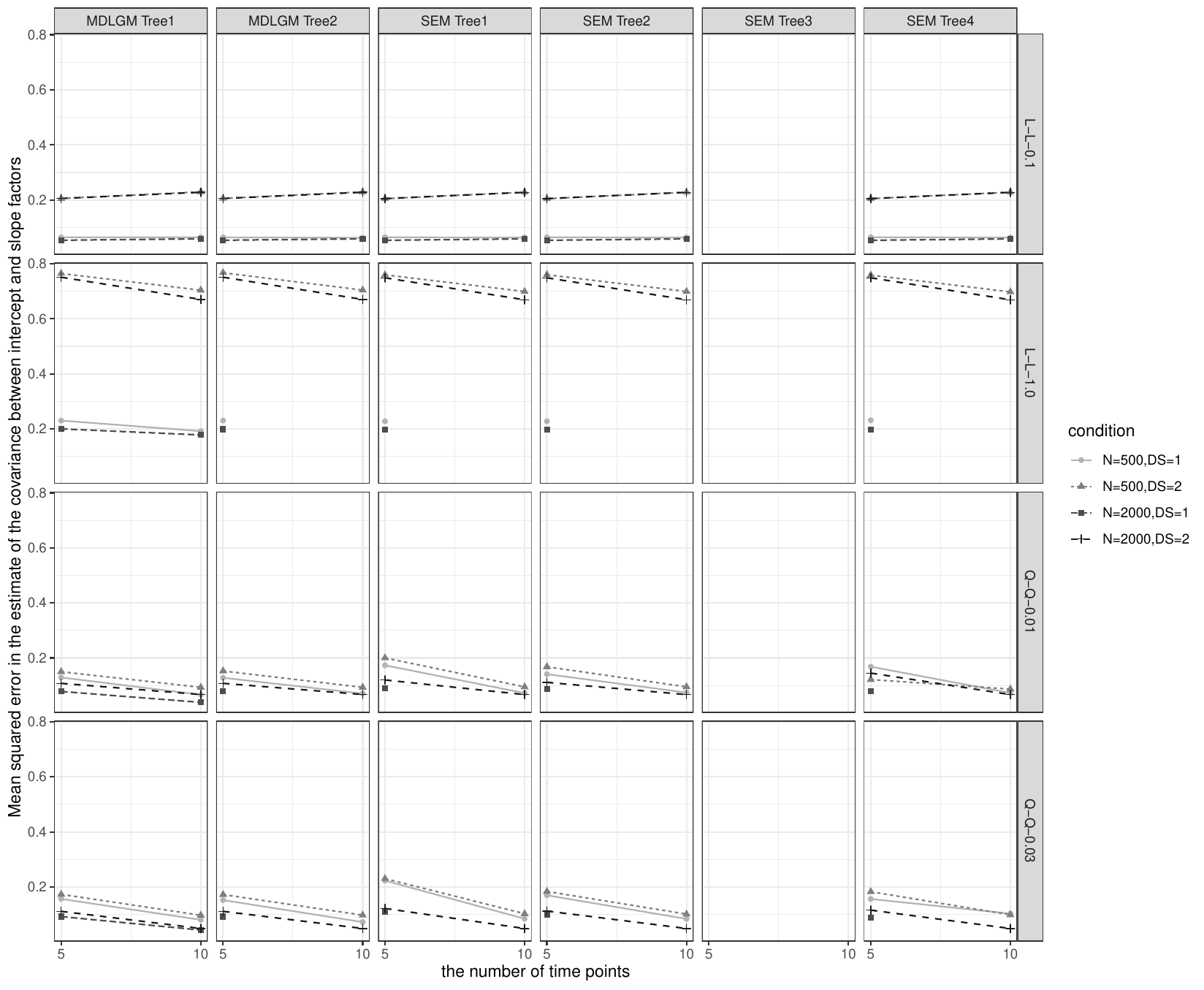}
    \caption{\\RMSE in $\phi_{IS}$ calculated using the data when the number of terminal nodes was correctly estimated under conditions in which the true model and the template model were identical. SEM Tree1 denotes ML estimation, SEM Tree2 denotes constrained ML (CML) estimation, SEM Tree3 denotes Bayesian estimation, and SEM Tree4 denotes ML estimation with an algorithm that avoids node splitting based on any estimated model that produces a warning. MDLGM Tree1 and MDLGM Tree2 denote splitting based on the Mahalanobis distance and deviance, respectively. $DS$ denotes the degree of separation, and the three-element labels on the vertical axis (e.g., Q-L-0.01) indicate, in order, the true model (linear or quadratic), the template model (linear or quadratic), and the specified value of the slope factor variance. The RMSE of the SEM Tree3 could not be calculated. }
    \label{fig:rmse_phi_is}
\end{figure}

\begin{figure}[h]
    \centering
    \includegraphics[width=0.9\linewidth]{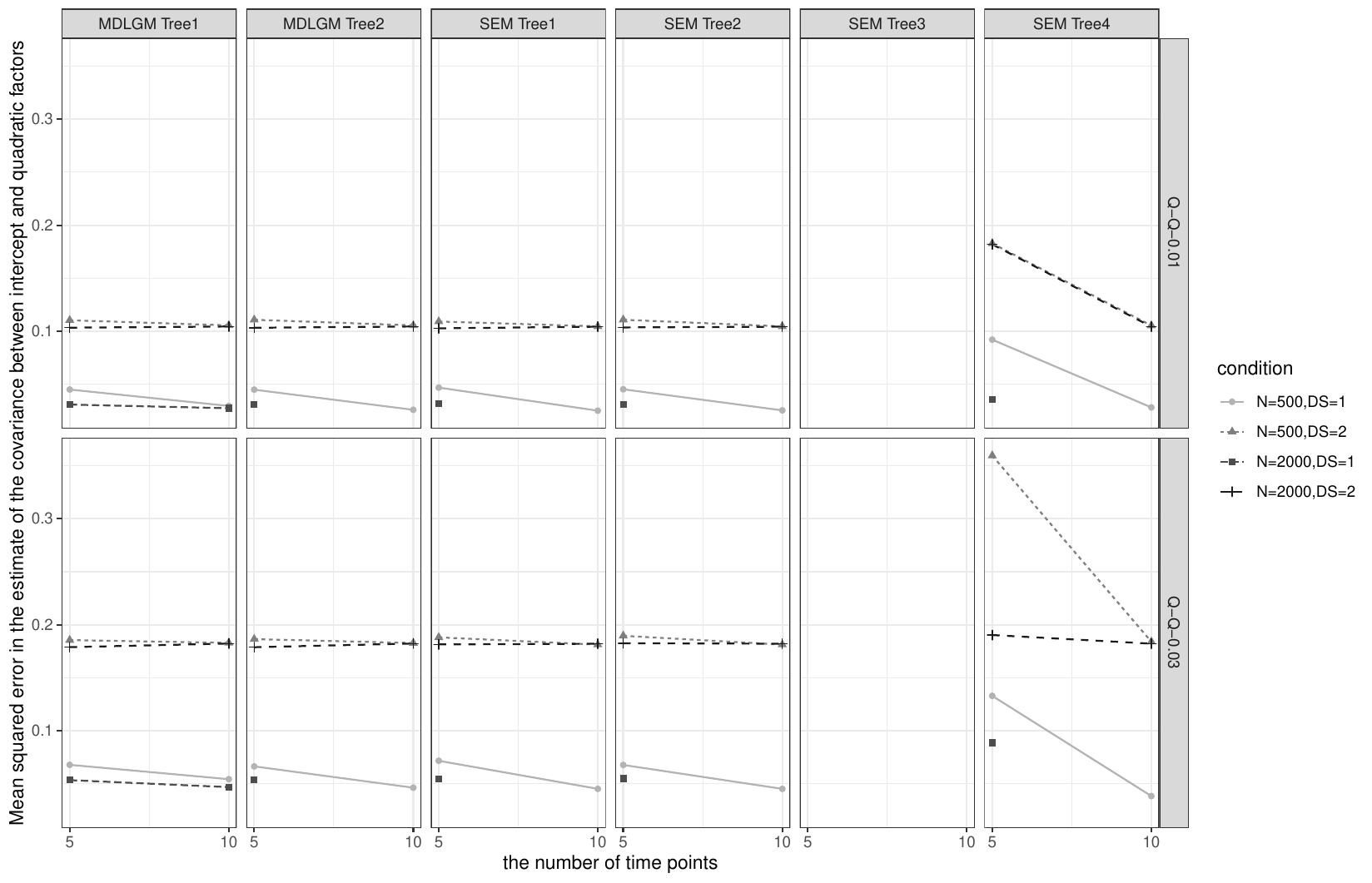}
    \caption{\\RMSE in $\phi_{IQ}$ calculated using the data when the number of terminal nodes was correctly estimated under conditions in which the true model and the template model were identical. SEM Tree1 denotes ML estimation, SEM Tree2 denotes constrained ML (CML) estimation, SEM Tree3 denotes Bayesian estimation, and SEM Tree4 denotes ML estimation with an algorithm that avoids node splitting based on any estimated model that produces a warning. MDLGM Tree1 and MDLGM Tree2 denote splitting based on the Mahalanobis distance and deviance, respectively. $DS$ denotes the degree of separation, and the three-element labels on the vertical axis (e.g., Q-L-0.01) indicate, in order, the true model (linear or quadratic), the template model (linear or quadratic), and the specified value of the slope factor variance. The RMSE of the SEM Tree3 could not be calculated. }
    \label{fig:rmse_phi_iq}
\end{figure}

\begin{figure}[h]
    \centering
    \includegraphics[width=0.9\linewidth]{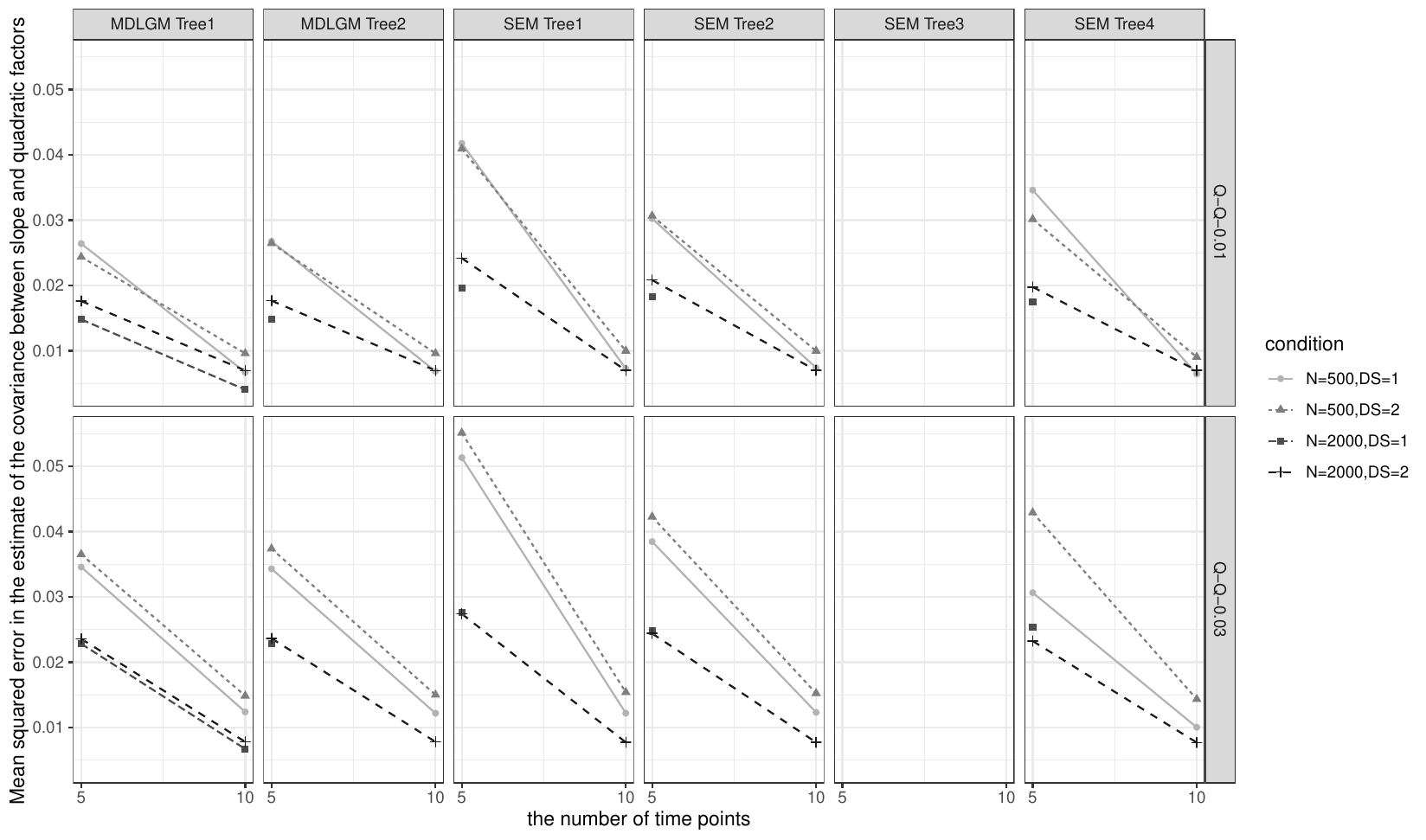}
    \caption{\\RMSE in $\phi_{SQ}$ calculated using the data when the number of terminal nodes was correctly estimated under conditions in which the true model and the template model were identical. SEM Tree1 denotes ML estimation, SEM Tree2 denotes constrained ML (CML) estimation, SEM Tree3 denotes Bayesian estimation, and SEM Tree4 denotes ML estimation with an algorithm that avoids node splitting based on any estimated model that produces a warning. MDLGM Tree1 and MDLGM Tree2 denote splitting based on the Mahalanobis distance and deviance, respectively. $DS$ denotes the degree of separation, and the three-element labels on the vertical axis (e.g., Q-L-0.01) indicate, in order, the true model (linear or quadratic), the template model (linear or quadratic), and the specified value of the slope factor variance. The RMSE of the SEM Tree3 could not be calculated. }
    \label{fig:rmse_phi_sq}
\end{figure}

Figure~\ref{fig:bias_phi_i}-\ref{fig:bias_phi_sq} show the biases of $\phi^2_I$, $\phi^2_S$, $\phi^2_Q$, $\phi_{IS}$, $\phi_{IQ}$, and $\phi_{SQ}$, respectively and Figure~\ref{fig:rmse_phi_i}-\ref{fig:rmse_phi_sq} show the RMSEs of those in each method and condition. As mentioned in the main text, these figures show that all methods except the SEM Tree4 exhibited basically similar patterns and somewhat large positive biases and RMSEs were observed for some parameters.  

 \clearpage

 \clearpage

 \clearpage

 \clearpage

 \clearpage

 \clearpage

 \clearpage

 \clearpage

 \clearpage

 \clearpage

 \clearpage

 \clearpage

\subsubsection*{Bias and RMSE of variances of the residuals}
\begin{figure}[h]
    \centering
    \includegraphics[width=0.9\linewidth]{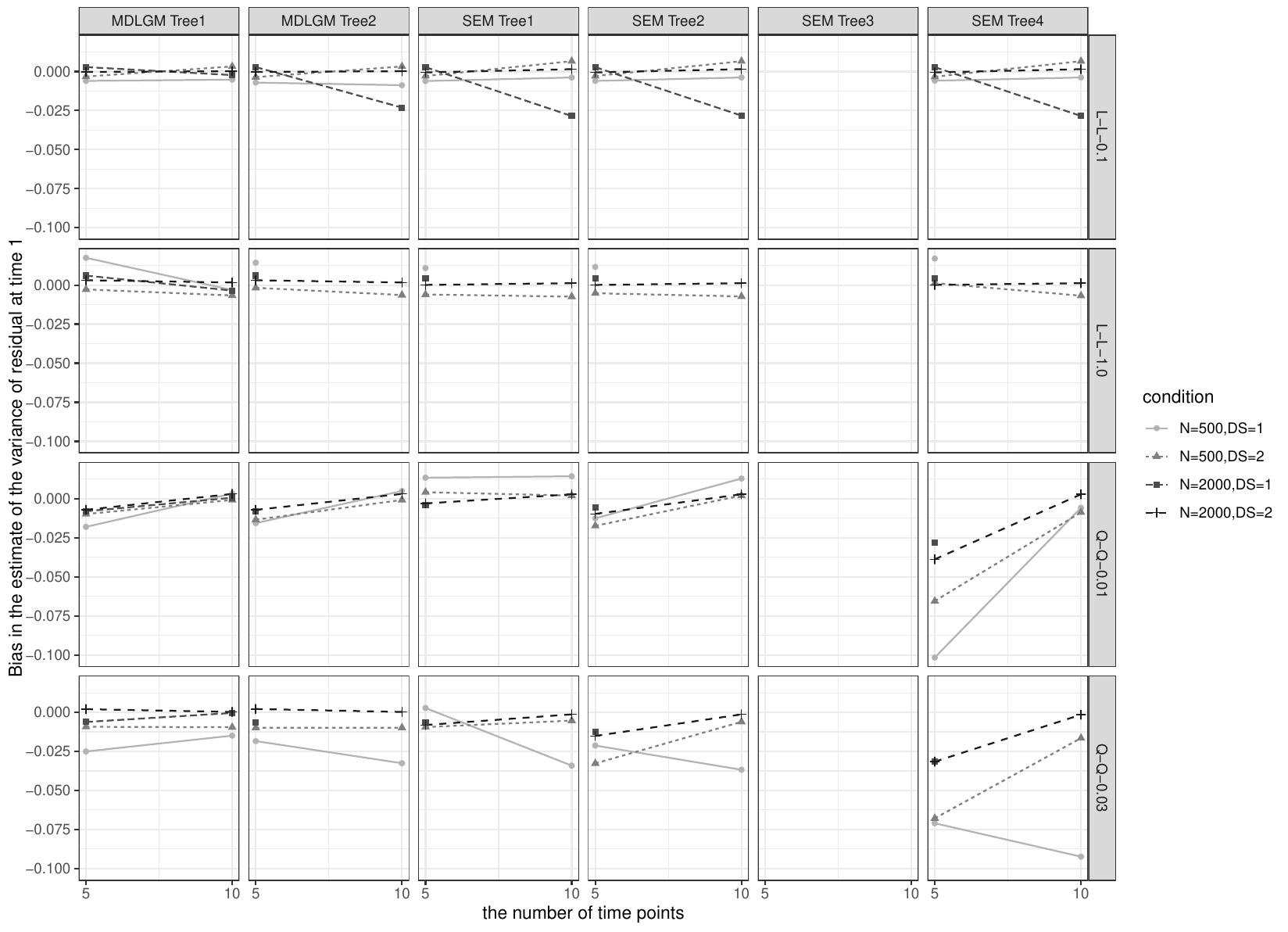}
    \caption{\\Bias in $\psi^2_1$ calculated using the data when the number of terminal nodes was correctly estimated under conditions in which the true model and the template model were identical. SEM Tree1 denotes ML estimation, SEM Tree2 denotes constrained ML (CML) estimation, SEM Tree3 denotes Bayesian estimation, and SEM Tree4 denotes ML estimation with an algorithm that avoids node splitting based on any estimated model that produces a warning. MDLGM Tree1 and MDLGM Tree2 denote splitting based on the Mahalanobis distance and deviance, respectively. $DS$ denotes the degree of separation, and the three-element labels on the vertical axis (e.g., Q-L-0.01) indicate, in order, the true model (linear or quadratic), the template model (linear or quadratic), and the specified value of the slope factor variance. The bias of the SEM Tree3 could not be calculated. }
    \label{fig:bias_psi_1}
\end{figure}

\begin{figure}[h]
    \centering
    \includegraphics[width=0.9\linewidth]{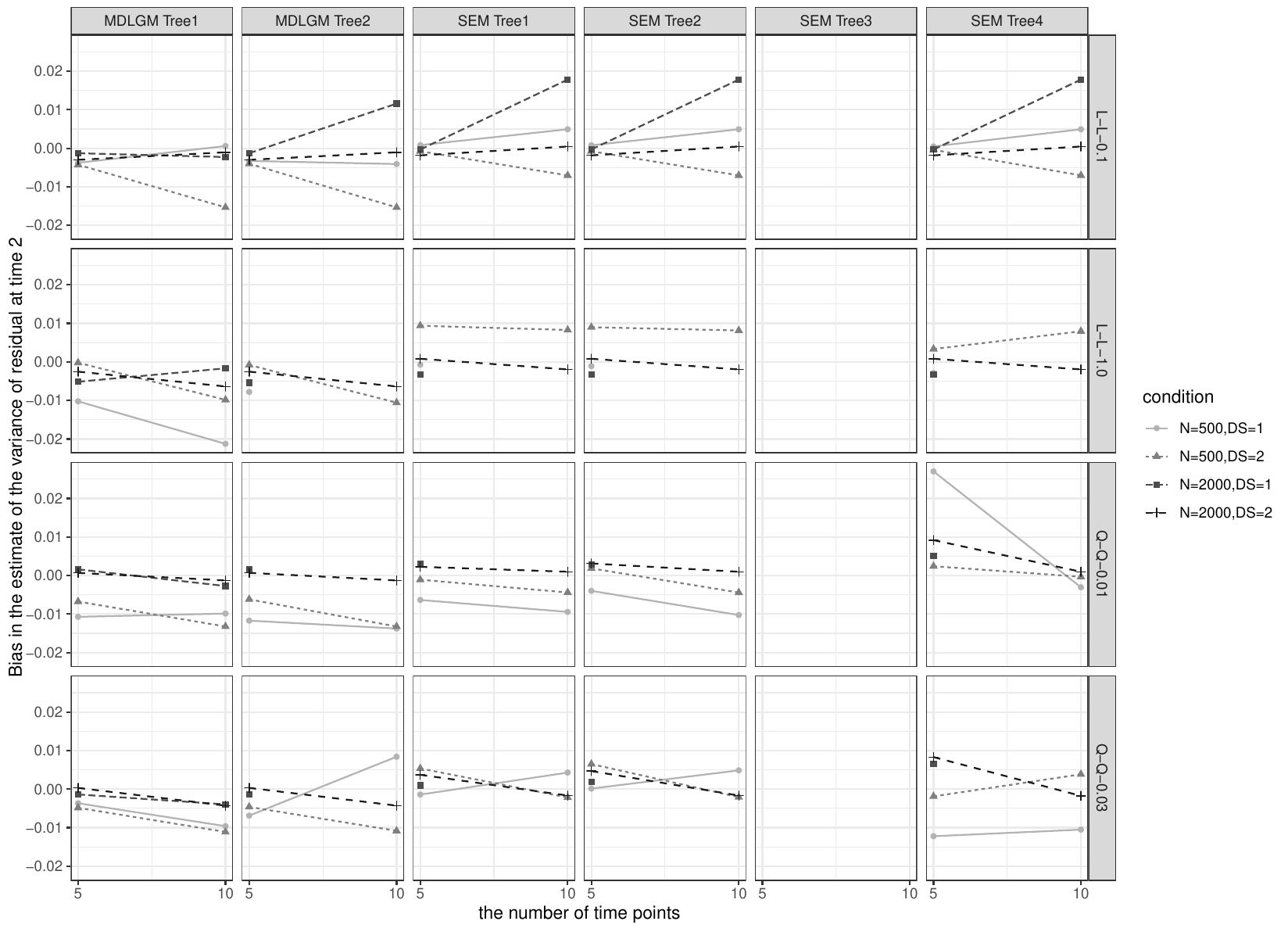}
    \caption{\\Bias in $\psi^2_2$ calculated using the data when the number of terminal nodes was correctly estimated under conditions in which the true model and the template model were identical. SEM Tree1 denotes ML estimation, SEM Tree2 denotes constrained ML (CML) estimation, SEM Tree3 denotes Bayesian estimation, and SEM Tree4 denotes ML estimation with an algorithm that avoids node splitting based on any estimated model that produces a warning. MDLGM Tree1 and MDLGM Tree2 denote splitting based on the Mahalanobis distance and deviance, respectively. $DS$ denotes the degree of separation, and the three-element labels on the vertical axis (e.g., Q-L-0.01) indicate, in order, the true model (linear or quadratic), the template model (linear or quadratic), and the specified value of the slope factor variance. The bias of the SEM Tree3 could not be calculated. }
    \label{fig:bias_psi_2}
\end{figure}

\begin{figure}[h]
    \centering
    \includegraphics[width=0.9\linewidth]{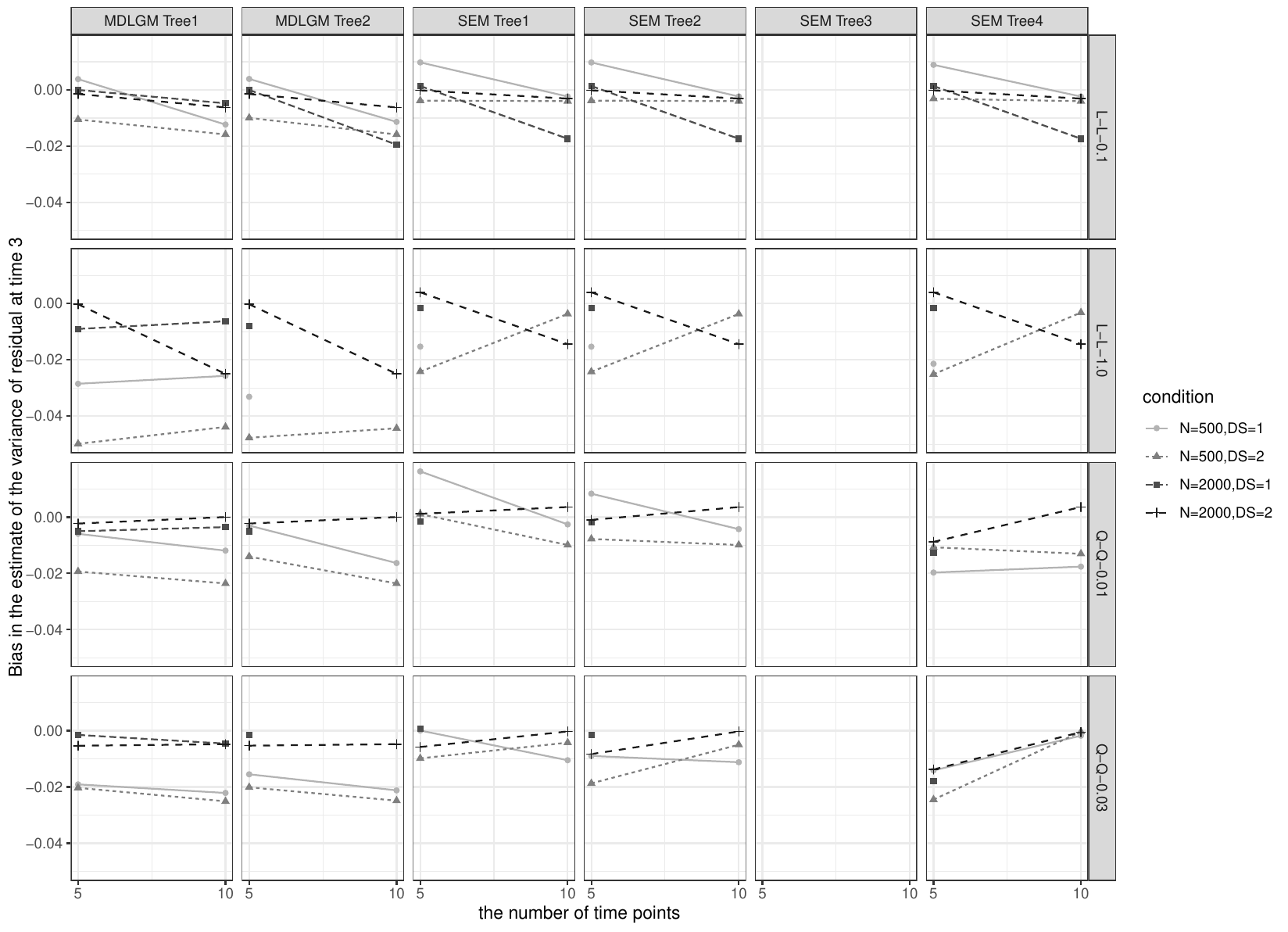}
    \caption{\\Bias in $\psi^2_3$ calculated using the data when the number of terminal nodes was correctly estimated under conditions in which the true model and the template model were identical. SEM Tree1 denotes ML estimation, SEM Tree2 denotes constrained ML (CML) estimation, SEM Tree3 denotes Bayesian estimation, and SEM Tree4 denotes ML estimation with an algorithm that avoids node splitting based on any estimated model that produces a warning. MDLGM Tree1 and MDLGM Tree2 denote splitting based on the Mahalanobis distance and deviance, respectively. $DS$ denotes the degree of separation, and the three-element labels on the vertical axis (e.g., Q-L-0.01) indicate, in order, the true model (linear or quadratic), the template model (linear or quadratic), and the specified value of the slope factor variance. The bias of the SEM Tree3 could not be calculated. }
    \label{fig:bias_psi_3}
\end{figure}

\begin{figure}[h]
    \centering
    \includegraphics[width=0.9\linewidth]{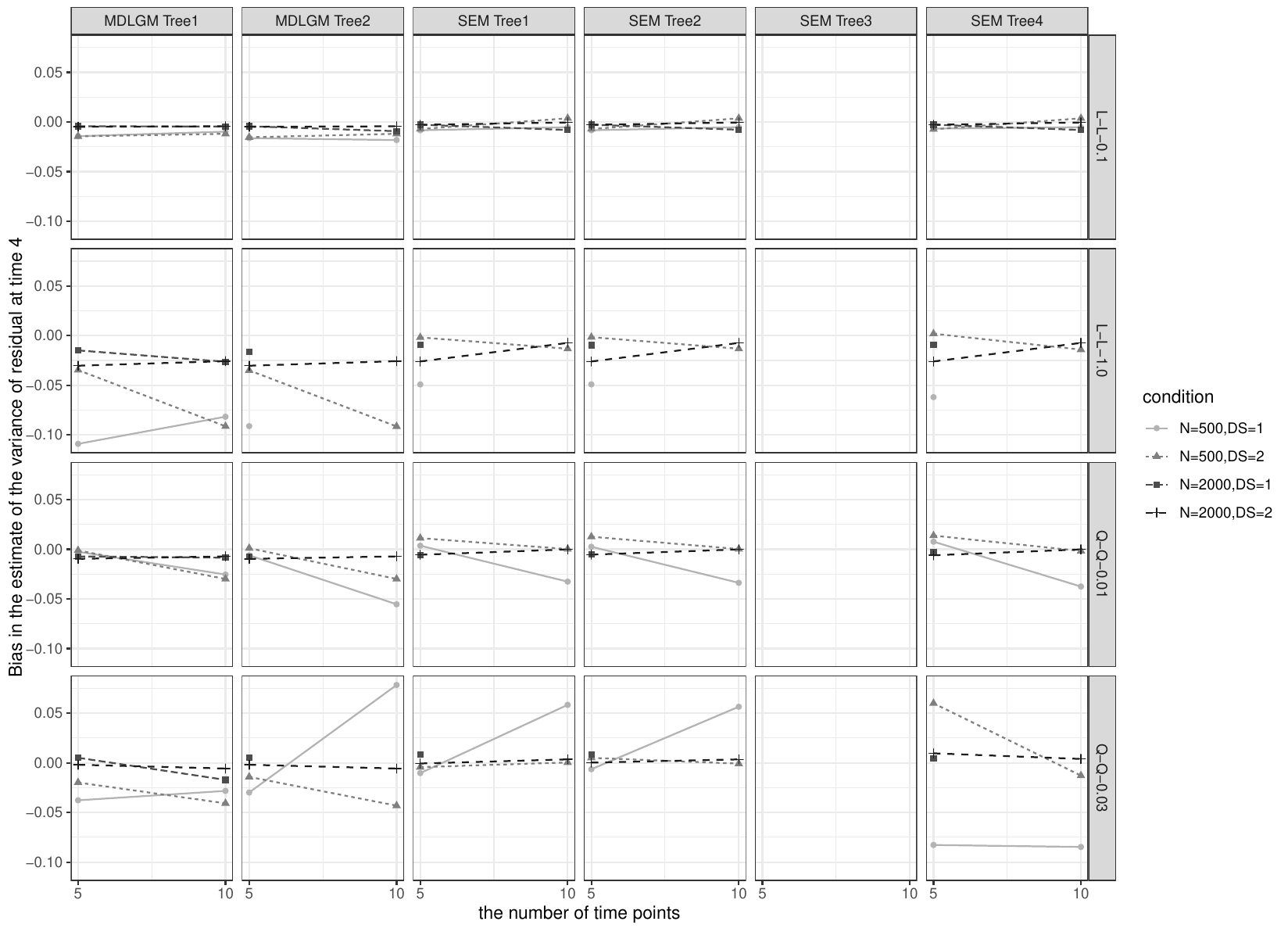}
    \caption{\\Bias in $\psi^2_4$ calculated using the data when the number of terminal nodes was correctly estimated under conditions in which the true model and the template model were identical. SEM Tree1 denotes ML estimation, SEM Tree2 denotes constrained ML (CML) estimation, SEM Tree3 denotes Bayesian estimation, and SEM Tree4 denotes ML estimation with an algorithm that avoids node splitting based on any estimated model that produces a warning. MDLGM Tree1 and MDLGM Tree2 denote splitting based on the Mahalanobis distance and deviance, respectively. $DS$ denotes the degree of separation, and the three-element labels on the vertical axis (e.g., Q-L-0.01) indicate, in order, the true model (linear or quadratic), the template model (linear or quadratic), and the specified value of the slope factor variance. The bias of the SEM Tree3 could not be calculated. }
    \label{fig:bias_psi_4}
\end{figure}

\begin{figure}[h]
    \centering
    \includegraphics[width=0.9\linewidth]{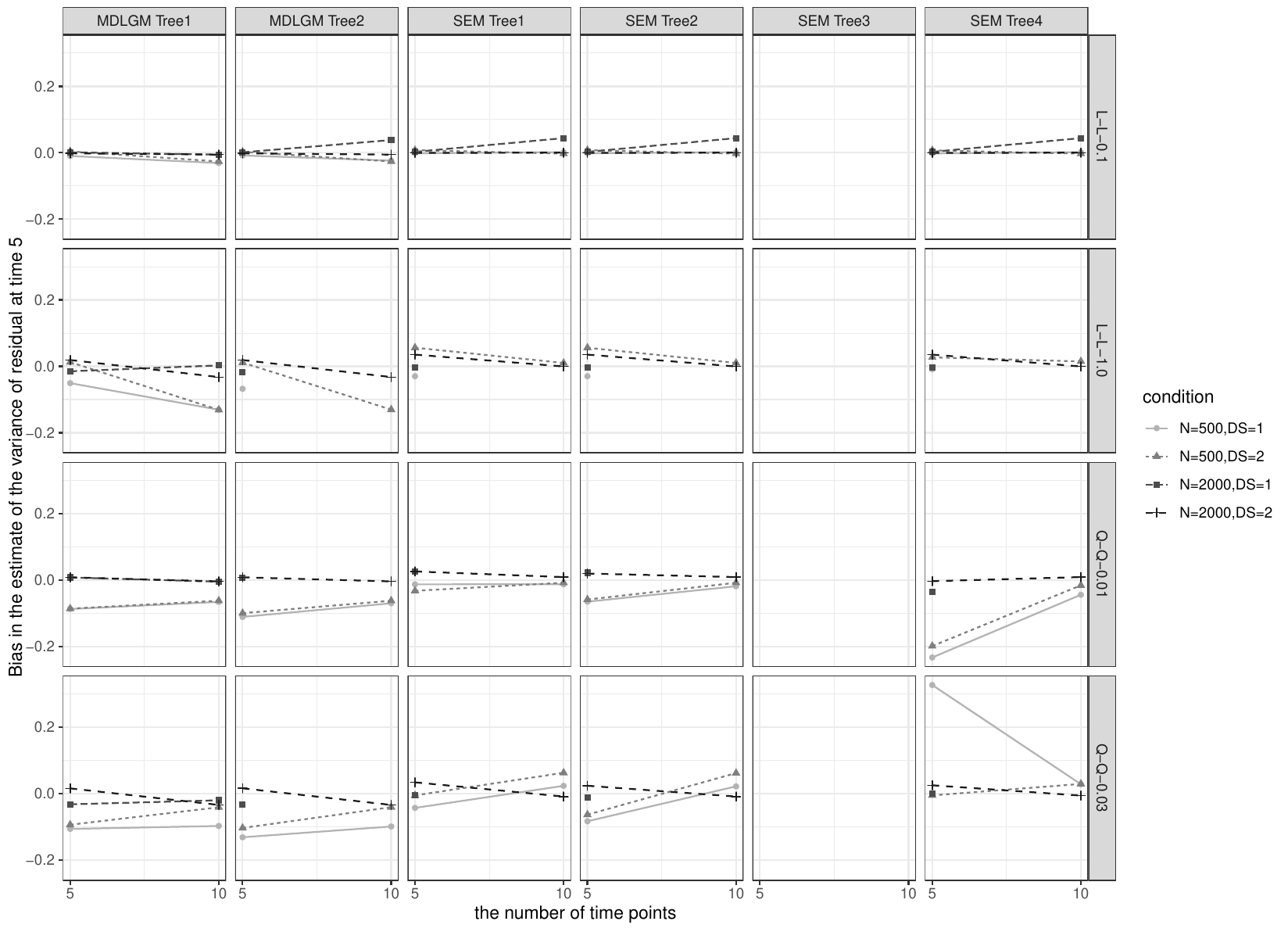}
    \caption{\\Bias in $\psi^2_5$ calculated using the data when the number of terminal nodes was correctly estimated under conditions in which the true model and the template model were identical. SEM Tree1 denotes ML estimation, SEM Tree2 denotes constrained ML (CML) estimation, SEM Tree3 denotes Bayesian estimation, and SEM Tree4 denotes ML estimation with an algorithm that avoids node splitting based on any estimated model that produces a warning. MDLGM Tree1 and MDLGM Tree2 denote splitting based on the Mahalanobis distance and deviance, respectively. $DS$ denotes the degree of separation, and the three-element labels on the vertical axis (e.g., Q-L-0.01) indicate, in order, the true model (linear or quadratic), the template model (linear or quadratic), and the specified value of the slope factor variance. The bias of the SEM Tree3 could not be calculated. }
    \label{fig:bias_psi_5}
\end{figure}

\begin{figure}[h]
    \centering
    \includegraphics[width=0.9\linewidth]{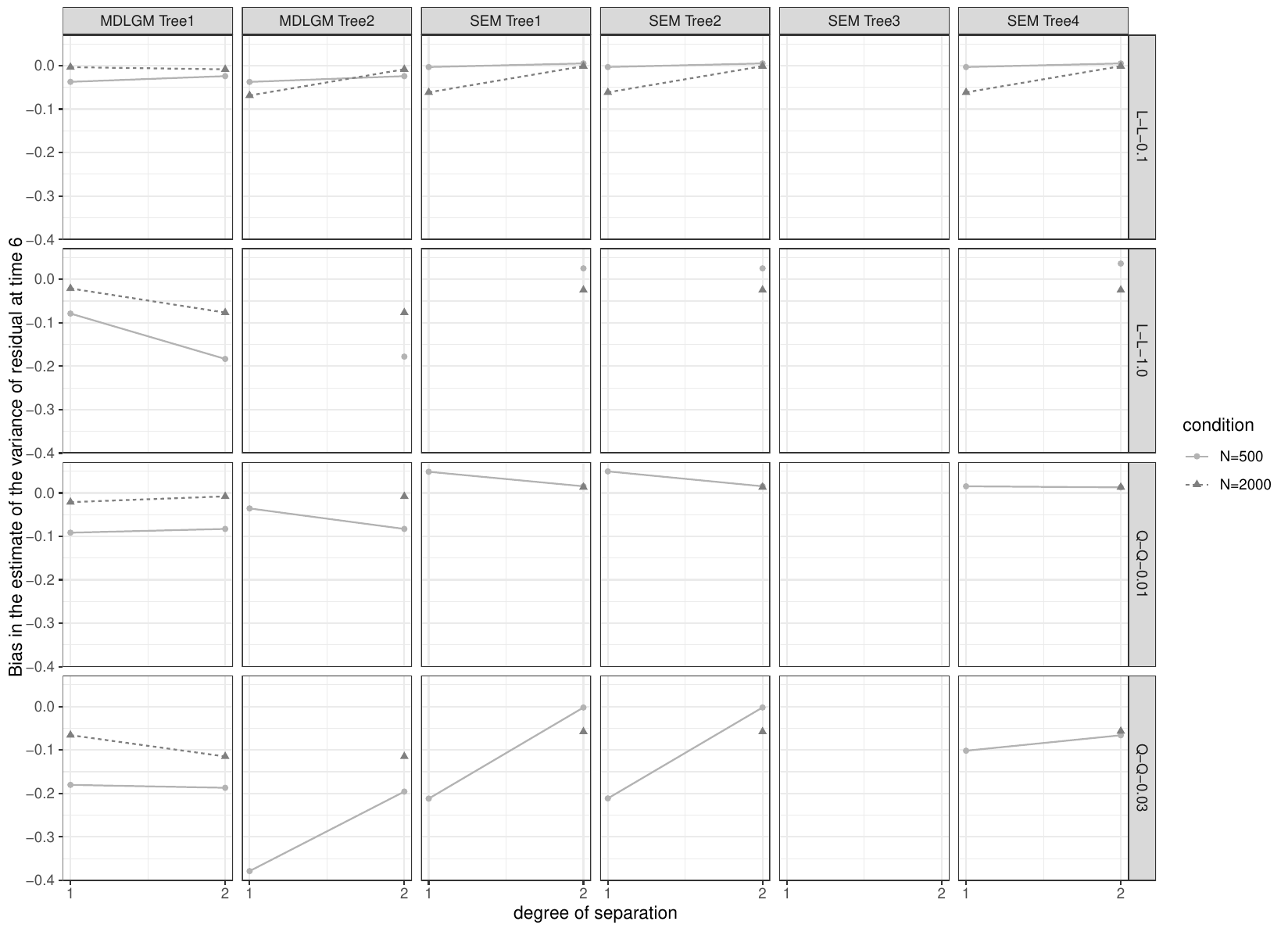}
    \caption{\\Bias in $\psi^2_6$ calculated using the data when the number of terminal nodes was correctly estimated under conditions in which the true model and the template model were identical. SEM Tree1 denotes ML estimation, SEM Tree2 denotes constrained ML (CML) estimation, SEM Tree3 denotes Bayesian estimation, and SEM Tree4 denotes ML estimation with an algorithm that avoids node splitting based on any estimated model that produces a warning. MDLGM Tree1 and MDLGM Tree2 denote splitting based on the Mahalanobis distance and deviance, respectively. The three-element labels on the vertical axis (e.g., Q-L-0.01) indicate, in order, the true model (linear or quadratic), the template model (linear or quadratic), and the specified value of the slope factor variance. The bias of the SEM Tree3 could not be calculated. }
    \label{fig:bias_psi_6}
\end{figure}

\begin{figure}[h]
    \centering
    \includegraphics[width=0.9\linewidth]{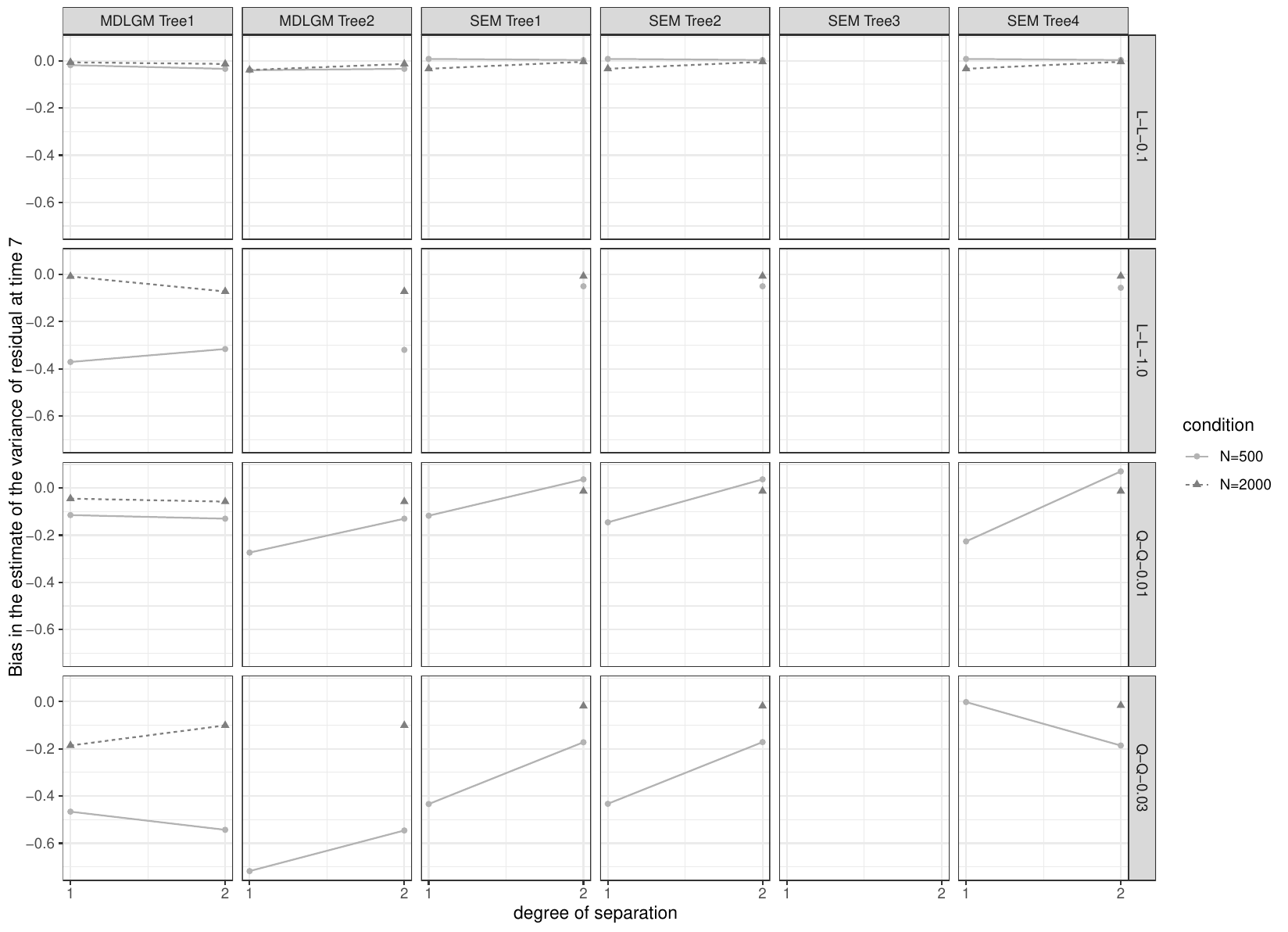}
    \caption{\\Bias in $\psi^2_7$ calculated using the data when the number of terminal nodes was correctly estimated under conditions in which the true model and the template model were identical. SEM Tree1 denotes ML estimation, SEM Tree2 denotes constrained ML (CML) estimation, SEM Tree3 denotes Bayesian estimation, and SEM Tree4 denotes ML estimation with an algorithm that avoids node splitting based on any estimated model that produces a warning. MDLGM Tree1 and MDLGM Tree2 denote splitting based on the Mahalanobis distance and deviance, respectively. The three-element labels on the vertical axis (e.g., Q-L-0.01) indicate, in order, the true model (linear or quadratic), the template model (linear or quadratic), and the specified value of the slope factor variance. The bias of the SEM Tree3 could not be calculated. }
    \label{fig:bias_psi_7}
\end{figure}

\begin{figure}[h]
    \centering
    \includegraphics[width=0.9\linewidth]{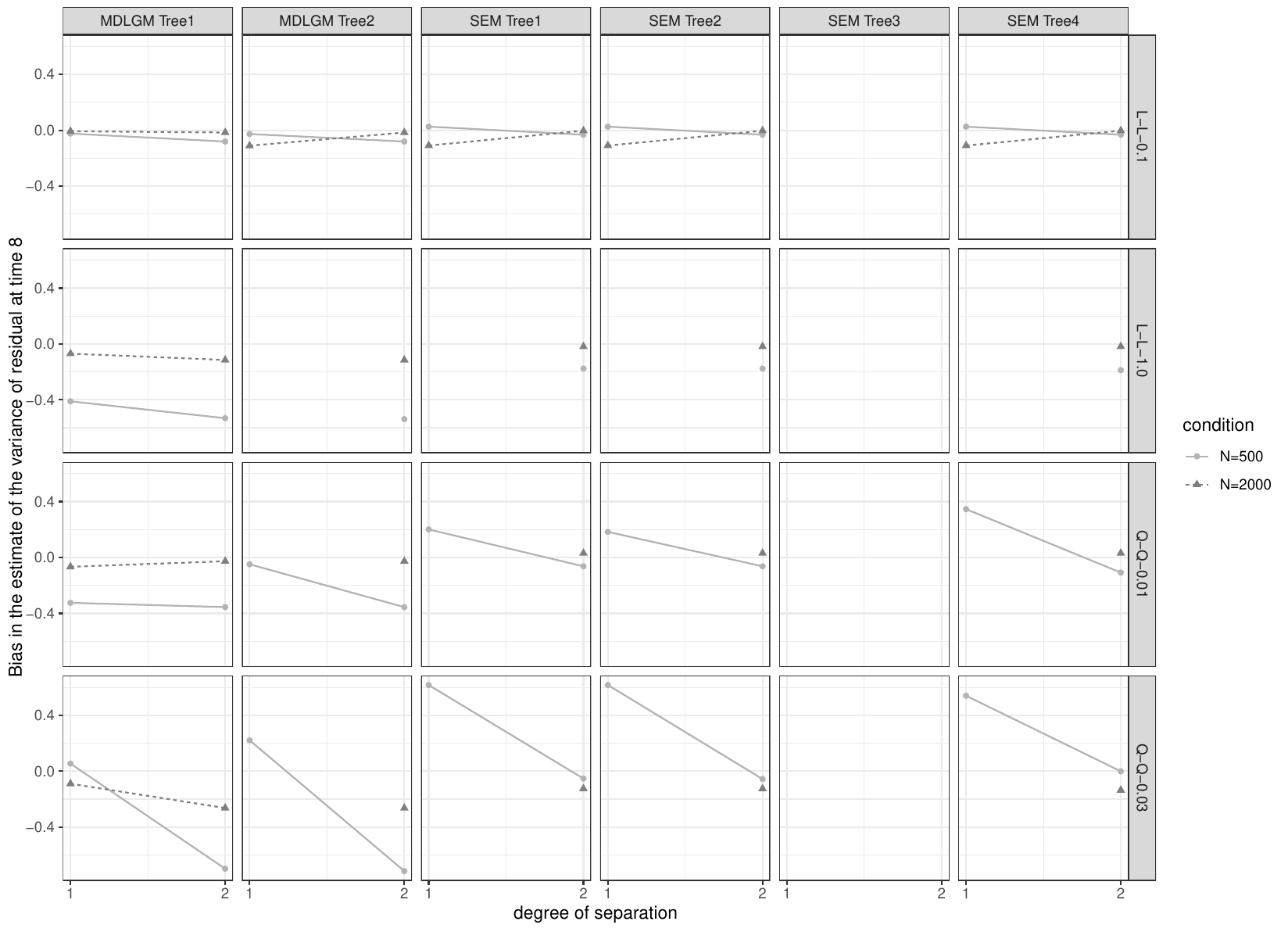}
    \caption{\\Bias in $\psi^2_8$ calculated using the data when the number of terminal nodes was correctly estimated under conditions in which the true model and the template model were identical. SEM Tree1 denotes ML estimation, SEM Tree2 denotes constrained ML (CML) estimation, SEM Tree3 denotes Bayesian estimation, and SEM Tree4 denotes ML estimation with an algorithm that avoids node splitting based on any estimated model that produces a warning. MDLGM Tree1 and MDLGM Tree2 denote splitting based on the Mahalanobis distance and deviance, respectively. The three-element labels on the vertical axis (e.g., Q-L-0.01) indicate, in order, the true model (linear or quadratic), the template model (linear or quadratic), and the specified value of the slope factor variance. The bias of the SEM Tree3 could not be calculated. }
    \label{fig:bias_psi_8}
\end{figure}

\begin{figure}[h]
    \centering
    \includegraphics[width=0.9\linewidth]{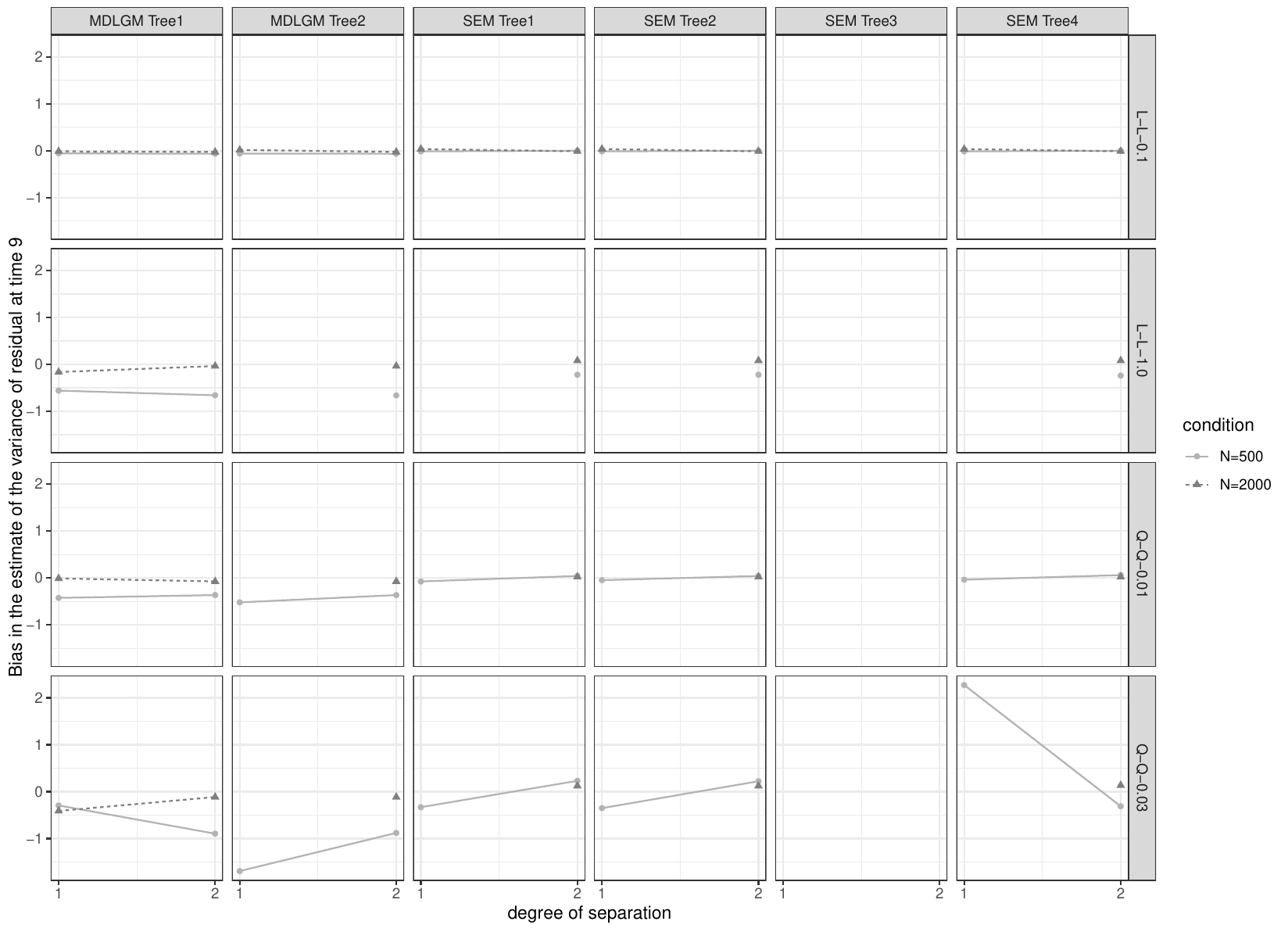}
    \caption{\\Bias in $\psi^2_9$ calculated using the data when the number of terminal nodes was correctly estimated under conditions in which the true model and the template model were identical. SEM Tree1 denotes ML estimation, SEM Tree2 denotes constrained ML (CML) estimation, SEM Tree3 denotes Bayesian estimation, and SEM Tree4 denotes ML estimation with an algorithm that avoids node splitting based on any estimated model that produces a warning. MDLGM Tree1 and MDLGM Tree2 denote splitting based on the Mahalanobis distance and deviance, respectively. The three-element labels on the vertical axis (e.g., Q-L-0.01) indicate, in order, the true model (linear or quadratic), the template model (linear or quadratic), and the specified value of the slope factor variance. The bias of the SEM Tree3 could not be calculated. }
    \label{fig:bias_psi_9}
\end{figure}

\begin{figure}[h]
    \centering
    \includegraphics[width=0.9\linewidth]{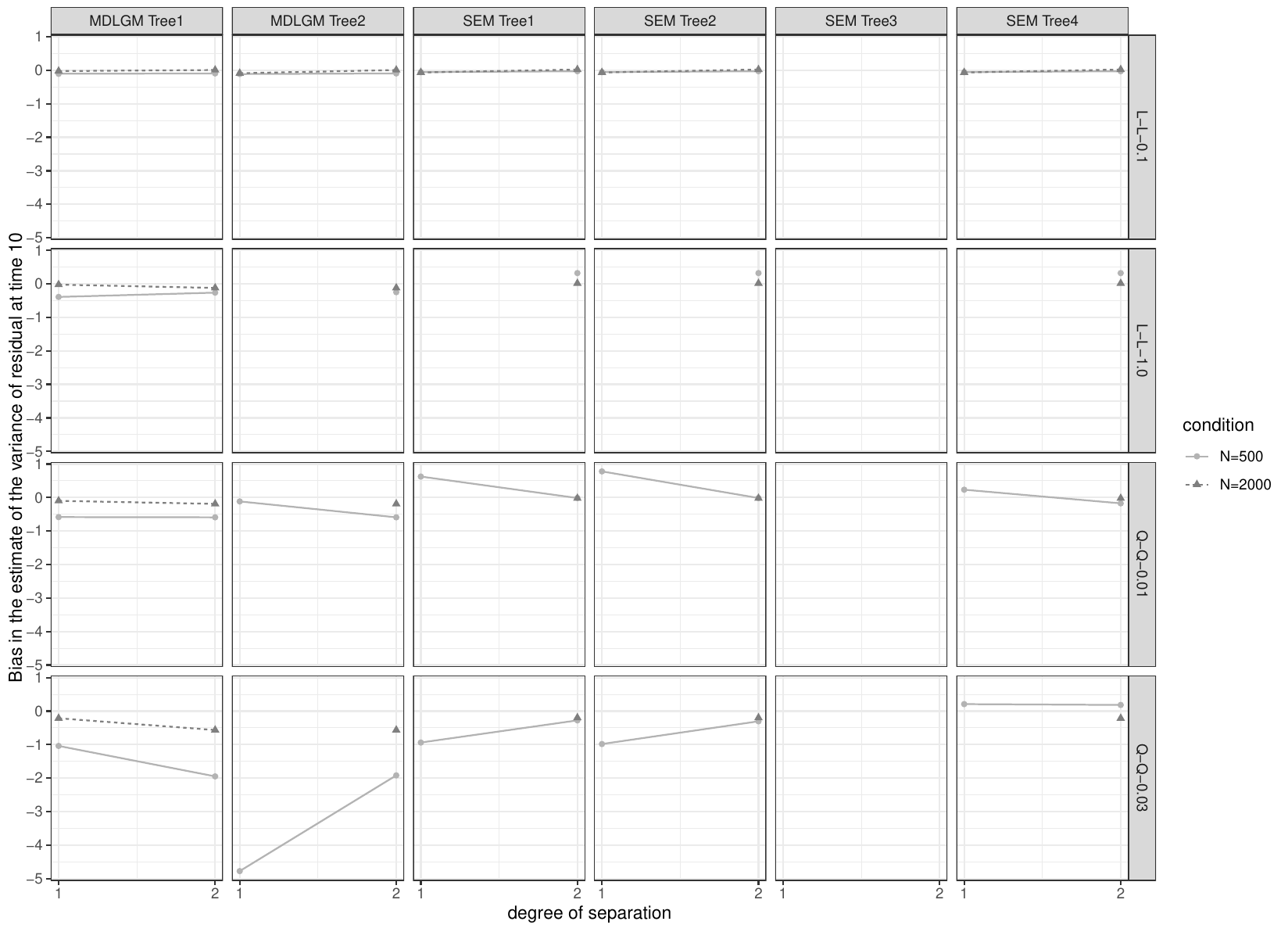}
    \caption{\\Bias in $\psi^2_{10}$ calculated using the data when the number of terminal nodes was correctly estimated under conditions in which the true model and the template model were identical. SEM Tree1 denotes ML estimation, SEM Tree2 denotes constrained ML (CML) estimation, SEM Tree3 denotes Bayesian estimation, and SEM Tree4 denotes ML estimation with an algorithm that avoids node splitting based on any estimated model that produces a warning. MDLGM Tree1 and MDLGM Tree2 denote splitting based on the Mahalanobis distance and deviance, respectively. The three-element labels on the vertical axis (e.g., Q-L-0.01) indicate, in order, the true model (linear or quadratic), the template model (linear or quadratic), and the specified value of the slope factor variance. The RMSE of the SEM Tree3 could not be calculated. }
    \label{fig:bias_psi_10}
\end{figure}

\begin{figure}[h]
    \centering
    \includegraphics[width=0.9\linewidth]{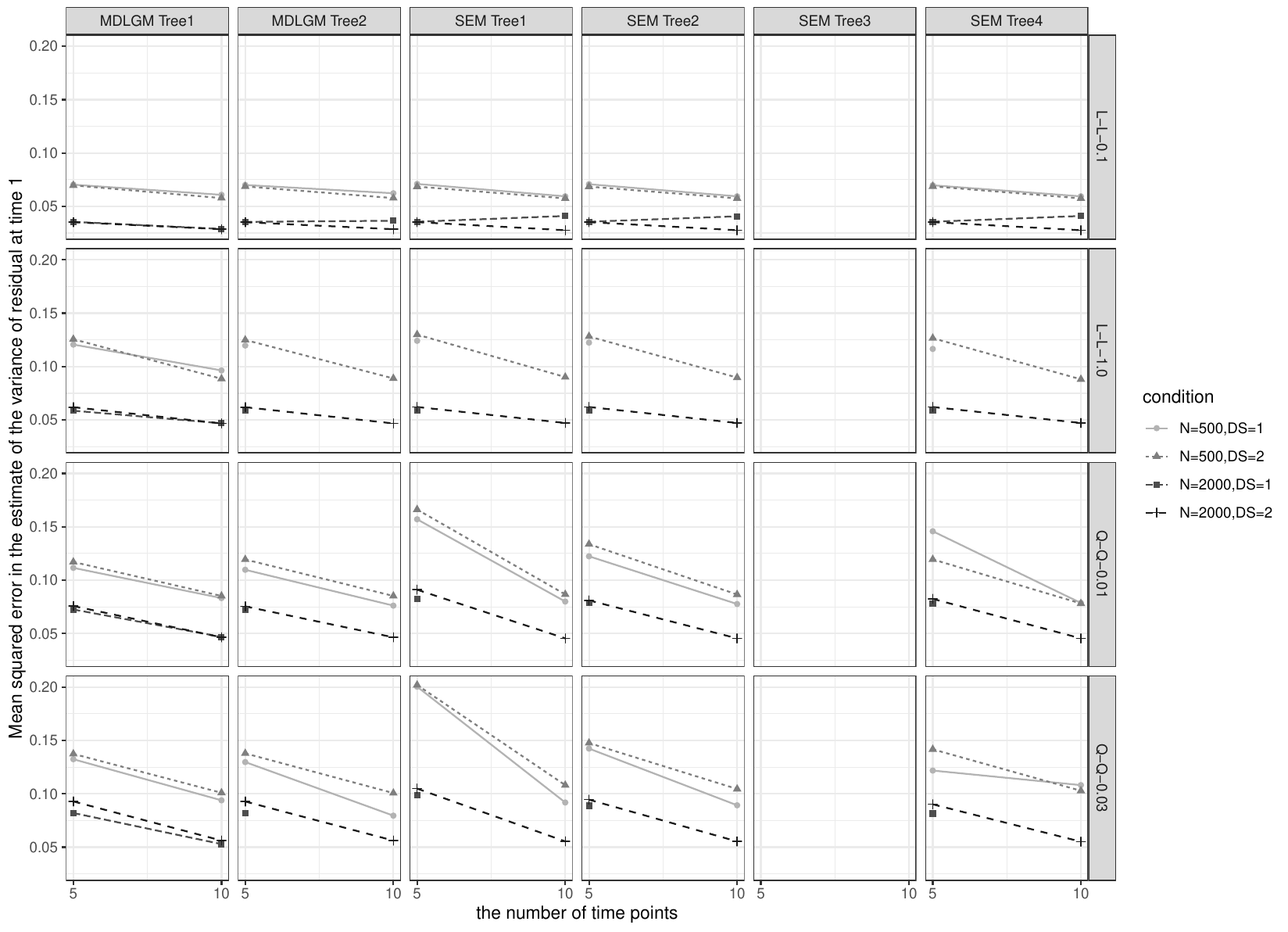}
    \caption{\\RMSE in $\psi^2_1$ calculated using the data when the number of terminal nodes was correctly estimated under conditions in which the true model and the template model were identical. SEM Tree1 denotes ML estimation, SEM Tree2 denotes constrained ML (CML) estimation, SEM Tree3 denotes Bayesian estimation, and SEM Tree4 denotes ML estimation with an algorithm that avoids node splitting based on any estimated model that produces a warning. MDLGM Tree1 and MDLGM Tree2 denote splitting based on the Mahalanobis distance and deviance, respectively. $DS$ denotes the degree of separation, and the three-element labels on the vertical axis (e.g., Q-L-0.01) indicate, in order, the true model (linear or quadratic), the template model (linear or quadratic), and the specified value of the slope factor variance. The RMSE of the SEM Tree3 could not be calculated. }
    \label{fig:rmse_psi_1}
\end{figure}

\begin{figure}[h]
    \centering
    \includegraphics[width=0.9\linewidth]{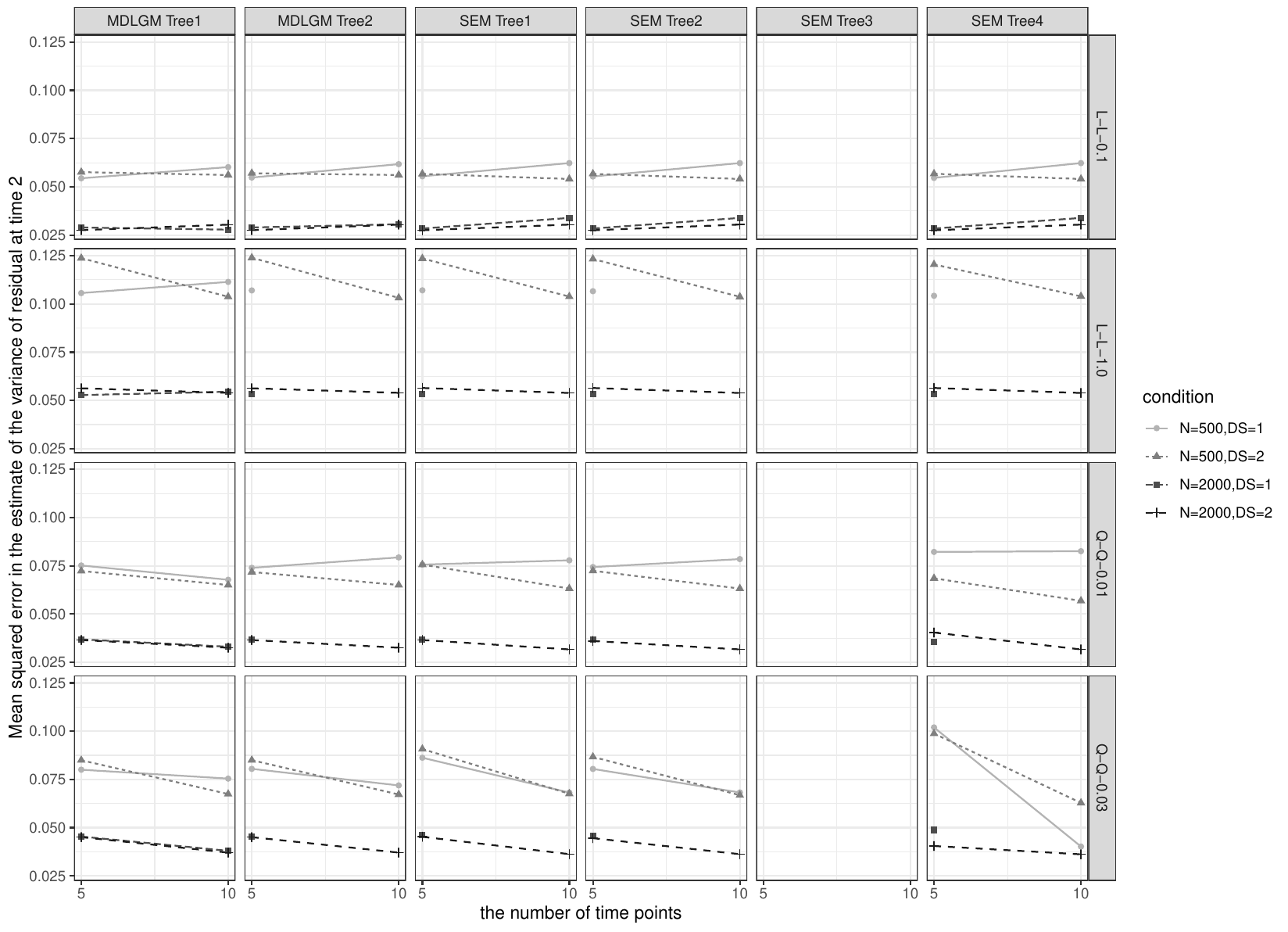}
    \caption{\\RMSE in $\psi^2_2$ calculated using the data when the number of terminal nodes was correctly estimated under conditions in which the true model and the template model were identical. SEM Tree1 denotes ML estimation, SEM Tree2 denotes constrained ML (CML) estimation, SEM Tree3 denotes Bayesian estimation, and SEM Tree4 denotes ML estimation with an algorithm that avoids node splitting based on any estimated model that produces a warning. MDLGM Tree1 and MDLGM Tree2 denote splitting based on the Mahalanobis distance and deviance, respectively. $DS$ denotes the degree of separation, and the three-element labels on the vertical axis (e.g., Q-L-0.01) indicate, in order, the true model (linear or quadratic), the template model (linear or quadratic), and the specified value of the slope factor variance. The RMSE of the SEM Tree3 could not be calculated. }
    \label{fig:rmse_psi_2}
\end{figure}

\begin{figure}[h]
    \centering
    \includegraphics[width=0.9\linewidth]{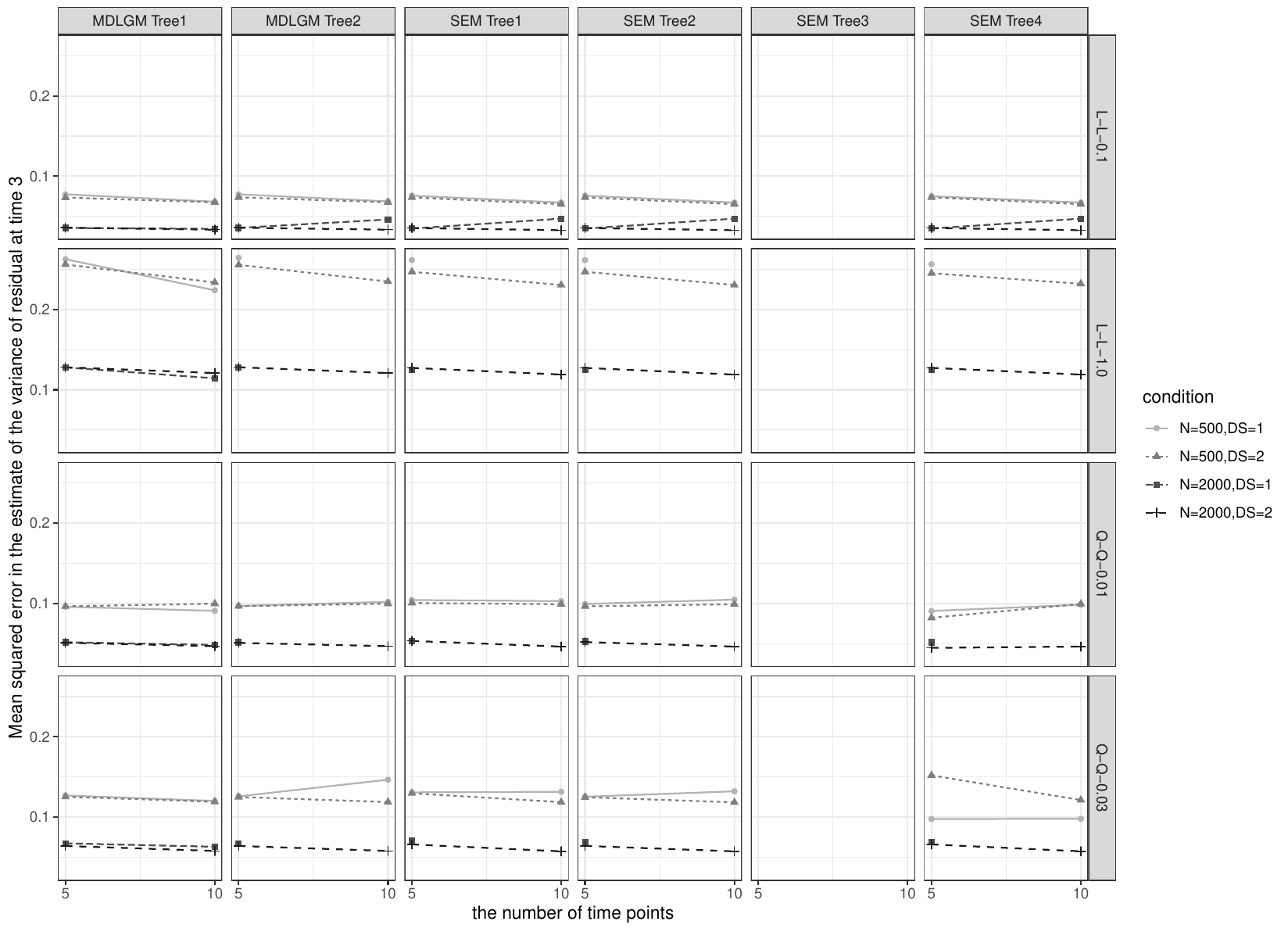}
    \caption{\\RMSE in $\psi^2_3$ calculated using the data when the number of terminal nodes was correctly estimated under conditions in which the true model and the template model were identical. SEM Tree1 denotes ML estimation, SEM Tree2 denotes constrained ML (CML) estimation, SEM Tree3 denotes Bayesian estimation, and SEM Tree4 denotes ML estimation with an algorithm that avoids node splitting based on any estimated model that produces a warning. MDLGM Tree1 and MDLGM Tree2 denote splitting based on the Mahalanobis distance and deviance, respectively. $DS$ denotes the degree of separation, and the three-element labels on the vertical axis (e.g., Q-L-0.01) indicate, in order, the true model (linear or quadratic), the template model (linear or quadratic), and the specified value of the slope factor variance. The RMSE of the SEM Tree3 could not be calculated. }
    \label{fig:rmse_psi_3}
\end{figure}

\begin{figure}[h]
    \centering
    \includegraphics[width=0.9\linewidth]{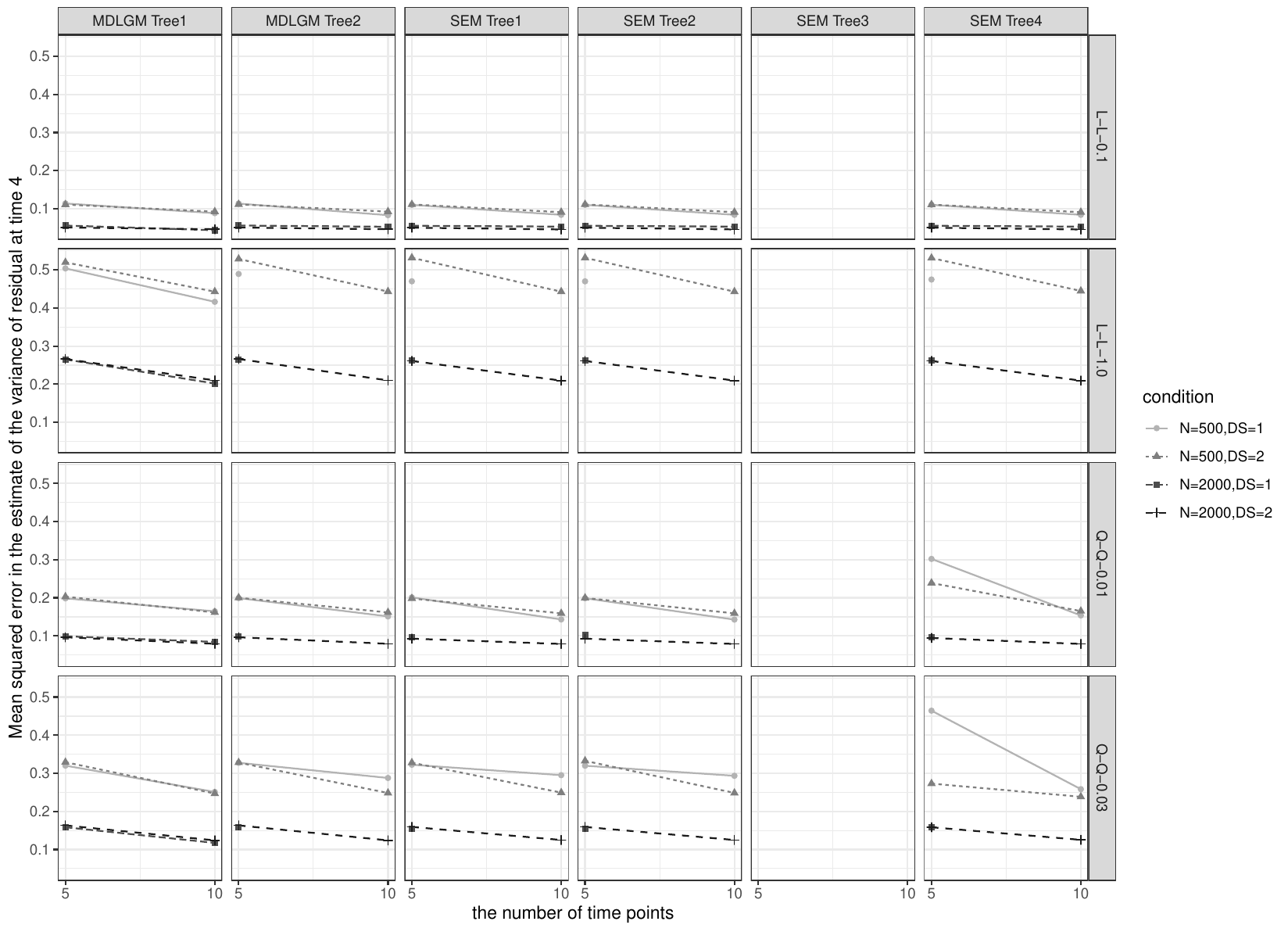}
    \caption{\\RMSE in $\psi^2_4$ calculated using the data when the number of terminal nodes was correctly estimated under conditions in which the true model and the template model were identical. SEM Tree1 denotes ML estimation, SEM Tree2 denotes constrained ML (CML) estimation, SEM Tree3 denotes Bayesian estimation, and SEM Tree4 denotes ML estimation with an algorithm that avoids node splitting based on any estimated model that produces a warning. MDLGM Tree1 and MDLGM Tree2 denote splitting based on the Mahalanobis distance and deviance, respectively. $DS$ denotes the degree of separation, and the three-element labels on the vertical axis (e.g., Q-L-0.01) indicate, in order, the true model (linear or quadratic), the template model (linear or quadratic), and the specified value of the slope factor variance. The RMSE of the SEM Tree3 could not be calculated. }
    \label{fig:rmse_psi_4}
\end{figure}

\begin{figure}[h]
    \centering
    \includegraphics[width=0.9\linewidth]{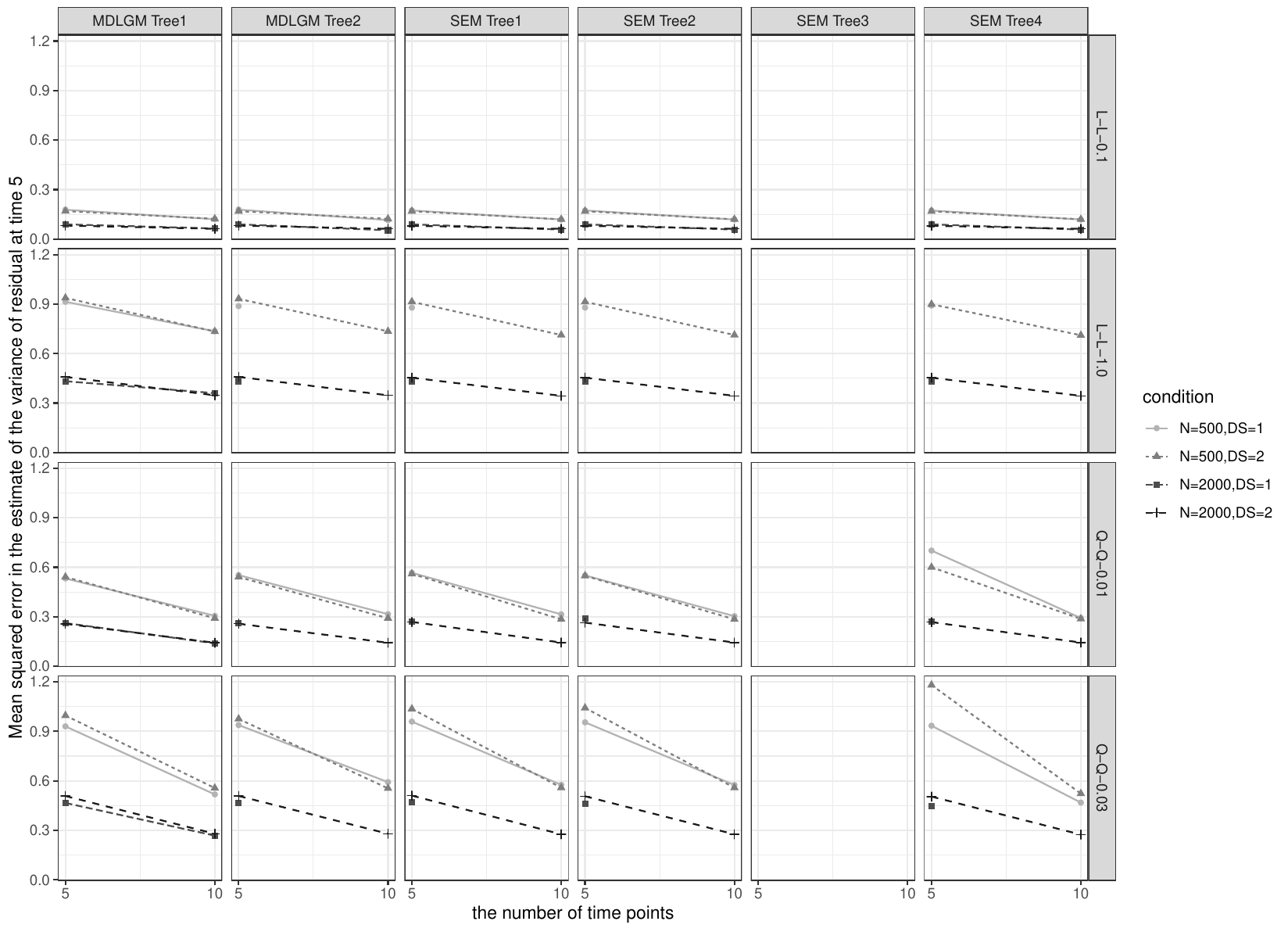}
    \caption{\\RMSE in $\psi^2_5$ calculated using the data when the number of terminal nodes was correctly estimated under conditions in which the true model and the template model were identical. SEM Tree1 denotes ML estimation, SEM Tree2 denotes constrained ML (CML) estimation, SEM Tree3 denotes Bayesian estimation, and SEM Tree4 denotes ML estimation with an algorithm that avoids node splitting based on any estimated model that produces a warning. MDLGM Tree1 and MDLGM Tree2 denote splitting based on the Mahalanobis distance and deviance, respectively. $DS$ denotes the degree of separation, and the three-element labels on the vertical axis (e.g., Q-L-0.01) indicate, in order, the true model (linear or quadratic), the template model (linear or quadratic), and the specified value of the slope factor variance. The RMSE of the SEM Tree3 could not be calculated. }
    \label{fig:rmse_psi_5}
\end{figure}

\begin{figure}[h]
    \centering
    \includegraphics[width=0.9\linewidth]{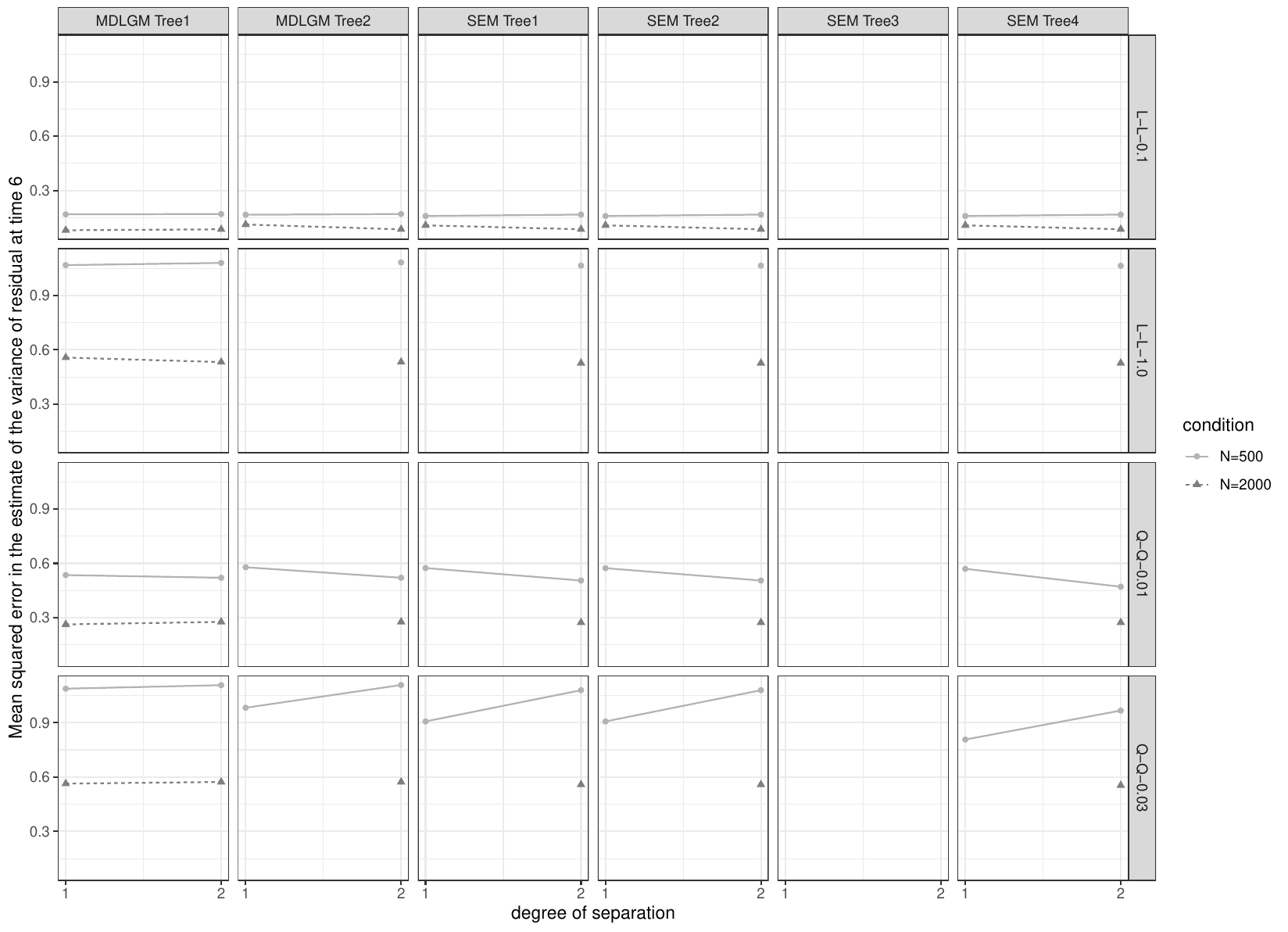}
    \caption{\\RMSE in $\psi^2_6$ calculated using the data when the number of terminal nodes was correctly estimated under conditions in which the true model and the template model were identical. SEM Tree1 denotes ML estimation, SEM Tree2 denotes constrained ML (CML) estimation, SEM Tree3 denotes Bayesian estimation, and SEM Tree4 denotes ML estimation with an algorithm that avoids node splitting based on any estimated model that produces a warning. MDLGM Tree1 and MDLGM Tree2 denote splitting based on the Mahalanobis distance and deviance, respectively. The three-element labels on the vertical axis (e.g., Q-L-0.01) indicate, in order, the true model (linear or quadratic), the template model (linear or quadratic), and the specified value of the slope factor variance. The RMSE of the SEM Tree3 could not be calculated. }
    \label{fig:rmse_psi_6}
\end{figure}

\begin{figure}[h]
    \centering
    \includegraphics[width=0.9\linewidth]{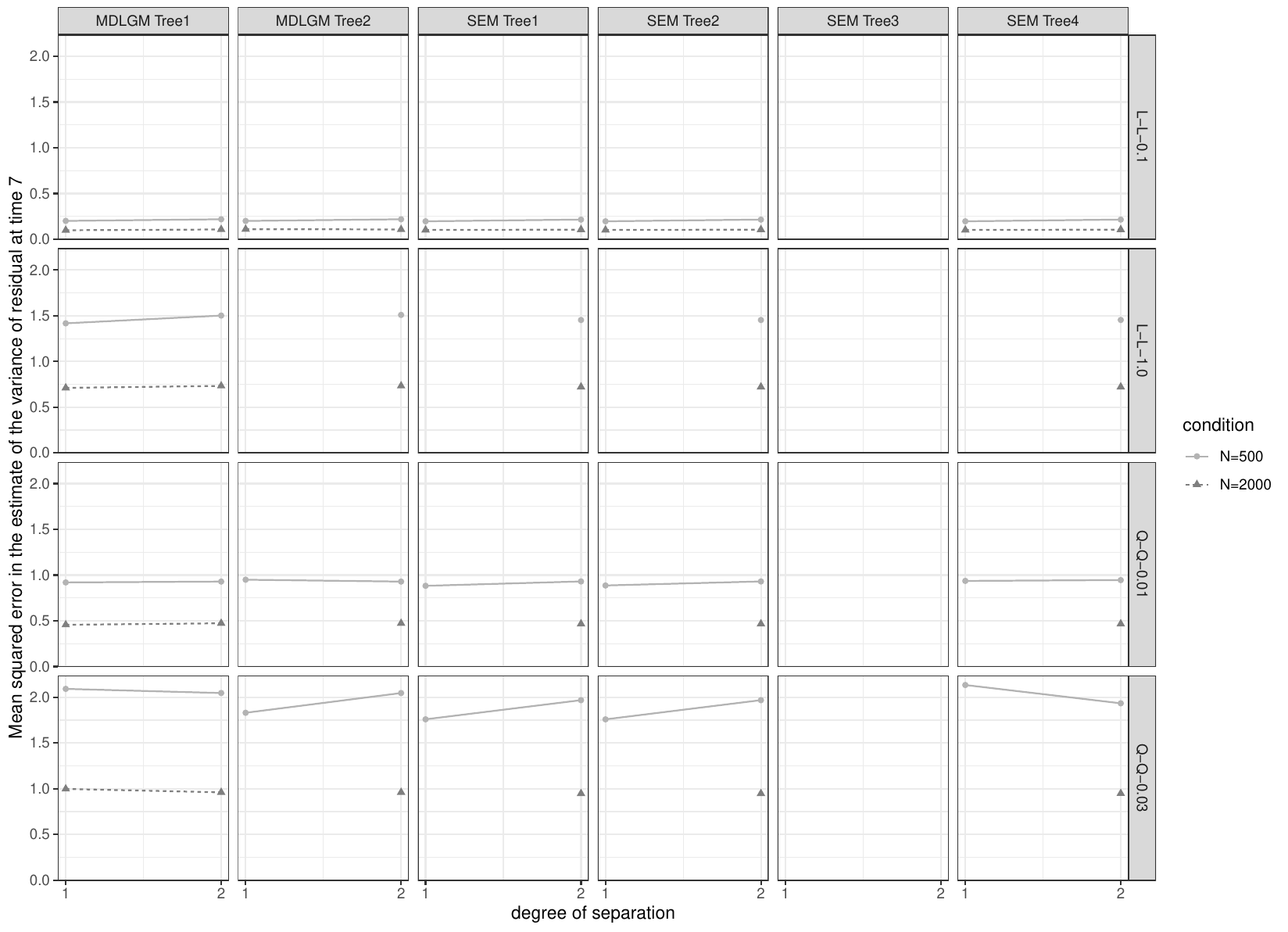}
    \caption{\\RMSE in $\psi^2_7$ calculated using the data when the number of terminal nodes was correctly estimated under conditions in which the true model and the template model were identical. SEM Tree1 denotes ML estimation, SEM Tree2 denotes constrained ML (CML) estimation, SEM Tree3 denotes Bayesian estimation, and SEM Tree4 denotes ML estimation with an algorithm that avoids node splitting based on any estimated model that produces a warning. MDLGM Tree1 and MDLGM Tree2 denote splitting based on the Mahalanobis distance and deviance, respectively. The three-element labels on the vertical axis (e.g., Q-L-0.01) indicate, in order, the true model (linear or quadratic), the template model (linear or quadratic), and the specified value of the slope factor variance. The RMSE of the SEM Tree3 could not be calculated. }
    \label{fig:rmse_psi_7}
\end{figure}

\begin{figure}[h]
    \centering
    \includegraphics[width=0.9\linewidth]{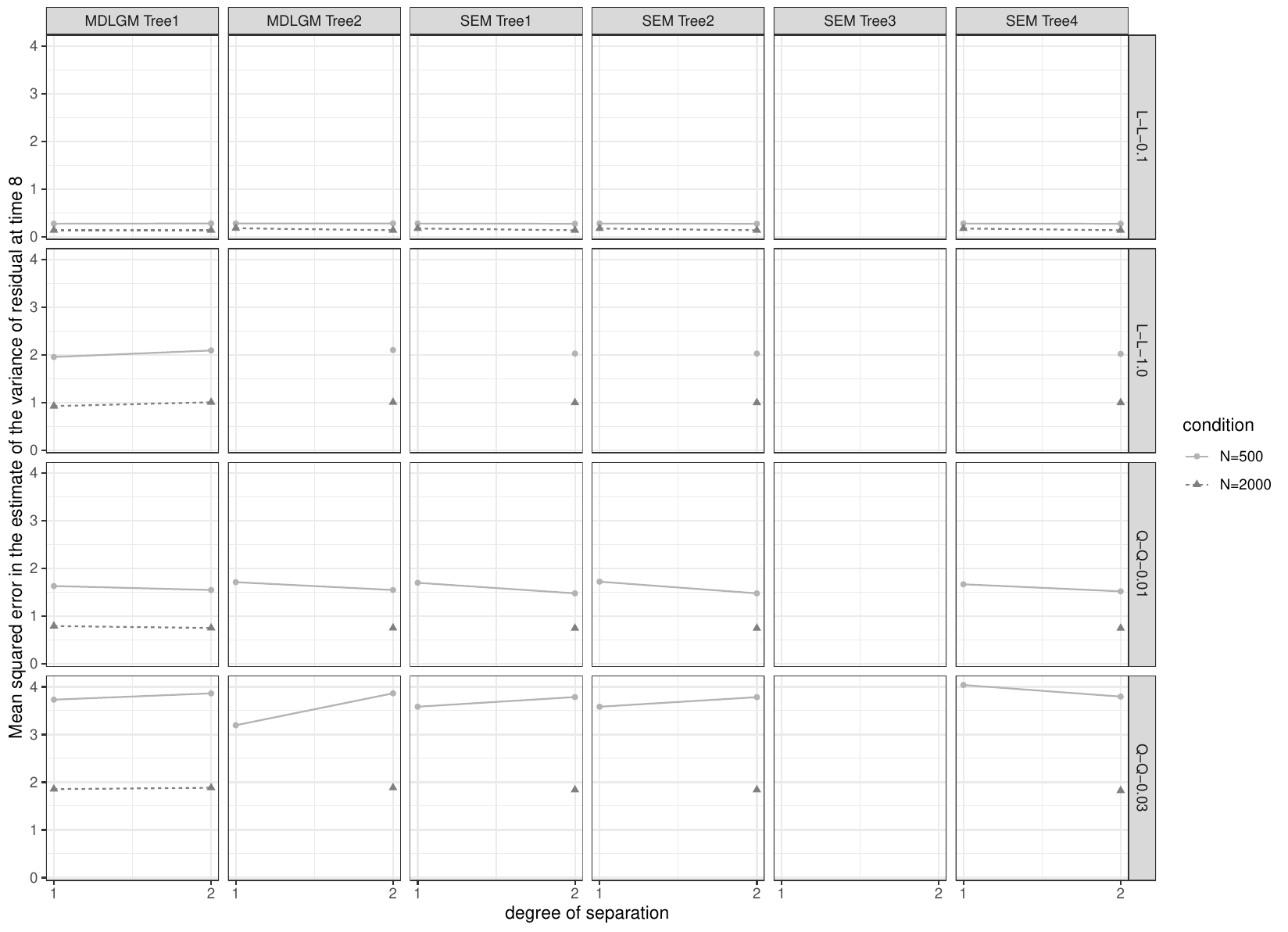}
    \caption{\\RMSE in $\psi^2_8$ calculated using the data when the number of terminal nodes was correctly estimated under conditions in which the true model and the template model were identical. SEM Tree1 denotes ML estimation, SEM Tree2 denotes constrained ML (CML) estimation, SEM Tree3 denotes Bayesian estimation, and SEM Tree4 denotes ML estimation with an algorithm that avoids node splitting based on any estimated model that produces a warning. MDLGM Tree1 and MDLGM Tree2 denote splitting based on the Mahalanobis distance and deviance, respectively. The three-element labels on the vertical axis (e.g., Q-L-0.01) indicate, in order, the true model (linear or quadratic), the template model (linear or quadratic), and the specified value of the slope factor variance. The RMSE of the SEM Tree3 could not be calculated. }
    \label{fig:rmse_psi_8}
\end{figure}

\begin{figure}[h]
    \centering
    \includegraphics[width=0.9\linewidth]{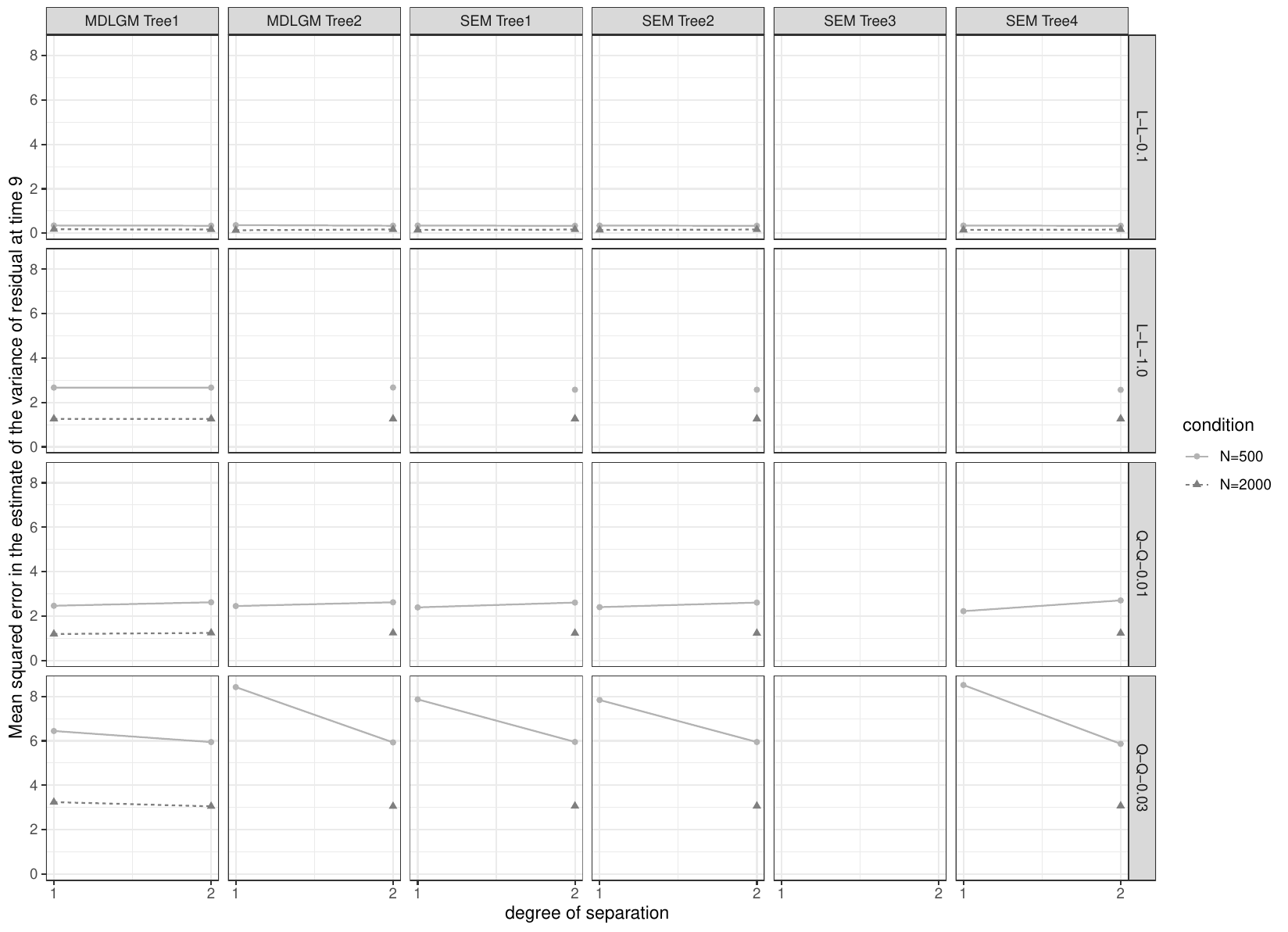}
    \caption{\\RMSE in $\psi^2_9$ calculated using the data when the number of terminal nodes was correctly estimated under conditions in which the true model and the template model were identical. SEM Tree1 denotes ML estimation, SEM Tree2 denotes constrained ML (CML) estimation, SEM Tree3 denotes Bayesian estimation, and SEM Tree4 denotes ML estimation with an algorithm that avoids node splitting based on any estimated model that produces a warning. MDLGM Tree1 and MDLGM Tree2 denote splitting based on the Mahalanobis distance and deviance, respectively. The three-element labels on the vertical axis (e.g., Q-L-0.01) indicate, in order, the true model (linear or quadratic), the template model (linear or quadratic), and the specified value of the slope factor variance. The RMSE of the SEM Tree3 could not be calculated. }
    \label{fig:rmse_psi_9}
\end{figure}

\begin{figure}[h]
    \centering
    \includegraphics[width=0.9\linewidth]{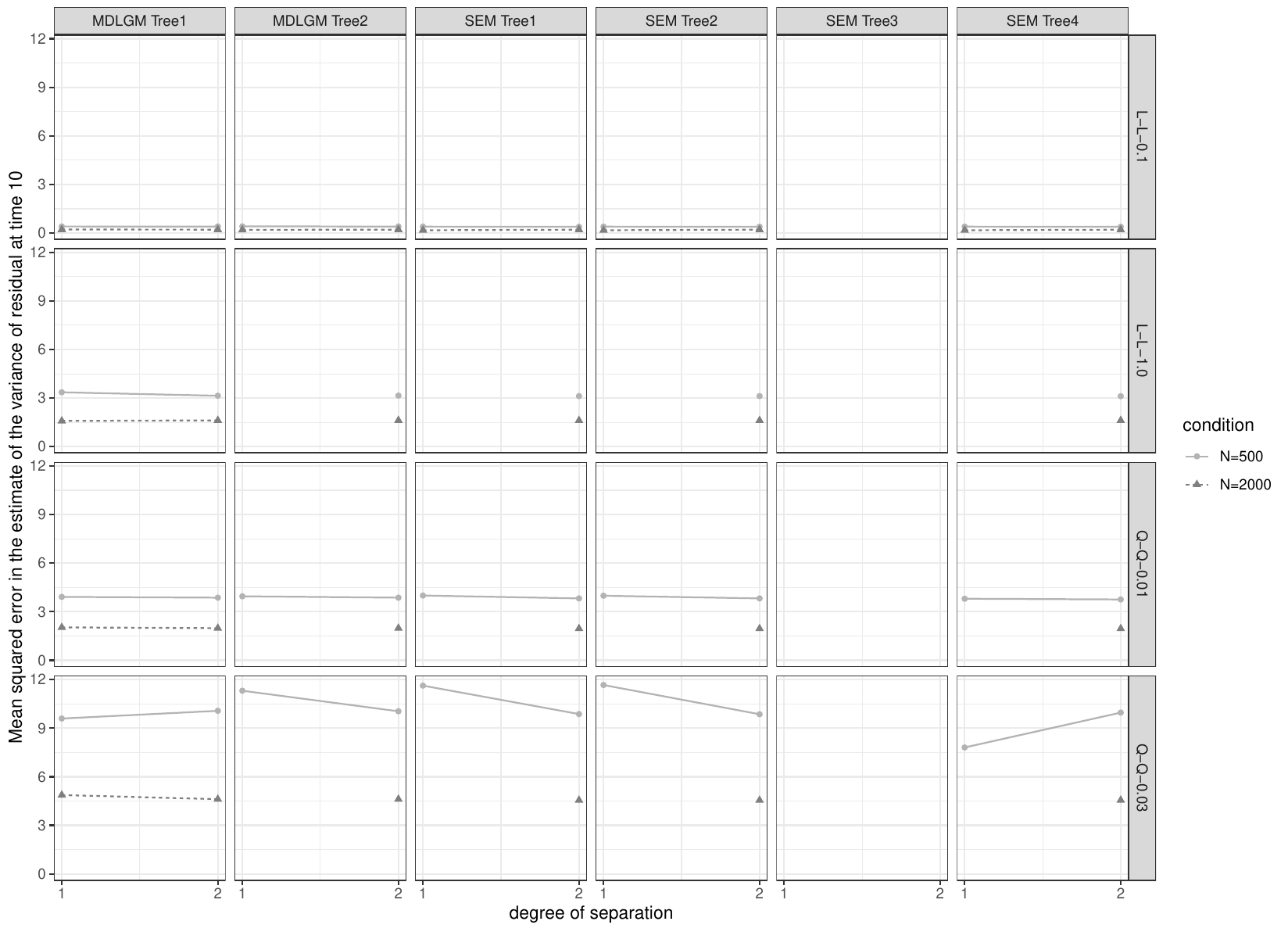}
    \caption{\\RMSE in $\psi^2_{10}$ calculated using the data when the number of terminal nodes was correctly estimated under conditions in which the true model and the template model were identical. SEM Tree1 denotes ML estimation, SEM Tree2 denotes constrained ML (CML) estimation, SEM Tree3 denotes Bayesian estimation, and SEM Tree4 denotes ML estimation with an algorithm that avoids node splitting based on any estimated model that produces a warning. MDLGM Tree1 and MDLGM Tree2 denote splitting based on the Mahalanobis distance and deviance, respectively. The three-element labels on the vertical axis (e.g., Q-L-0.01) indicate, in order, the true model (linear or quadratic), the template model (linear or quadratic), and the specified value of the slope factor variance. The RMSE of the SEM Tree3 could not be calculated. }
    \label{fig:rmse_psi_10}
\end{figure}

Figure~\ref{fig:bias_psi_1}-\ref{fig:bias_psi_10} show the biases of $\psi^2_1$ to $\psi^2_{10}$, respectively and Figure~\ref{fig:rmse_psi_1}-\ref{fig:rmse_psi_10} show the RMSEs of those in each method and condition. As mentioned in the main text, these figures show that all methods exhibited slightly different patterns for the elements. 

 \clearpage

 \clearpage

 \clearpage

 \clearpage

 \clearpage

 \clearpage

 \clearpage

 \clearpage

 \clearpage

 \clearpage

 \clearpage

 \clearpage

 \clearpage

 \clearpage

 \clearpage

 \clearpage

 \clearpage

 \clearpage

 \clearpage

 \clearpage

\subsection{Correlations among the estimates from different estimation methods}
\begin{figure}[h]
    \centering
    \includegraphics[width=0.9\linewidth]{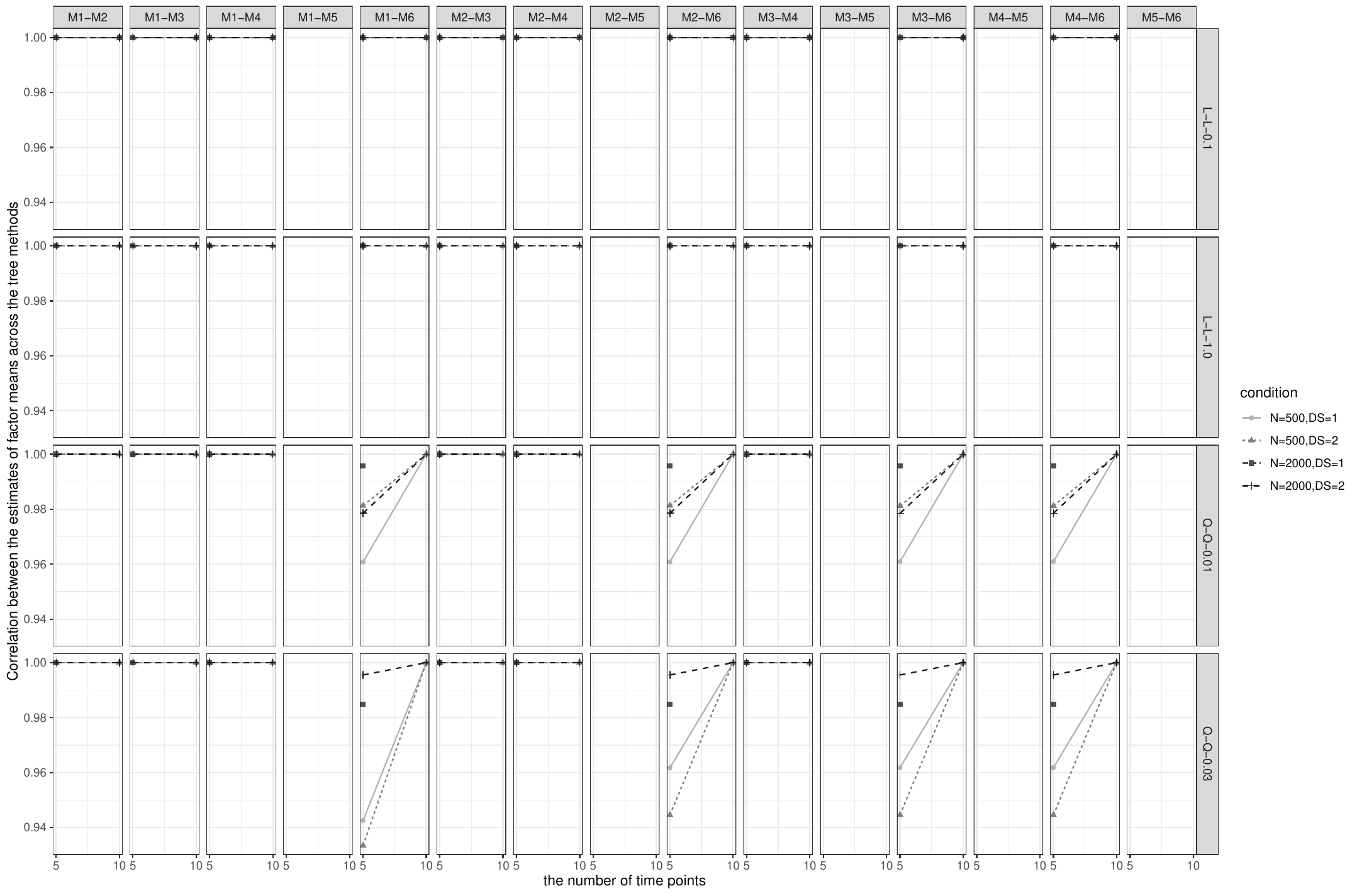}
    \caption{\\Average correlation among the estimates of $\boldsymbol{\mu}$ of the 4 nodes between the estimation methods when the methods estimated the true number of nodes correctly under conditions in which the true and template model were identical. M1 and M2 denote the MDLGM Tree methods which split nodes based on the Mahalanobis distance and decviance, respectively. M3-M5 denote the SEM Tree methods which estimate parameters by ML estimation, constrained ML (CML) estimation, Bayesian estimation, respectively. M6 denotes the SEM Tree method which is based on ML estimation with an algorithm that avoids node splitting based on any estimated model that produces a warning. The two-element labels on the horizontal axis (e.g., M2-M3) indicate between which the correlation was calculated. $DS$ denotes the degree of separation, and the three-element labels on the vertical axis (e.g., Q-L-0.01) indicate, in order, the true model (linear or quadratic), the template model (linear or quadratic), and the specified value of the slope factor variance. The correlations for the SEM Tree3 could not be calculated. }
    \label{fig:cor_mu}
\end{figure}

\begin{figure}[h]
    \centering
    \includegraphics[width=0.9\linewidth]{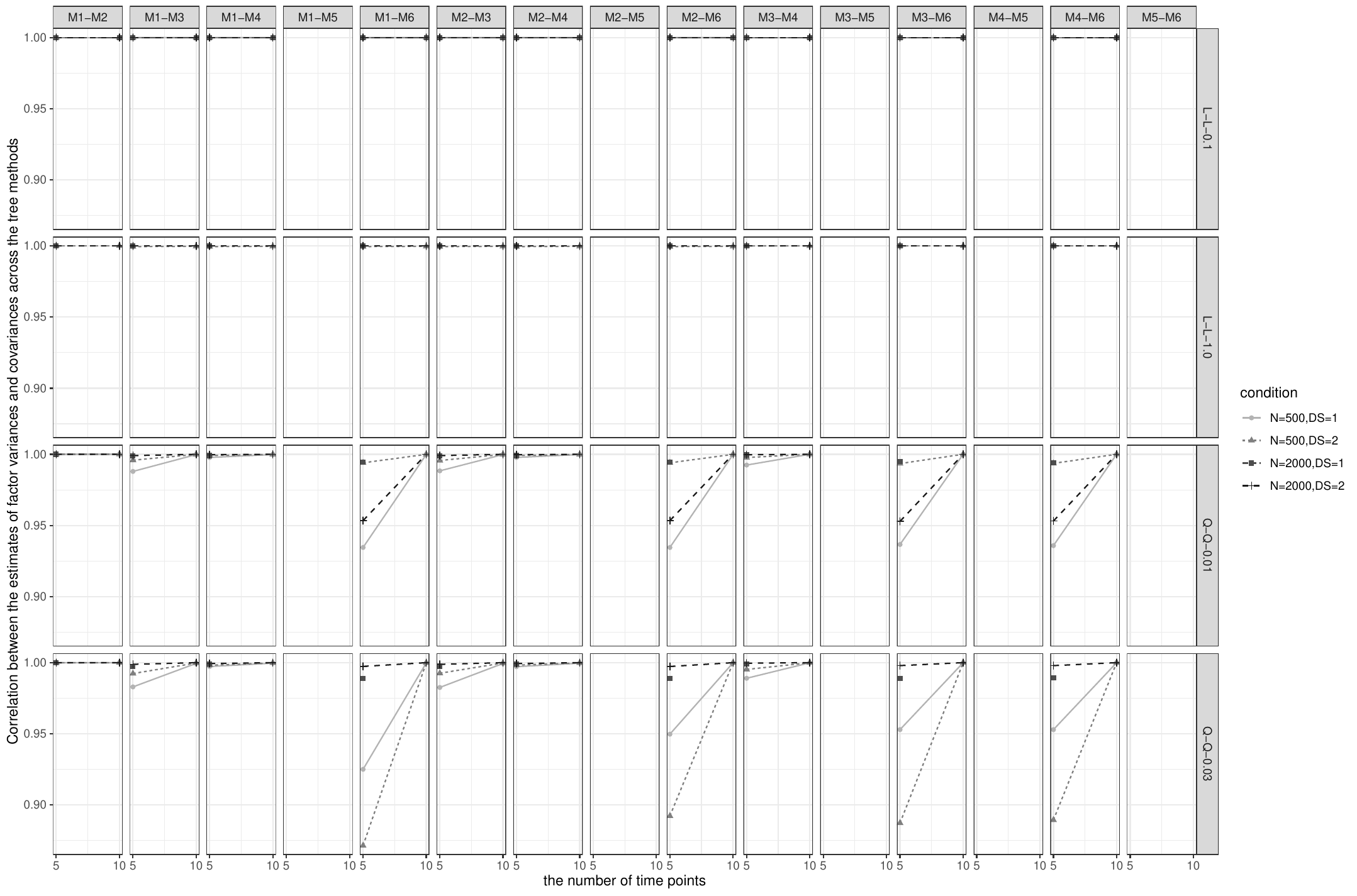}
    \caption{\\Average correlation among the estimates of unique element of $\boldsymbol{\Phi}$ of the 4 nodes between the estimation methods when the methods estimated the true number of nodes correctly under conditions in which the true and template model were identical. M1 and M2 denote the MDLGM Tree methods which split nodes based on the Mahalanobis distance and decviance, respectively. M3-M5 denote the SEM Tree methods which estimate parameters by ML estimation, constrained ML (CML) estimation, Bayesian estimation, respectively. M6 denotes the SEM Tree method which is based on ML estimation with an algorithm that avoids node splitting based on any estimated model that produces a warning. The two-element labels on the horizontal axis (e.g., M2-M3) indicate between which the correlation was calculated. $DS$ denotes the degree of separation, and the three-element labels on the vertical axis (e.g., Q-L-0.01) indicate, in order, the true model (linear or quadratic), the template model (linear or quadratic), and the specified value of the slope factor variance. The correlations for the SEM Tree3 could not be calculated. }
    \label{fig:cor_phi}
\end{figure}

\begin{figure}[h]
    \centering
    \includegraphics[width=0.9\linewidth]{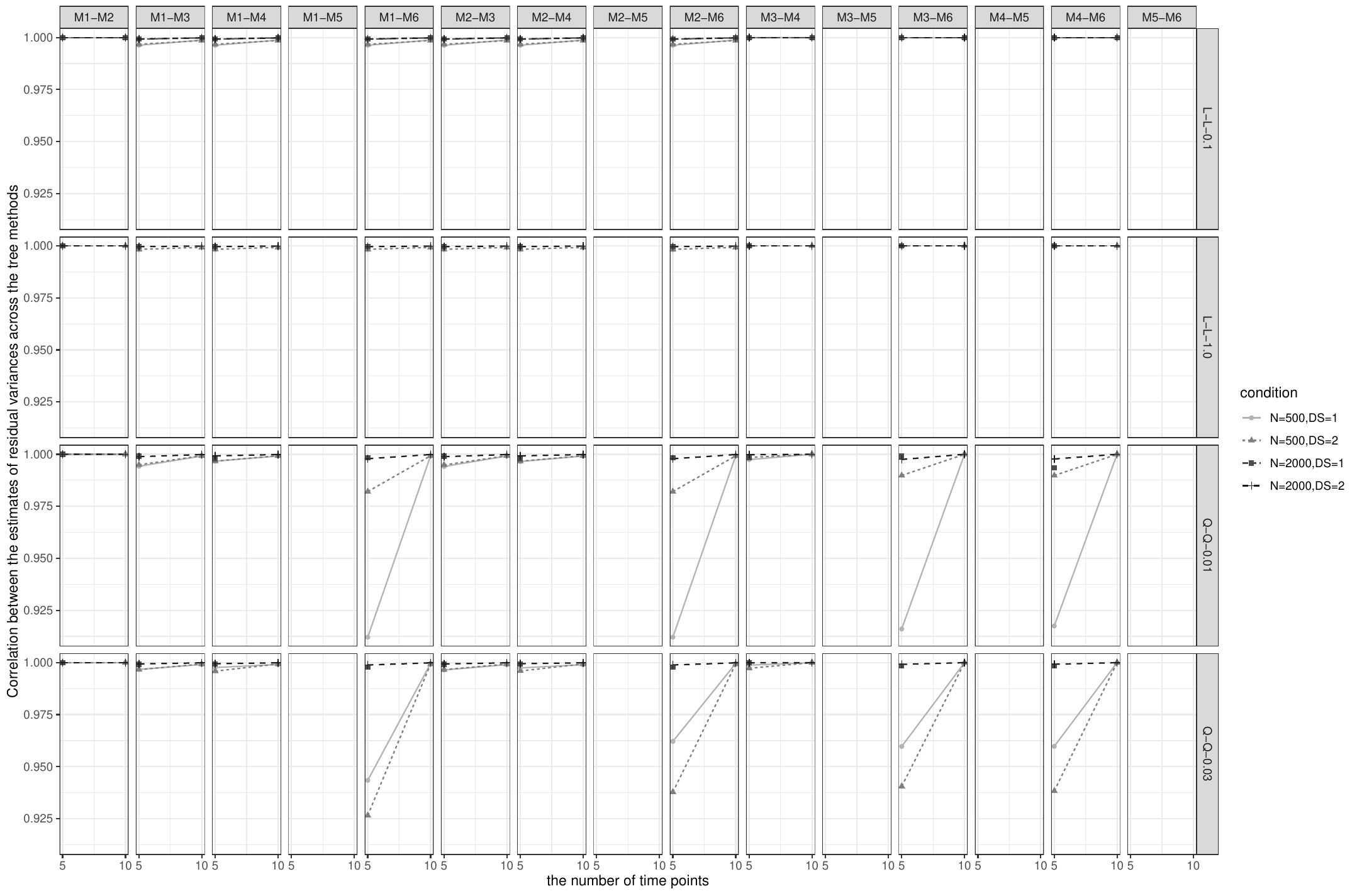}
    \caption{\\Average correlation among the estimates of residual variances of the 4 nodes between the estimation methods when the methods estimated the true number of nodes correctly under conditions in which the true and template model were identical. M1 and M2 denote the MDLGM Tree methods which split nodes based on the Mahalanobis distance and decviance, respectively. M3-M5 denote the SEM Tree methods which estimate parameters by ML estimation, constrained ML (CML) estimation, Bayesian estimation, respectively. M6 denotes the SEM Tree method which is based on ML estimation with an algorithm that avoids node splitting based on any estimated model that produces a warning. The two-element labels on the horizontal axis (e.g., M2-M3) indicate between which the correlation was calculated. $DS$ denotes the degree of separation, and the three-element labels on the vertical axis (e.g., Q-L-0.01) indicate, in order, the true model (linear or quadratic), the template model (linear or quadratic), and the specified value of the slope factor variance. The correlations for the SEM Tree3  could not be calculated. }
    \label{fig:cor_psi}
\end{figure}

Figure~\ref{fig:cor_mu}-\ref{fig:cor_psi} show the average correlation among the estimate of $\boldsymbol{\mu}$, unique elements of $\boldsymbol{\Phi}$, and $\boldsymbol{\Psi}$ of 4 nodes between the estimation methods when the methods estimated the true number of nodes correctly under conditions in which the true and template model were identical respectively. As mentioned in the main text, regardless of the type of parameter, the correlations between methods were almost 1. 

 \clearpage

 \clearpage

 \clearpage


\def\spacingset#1{\renewcommand{\baselinestretch}{#1}\small\normalsize}
\spacingset{1.75}

\if0\blind
{
  \bigskip
  \begin{center}\vspace{1.5em}
    {\Large\bfseries Matrix Decomposition Latent Growth Model Tree\par}
    \vspace{2.5em}
  \end{center}
}
\fi

\begin{abstract}
Latent growth models (LGMs) have been widely used to describe individual trajectories of change, average change patterns, and individual differences in change within a group. Structural equation model tree (SEM Tree) partitions samples through recursive partitioning using external covariates and characterizes between-group differences via parameters of prespecified SEM template models. An SEM Tree with an LGM as its template model may therefore serve as a useful classification method for longitudinal data. However, current SEM Tree implementations may face important practical obstacles, including high computational cost and frequent improper solutions caused by model misspecification or data characteristics (e.g., a limited number of measurement occasions). In the present study, we developed a matrix decomposition LGM tree (MDLGM Tree), which applies the matrix decomposition factor analysis (MDFA) approach, an eigenvalue-decomposition-based method for model estimation. Previous studies on MDFA and its extensions to SEM have shown that this approach can reduce the risk of improper solutions while yielding estimates highly similar to those obtained by maximum likelihood estimation. Through a series of simulations, we demonstrated that, compared with conventional SEM Tree, the MDLGM Tree greatly reduces the occurrence of improper solutions and is also highly efficient in terms of computation time.
\end{abstract}

\noindent Key words: latent growth model, structural equation model tree, matrix decomposition factor analysis, classification, longitudinal data

\newpage

\section{Introduction}
Latent growth models (LGMs; \citealp{McArdleEpstein1987, McArdle1988, MeredithTisak1984, MeredithTisak1990}) have been widely used to describe individual trajectories of change, as well as the average pattern of change and individual differences in change within a group. Parameter estimates in LGM are often obtained by maximum likelihood (ML) estimation in structural equation modeling (SEM) based on the framework of mean and covariance structure analysis. By drawing on the SEM framework, it becomes possible to implement, flexible modeling of, for example, common factors based on multiple indicators and residual structures, as well as models involving multiple outcomes and measurement errors in a coherent and straightforward manner (see also \citealp{McNeishMatta2018, Usami2026a}).

In many substantive applications, however, the population may be heterogeneous, and a single LGM may not adequately characterize all individuals. For example, \cite{Sakaki2024} used LGM and showed that adolescents differed across groups in the developmental trajectories of mathematics-related achievement emotions, with heterogeneity in the LGM intercepts and slopes observed as a function of background characteristics such as gender, family SES, school track, and initial mathematics achievement.

SEM Tree (\citealp{Brandmaier2013, Brandmaier2014}) addresses this issue by combining SEM with recursive partitioning so that subgroups can be identified on the basis of observed covariates. More specifically, SEM Tree partitions samples based on recursive partitioning using external covariates and can characterize between-group differences through differences in the parameters of prespecified template models specified as SEMs. Although several recursive partitioning methods based on SEM other than SEM Tree have been proposed (e.g., \citealp{MerkleZeileis2011, Sanchez2009}), SEM Tree performs the kind of partitioning primarily based on ML (\citealp[p.73]{Brandmaier2013}). SEM Tree analyses can be conducted using the \texttt{semtree} package in R \citep{BrandmaierPrindle2018} and the \texttt{MplusTrees} package (\citealp{Serang2021, Serang2025}). In addition, SEM forests, which incorporate the random forest approach \citep{Breiman2001}, have been proposed as an ensemble-based extension \citep{BrandmaierPrindleMcArdleLindenberger2016}, and more recently, integrative methods combining continuous-time modeling \citep{Voelkle2012} with SEM forests have also been developed \citep{CancerArnoldEstradaVoelkle2025}.

An SEM Tree with an LGM as its template model may be regarded as a useful classification method for longitudinal data. At the same time, however, current implementations of SEM Tree may face important practical obstacles, including extremely high computational cost and the frequent occurrence of improper solutions arising from model misspecification and characteristics of the data themselves (e.g., a limited number of measurement occasions and sample sizes). Regarding the former point, estimation can be computationally demanding because the template model must be fitted repeatedly across a large number of candidate splits. Regarding the latter point, when the template model is misspecified or when sample sizes within nodes are limited, improper solutions, such as negative residual variances (or boundary solutions) and non-positive definite covariance matrices among growth factors, may occur frequently (e.g., \citealp{Usami2026a}). 

In analyses involving iterative computations, such as tree-based methods, it is currently very difficult to completely avoid the problem of improper solutions. In the \texttt{semtree} package and the \texttt{MplusTrees} package, nodes appear to be extracted even when parameter estimate(s) in template model exhibit improper solutions\footnote{The \texttt{semtree} package provides control options such as \texttt{exclude.code} for excluding models with specified fitting error codes from split evaluation; however, by default, models with fitting errors are retained. Moreover, this option appears to operate on the basis of optimizer error codes rather than by comprehensively detecting and excluding all improper solutions, such as negative variance estimates, or non-positive definite matrices. In addition, \texttt{exclude.heywood} is described as reporting identification problems in the covariance structure of an SEM, rather than as automatically pruning such nodes.}. Likewise, in software developed mainly in the context of mixed-effects-model-based recursive partitioning (\citealp[e.g.,][]{AbdolellLeBlancStephensHarrison2002, HajjemBellavanceLarocque2014, SelaSimonoff2012, FokkemaGreiff2017}; see also \citealp{JacobucciGrimmZhang2023})—such as \texttt{longRPart2} \citep{JacobucciStewartAbdolellSerangStegmann2017}, \texttt{glmertree}, and \texttt{REEMtree}—, there does not appear to be an explicit built-in mechanism for automatically excluding nodes even when fitting the mixed-effects model yields improper solutions; rather, such tree structures seem generally to be returned unless the analyst imposes additional constraints. To the best of the present author’s knowledge, explicit built-in mechanisms for automatically excluding nodes solely because the fitted model in the node showed improper solutions are not common in currently available model-based tree software.

In the present study, with the aim of proposing a more reasonable algorithm from the viewpoints of computational burden and the handling of improper solutions, and focusing on relatively simple LGMs such as the linear and quadratic LGM, we develop a matrix decomposition latent growth model tree (MDLGM Tree), which combines a decision-tree algorithm with an LGM estimated using a matrix decomposition factor analysis (MDFA) procedure \citep{Adachi2018}. MDFA is an eigenvalue-decomposition-based approach that estimates parameters iteratively using a least squares criterion defined in the data space rather than in the covariance space like standard SEM. Although MDFA was originally proposed in the context of FA, recent developments have extended it to SEMs involving structural relations among latent variables \citep{Yamashita2024}, and its statistical properties have also begun to be examined theoretically \citep{Terada2024}. One important feature of MDFA is that negative variance estimates do not arise in principle during the iterative updating process. In the context of longitudinal modeling aimed at disentangling within-person relations from between-person differences (stable traits), MDFA has been used for parameter estimation and has been shown to entail a substantially lower risk of improper solutions than commonly used ML, conditional ML, and unweighted least squares estimators, while tending to yield solutions similar to those obtained by ML \citep{Usami2026b}. Through a series of simulation studies, we will show that, compared with the SEM Tree, the proposed MDLGM Tree substantially reduces the occurrence of improper solutions while also improving computational efficiency.

The remainder of this paper is organized as follows. Section 2 provides an overview of LGM and MDFA. Section 3 introduces the MDLGM and describes its estimation procedure. Section 4 describes the procedure of the MDLGM Tree, and Section 5 presents simulation studies evaluating the frequency of improper solutions, estimated number of (terminal) nodes, the bias and RMSE of parameter estimates in LGM, and computational cost across methods, under both correctly specified and misspecified LGMs. Section 6 concludes with a summary of the main findings and directions for future research.

\section{Latent Growth Models and Matrix Decomposition Approach}
\subsection{Latent growth models}
In psychological and behavioral research, a variety of LGM specifications have been proposed and applied (e.g., \citealp{Grimm2016}; \citealp{Usami2026a}). In the present study, we focus on the linear LGM and the quadratic LGM, mainly because of its popularity. Let \(y_{it}\) denote the measurement for individual \(i\) (\(i=1,\ldots,N\)) at time point \(t\) (\(t=1,\ldots,T\)). In the linear LGM, \(y_{it}\) is expressed as
\begin{equation}
  y_{it} = f_{Ii} + (t-1) f_{Si} + e_{it}.
  \label{eq:linear LGM_scalar}
\end{equation}
Here, \(f_{Ii}\) is the intercept factor, representing the value at the first time point (\(t=1\)) of the linear trajectory for individual \(i\). \(f_{Si}\) is the slope factor, representing the amount of change between adjacent time points in the trajectory. These factors are collectively referred to as growth factors. The means and (co)variances of the growth factors, namely
\[
  \mathrm{E}[f_{Ii}]=\mu_I,\quad
  \mathrm{E}[f_{Si}]=\mu_S,\quad
  \operatorname{Var}[f_{Ii}]=\phi_I^2,\quad
  \operatorname{Var}[f_{Si}]=\phi_S^2,\quad
  \operatorname{Cov}[f_{Ii},f_{Si}]=\phi_{IS},
\]
are estimated as free parameters. And let $
  \bd{\mu}=(\mu_I,\mu_S)^\top, \bd{\Phi}=\left(\begin{smallmatrix}
\phi_I^2 & \phi_{IS}\\
\phi_{IS} & \phi_S^2
\end{smallmatrix}\right)$.

The term \(e_{it}\) is a residual, with \(\mathrm{E}[e_{it}]=0\) and \(\operatorname{Var}[e_{it}]=\psi_t^2\). The residuals are assumed to be uncorrelated with the growth factors, and it is also assumed that residuals are uncorrelated across different individuals. Furthermore, residuals at different time points are usually assumed to be uncorrelated ($\operatorname{Cov}[e_{it},e_{it'}]=0\quad (t\neq t')$) (however, see also, \citealp{Bollen2006}, p.~22; \citealp{Kline2016}, pp.~378--380). Therefore, totally \(T\) parameters are included in the residual covariance matrix $\bd{\Psi}=\operatorname{diag}(\psi_1^2,\ldots,\psi_T^2)$. Consequently, the total number of parameters in the model, that is, the dimension of model parameters $\bd{\theta} =
  \left(
    \bd{\mu}^\top,
    \operatorname{vech}(\bd{\Phi})^\top,
    \operatorname{diag}(\bd{\Psi})^\top
  \right)^\top
$ becomes \(T+5\).

The linear LGM can be regarded as a confirmatory FA model with two common (growth) factors and mean structure. In this model, the factor loadings are predetermined according to the type of growth factor. In addition, instead of estimating intercepts, which are fixed to zero, the means and (co)variances of the growth factors are estimated.

The quadratic LGM is expressed as
\begin{equation}
  y_{it} = f_{Ii} + (t-1) f_{S1i} + (t-1)^2 f_{S2i} + e_{it}.
  \label{eq:quadratic LGM_scalar}
\end{equation}
This model can be regarded as a three-factor model that includes three growth factors, \(f_{Ii}\), \(f_{S1i}\), and \(f_{S2i}\). Specifically, in addition to the linear slope factor \(S_1\), whose corresponding factor loadings are fixed at \(t-1\), the model includes a second-order slope factor \(S_2\), whose corresponding factor loadings are fixed at \((t-1)^2\). The factor \(S_1\) represents the slope of the tangent to the trajectory at \(t=1\), whereas \(S_2\) represents (half of) the increment in the tangent slope (\citealp{Bollen2006}; \citealp{Usami2026a}).

Except for the inclusion of such two slope factors, the basic model assumptions are the same as those of the linear LGM. The mean and individual differences in the trajectory are represented by a total of nine free parameters: three factor means contained in \(\bd{\mu}\) and six (co)variances contained in \(\bd{\Phi}\), where \(\bd{\mu}\) and \(\bd{\Phi}\) are defined analogously to the linear LGM. Because the residual covariance matrix \(\bd{\Psi}\) contains \(T\) parameters, the total number of free parameters becomes \(T+9\). Although the present specification assumes a trajectory that increases over time, different time point for vertex of trajectory can be specified by modifying the values of factor loadings.

In the ML estimation for LGM (or SEM), under the assumption of multivariate normality, model parameters \(\bd{\theta}\) are estimated by minimizing the following loss function:
\begin{align}
  f_{ML}
  &=
  \log\left|\bd{\Sigma}(\bd{\theta})\right|
  -
  \log\left|\bd{S}\right|
  +
  \operatorname{tr}\left\{
    \bd{S}\bd{\Sigma}(\bd{\theta})^{-1}
  \right\}   +
  \left\{
    \bar{\bd{y}}-\bd{\mu}(\bd{\theta})
  \right\}^{\top}
  \bd{\Sigma}(\bd{\theta})^{-1}
  \left\{
    \bar{\bd{y}}-\bd{\mu}(\bd{\theta})
  \right\}
  -
  T.
  \label{fML}
\end{align}
Here, \(|\cdot|\) denotes the determinant of a matrix, \(\operatorname{tr}(\cdot)\) denotes the trace of a matrix, \(\bar{\bd{y}}\) is the sample mean vector, and \(\bd{S}\) is the sample covariance matrix. In addition, \(\bd{\mu}(\bd{\theta})\) and \(\bd{\Sigma}(\bd{\theta})\) denote the mean structure and covariance structure implied by the LGM, respectively. Parameter estimates \(\hat{\bd{\theta}}\) are typically obtained through iterative numerical procedures, such as the Newton--Raphson method. Although ML estimation is widely used in LGM (or SEM), it is empirically known that estimation difficulties, including nonconvergence and improper solutions, may occur, particularly when $T$ or $N$ is small, or when the model is misspecified, for example, by assuming a wrong functional form for the trajectory.

\subsection{Matrix decomposition factor analysis}
MDFA obtains solutions in FA models by exploiting eigenvalue decomposition. Let \(\bd{Y}\) (\(N\times T\)) be a column-centered data matrix consisting of \(N\) observations on \(T\) observed variables, where \(T\) is equal to the number of time points in the present context. Using matrix notation, the MDFA model can be expressed as
\begin{gather}
  \bd{Y}=\bd{F}\bd{\Lambda}^{\top}+\bd{U}\bd{\Psi}+\bd{E}
  =\bd{Z}\bd{B}^{\top}+\bd{E},
  \label{eq:mdfa_model}
\end{gather}
where \(\bd{F}\) (\(N\times r\)) is the common factor score matrix, \(\bd{U}\) (\(N\times T\)) is the unique factor score matrix, \(\bd{\Lambda}\) (\(T\times r\)) is the factor loading matrix, \(\bd{\Psi}\) (\(T\times T\)) is a diagonal matrix whose diagonal elements contain the square roots of the unique variances, and \(\bd{E}\) (\(N\times T\)) is an error matrix whose elements represent discrepancies between the implied structure and the observed data. Here, \(r\) denotes the number of assumed common factors.

Here, the following constraints are imposed on the factor scores:
\begin{gather}
  \frac{1}{N}\bd{F}^{\top}\bd{F}=\bd{I}_r,\qquad
  \frac{1}{N}\bd{U}^{\top}\bd{U}=\bd{I}_T,\qquad
  \bd{F}^{\top}\bd{U}=\bd{0}_{r\times T}.\label{MDFAassumption}
\end{gather}
Furthermore, both \(\bd{F}\) and \(\bd{U}\) are now assumed to be column-centered.

Unlike conventional FA model, MDFA treats \(\bd{F}\) and \(\bd{U}\) as fixed parameters and considers the following Frobenius norm-based least squares criterion for parameter estimation:
\begin{equation}
  f(\bd{Z},\bd{B})=\lVert \bd{E} \rVert^2=
  \left\lVert
    \bd{Y}-\left(\bd{F}\bd{\Lambda}^{\top}+\bd{U}\bd{\Psi}
    \right)
  \right\rVert^2
  =
  \left\lVert
    \bd{Y}
    -
    \bd{Z}\bd{B}^{\top}
  \right\rVert^2.
  \label{eq:mdfa_loss}
\end{equation}
Here, \(\bd{Z}=[\bd{F},\bd{U}]\) and \(\bd{B}=[\bd{\Lambda},\bd{\Psi}]\). This criterion can be rewritten as
\begin{gather}
  \left\lVert
    \bd{Y}
    -
    \bd{Z}\bd{B}^{\top}
  \right\rVert^2
  =
  \left\lVert
    \bd{Y}
    -
    \bd{Z}\bd{S_{YZ}}^{\top}
  \right\rVert^2
  +
  N
  \left\lVert
    \bd{S_{YZ}}
    -
    \bd{B}
  \right\rVert^2,
\end{gather}
which becomes a function of
$\bd{S_{YZ}}=\frac{1}{N}\bd{Y}^{\top}\bd{Z}=
  \left[
    \frac{1}{N}\bd{Y}^{\top}\bd{F},
    \frac{1}{N}\bd{Y}^{\top}\bd{U}
  \right]
  =[\bd{S_{YF}},\bd{S_{YU}}]$
(\citealp{Adachi2018}, Theorem~2.1). This expression indicates that only the second term $N\left\lVert \bd{S_{YZ}}-\bd{B} \right\rVert^2$ depends on \(\bd{B}\), and that the solution can be obtained as
\begin{gather}
  \hat{\bd{\Lambda}}=\bd{S_{YF}},
  \hspace{5mm}
  \hat{\bd{\Psi}}=\operatorname{diag}(\bd{S_{YU}}).
  \label{estB}
\end{gather}

Let \(\bd{L}\) be the orthonormal matrix consisting of eigenvectors, and let \(\boldsymbol{\Delta}^2\) be the diagonal matrix consisting of the positive eigenvalues obtained from the eigenvalue decomposition of
\begin{gather}
  \bd{B}^{\top}\bd{S}\bd{B}
  =
  \bd{L}\bd{\Delta}^{2}\bd{L}^{\top}.
  \label{eigendecomp}
\end{gather}
Adachi \& Trendafilov (2018, Lemma~3.1) showed that \(\bd{S_{YZ}}\) can be calculated as
\begin{gather}
  \bd{S_{YZ}}
  =
  \bd{S}\bd{B}\bd{L}\bd{\Delta}^{-1}\bd{L}^{\top}.
  \label{estSYZ}
\end{gather}
Furthermore, the loss function in Equation~\eqref{eq:mdfa_loss} can also be written as
$N\,\operatorname{tr}\left\{
    \bd{S}-\bd{B}\bd{B}^{\top}
  \right\}$ (\citealp{Adachi2019}).

Using these relations, the estimate of \(\bd{B}\) is obtained through the following procedure:
\begin{enumerate}
  \item Prepare initial values \(\bd{\Lambda}^{(0)}\) and \(\bd{\Psi}^{(0)}\), and let
 $\bd{B}^{(0)}=[\bd{\Lambda}^{(0)},\bd{\Psi}^{(0)}].
  $

  \item Using \(\bd{B}^{(0)}\) and \(\bd{S}\), perform the eigenvalue decomposition
  ${\bd{B}^{(0)}}^{\top}\bd{S}\bd{B}^{(0)}
    =\bd{L}\bd{\Delta}^{2}\bd{L}^{\top}.$

  \item Compute \(\bd{S_{YZ}}\) as
  $\bd{S_{YZ}}=[\bd{S_{YF}},\bd{S_{YU}}]=
    \bd{S}\bd{B}^{(0)}\bd{L}\bd{\Delta}^{-1}\bd{L}^{\top}.
  $

  \item Update \(\bd{B}^{(0)}\) to
  $\bd{B}^{(1)}=[\bd{S_{YF}},\operatorname{diag}(\bd{S_{YU}})].
  $

  \item Using \(\bd{B}^{(1)}\), compute the loss function as
  $\lVert \bd{E} \rVert^2=
    N\,\operatorname{tr}
    \left\{\bd{S}-\bd{B}^{(1)}\left(\bd{B}^{(1)}\right)^{\top}
    \right\}.
  $

  \item If the change in the value of the loss function satisfies a prespecified convergence criterion, \(\bd{B}^{(1)}\) is returned as the final estimate. Otherwise, set \(\bd{B}^{(0)}=\bd{B}^{(1)}\) and repeat Steps~2--5 until convergence is achieved.
\end{enumerate}

Reflecting the fact that \bd{S_{YU}} is computed based on the eigenvalue decomposition, the estimates of unique variance parameters obtained from \(\operatorname{diag}(\bd{S_{YU}})\) are, in principle, prevented from taking negative values. Moreover, reflecting the nonconvexity of the loss function in Equation ~\eqref{eq:mdfa_loss}, MDFA involves the risk of converging to local optima. To address this issue, multiple starts are used, in which the same algorithm described above is applied to each set of initial values, and the solution yielding the smallest value of \(\lVert \bd{E} \rVert^2\) is selected as the final solution.

\section{Matrix Decomposition Latent Growth Model}
In this section, we describe MDLGM, which serves as the basis for estimating the LGM used as the template model in the proposed MDLGM Tree. In MDLGM, LGM is estimated using the idea of matrix decomposition, as in the MDFA described in the previous section. As explained, the linear LGM and the quadratic LGM can be represented as FA models with \(r=2\) and \(r=3\) common factors, respectively. However, because LGM includes a mean structure, it differs from the original MDFA setting (Equation~\eqref{MDFAassumption}) in that the factor means are not zero and the common (growth) factors are not mutually uncorrelated. In this section, we propose MDLGM as an extension of the original MDFA and describe its estimation procedure.

\subsection{Formulation of MDLGM}
Using the fact that both the linear LGM and the quadratic LGM shown in Equations~\eqref{eq:linear LGM_scalar}--\eqref{eq:quadratic LGM_scalar} can be commonly represented as FA models, MDLGM first reformulates these LGMs using the following MDFA-like matrix form:
\begin{equation}
  \bd{Y^*}
  =
  \bd{F^*}\bd{\Lambda}^{\top}
  +
  \bd{U}\bd{\Psi}
  +
  \bd{E}.
  \label{eq:LGM_matrix}
\end{equation}
Let \(\bd{Z}=[\bd{F^*},\bd{U}]\) and \(\bd{B}=[\bd{\Lambda},\bd{\Psi}]\). Here, \(\bd{Y^*}\) is an \(N\times T\) observation matrix, which is not column-centered. For example, in the case of the linear LGM, \(\bd{F^*}\) is an \(N\times 2\) matrix whose first and second columns correspond to the intercept factor \(f_I\) and the slope factor \(f_S\), respectively, and this matrix is also not column-centered. The matrix \(\bd{\Lambda}\) is a known \(T\times 2\) matrix whose \((t,1)\) and \((t,2)\) elements are fixed to \(1\) and \(t-1\), respectively. The same formulation applies to the quadratic LGM. The matrix \(\bd{U}\) is an \(N\times T\) matrix of residual scores, where each column is standardized to have zero mean and unit variance. \(\bd{\Psi}\) is a \(T\times T\) diagonal matrix whose diagonal elements contain information on the residual standard deviations at the corresponding time points. The matrix \(\bd{E}\) is an \(N\times T\) error matrix that captures discrepancies between the implied structure and the observed data. In the MDLGM, the following constraints are imposed on $\boldsymbol{F^*}$ and $\boldsymbol{U}$:
\begin{gather}
  \frac{1}{N}\boldsymbol{F}^{*\top}\boldsymbol{F^*}=\boldsymbol{\Phi},\qquad
  \frac{1}{N}\boldsymbol{U}^{\top}\boldsymbol{U}=\boldsymbol{I}_T,\qquad
  \boldsymbol{F}^{*\top}\boldsymbol{U}=\boldsymbol{0}_{r\times T}. \label{MDLGMAassumption}
\end{gather}

That is, unlike the Equation~(\ref{MDFAassumption}) in MDFA, in the present MDLGM, \(\boldsymbol{\Lambda}\) is known, whereas the covariance matrix of the common (growth) factors, \(\bd{\Psi}\), must be estimated. Here again, MDLGM treats \(\bd{F^*}\) and \(\bd{U}\) as fixed parameters and estimates \(\boldsymbol{\Phi}\) and \(\bd{\Psi}\) by minimizing the loss function $\lVert \bd{E} \rVert^2 = \left\lVert \bd{Y^*} - \left( \bd{F^*}\bd{\Lambda}^{\top} + \bd{U}\bd{\Psi} \right) \right\rVert^2.$

This loss function can be decomposed as
\begin{align}
  \left\lVert
    \bd{Y^*}
    -
    \left(
      \bd{F^*}\bd{\Lambda}^{\top}
      +
      \bd{U}\bd{\Psi}
    \right)
  \right\rVert^2
  &=
  \left\lVert
    \left\{
      \bd{Y}
      -
      \left(
        \bd{F}\bd{\Lambda}^{\top}
        +
        \bd{U}\bd{\Psi}
      \right)
    \right\}
    +
    \bd{1_N}
    \left(
      \bar{\bd{y}}^{\top}
      -
      \bd{\mu}^{\top}\bd{\Lambda}^{\top}
    \right)
  \right\rVert^2
  \notag\\
&=
  \left\lVert
    \bd{Y}
    -
    \left(
      \bd{F}\bd{\Lambda}^{\top}
      +
      \bd{U}\bd{\Psi}
    \right)
  \right\rVert^2
  +
  N
  \left\lVert
    \bar{\bd{y}}^{\top}
    -
    \bd{\mu}^{\top}\bd{\Lambda}^{\top}
  \right\rVert^2
  \notag\\
&=
  \left\lVert
    \bd{Y}
    -
    \bd{Z}\bd{B}^{\top}
  \right\rVert^2
  +
  N
  \left\lVert
    \bar{\bd{y}}^{\top}
    -
    \bd{\mu}^{\top}\bd{\Lambda}^{\top}
  \right\rVert^2,
  \label{eq:mdLGM_loss_decomp}
\end{align}
where \(\bar{\bd{y}}\) and \(\bd{\mu}\) denote the vectors of sample means and factor means, respectively. Accordingly,
$\bd{Y}=\bd{Y^*}-\bd{1_N}\bar{\bd{y}}^{\top}, \bd{F}=\bd{F^*}
-\bd{1_N}\bd{\mu}^{\top}$ are column-centered matrices. Equation~\eqref{eq:mdLGM_loss_decomp} shows that the MDLGM loss function decomposes into two independent components: one associated with the estimation of \(\bd{\Phi}\) and \(\bd{\Psi}\), and the other associated with the estimation of \(\bd{\mu}\). The former has the same form as the MDFA loss function in Equation~\eqref{eq:mdfa_loss} and therefore it may appear that this component can be minimized using the original MDFA algorithm. However, unlike the MDFA, MDLGM needs to treat the common (growth) factor (co)variances as free parameters and \(\bd{\Lambda}\) as fixed parameters a priori, so the original MDFA algorithm cannot be directly used to minimize this loss function. 

\subsection{Reparameterization and estimation}
We now reparameterize the common factor scores as
\begin{gather}
\bd{F} =\widetilde{\bd{F}}\bd{\Phi}^{1/2},  
\end{gather}
where $N^{-1}\widetilde{\bd{F}}^{\top}\widetilde{\bd{F}}
  =\bd{I}_r$, and \(\bd{\Phi}^{1/2}\) is a matrix satisfying
$\bd{\Phi}=\bd{\Phi}^{1/2}\bd{\Phi}^{1/2}$. Using \(\widetilde{\bd{F}}\), we obtain the following relationship:
\begin{equation}
  \bd{F}\bd{\Lambda}^{\top}+\bd{U}\bd{\Psi}
  =\widetilde{\bd{F}}\bd{\Phi}^{1/2}\bd{\Lambda}^{\top}
  +\bd{U}\bd{\Psi}
  =\widetilde{\bd{F}}\widetilde{\bd{\Lambda}}^{\top}
  +\bd{U}\bd{\Psi}
  =\widetilde{\bd{Z}}\widetilde{\bd{B}}^{\top}.
  \label{eq:reparam}
\end{equation}
Here,
$\widetilde{\bd{\Lambda}}^{\top}=\bd{\Phi}^{1/2}\bd{\Lambda}^{\top}$, and $\widetilde{\bd{Z}}=[\widetilde{\bd{F}},\bd{U}]$, $
\widetilde{\bd{B}}=[\widetilde{\bd{\Lambda}},\bd{\Psi}].$
Thus, if \(\widetilde{\bd{\Lambda}}\) is regarded as a coefficient matrix that contains information on the (co)variances of the common (growth) factors as well as information on the (prespecified) factor loadings, the first term in Equation~\eqref{eq:mdLGM_loss_decomp},
$\left\lVert\bd{Y}-\bd{Z}\bd{B}^{\top}\right\rVert^2$,
can be expressed as
$\left\lVert
    \bd{Y}-\widetilde{\bd{Z}}\widetilde{\bd{B}}^{\top}
  \right\rVert^2.
$
This means that the loss function of MDLGM can now be expressed in a form corresponding to the original MDFA parametrization in which the common factor variances are fixed to one. Because the parameter \(\widetilde{\bd{B}}\) appears only in this term, \(\widetilde{\bd{B}}\) can be estimated using the MDFA algorithm.

Specifically,
$\left\lVert\bd{Y}-\widetilde{\bd{Z}}\widetilde{\bd{B}}^{\top} \right\rVert^2
$
is minimized and \(\widetilde{\bd{B}}\) is estimated through the following procedure:
\begin{enumerate}
  \item Specify initial values for \(\bd{\Phi}^{1/2}\) and \(\bd{\Psi}\), denoted by \(\bd{\Phi}^{1/2(0)}\) and \(\bd{\Psi}^{(0)}\), respectively. Then obtain the initial value of \(\widetilde{\bd{\Lambda}}\) as
  $\widetilde{\bd{\Lambda}}^{(0)}=\bd{\Lambda}\bd{\Phi}^{1/2(0)}$ and set $\widetilde{\bd{B}}^{(0)}
    = [\widetilde{\bd{\Lambda}}^{(0)},\bd{\Psi}^{(0)}].
  $

  \item Perform the eigenvalue decomposition
  $\left(\widetilde{\bd{B}}^{(0)}\right)^{\top}
    \bd{S}
    \widetilde{\bd{B}}^{(0)}=
    \bd{L}\bd{\Delta}^{2}\bd{L}^{\top}.
  $

  \item Compute \(\bd{S_{Y\widetilde{Z}}}\) as
  $\bd{S_{Y\widetilde{Z}}}=[\bd{S_{Y\widetilde{F}}},\bd{S_{YU}}]
    =\bd{S}\widetilde{\bd{B}}^{(0)}\bd{L}\bd{\Delta}^{-1}\bd{L}^{\top}.
  $

  \item Update \(\widetilde{\bd{B}}^{(0)}\) to
  $\widetilde{\bd{B}}^{(1)}=
    [\widetilde{\bd{\Lambda}}^{(1)},\bd{\Psi}^{(1)}].$
  Here,
  $\bd{\Psi}^{(1)}=\operatorname{diag}(\bd{S_{YU}})$,
  and \(\bd{\Phi}^{1/2(1)}\), which is needed to compute
  $\widetilde{\bd{\Lambda}}^{(1)}=\bd{\Lambda}\bd{\Phi}^{1/2(1)}$, is obtained as the quantity that minimizes the criterion
  $\left\lVert\bd{S_{Y\widetilde{F}}}-\bd{\Lambda}\bd{\Phi}^{1/2}\right\rVert^2.$ More
  specifically, using generalized inverse of $\boldsymbol{\Lambda}$ (denoted as $\boldsymbol{\Lambda}^{+}$), update $\boldsymbol{\Phi}^{1/2}$ as $\boldsymbol{\Phi}^{1/2(1)}=\boldsymbol{\Lambda}^{+}\bm{S_{Y\widetilde{F}}}$.

  \item Compute $ \lVert \bd{E} \rVert^2
    =N\,\operatorname{tr} \left\{\bd{S}-\widetilde{\bd{B}}^{(1)}
      \left(\widetilde{\bd{B}}^{(1)}\right)^{\top}
    \right\}.$

  \item If the change in the value of the loss function satisfies a prespecified convergence criterion, return \(\widetilde{\bd{B}}^{(1)}\) as the final estimate and estimate \(\bd{\Phi}\) as
  $\bd{\Phi}=\bd{\Phi}^{1/2(1)}\bd{\Phi}^{1/2(1)}$. 
  Otherwise, set \(\widetilde{\bd{B}}^{(0)}=\widetilde{\bd{B}}^{(1)}\) and repeat Steps~2--5 until convergence is achieved.
\end{enumerate}

For the same reason as in the MDFA described in the previous section, MDLGM involves the risk of converging to local optima. When multiple starts are used, the same algorithm described above is applied to each set of initial values, and the solution yielding the smallest value of \(\lVert \bd{E} \rVert^2\) is selected as the final solution.

The parameter included only in the second term of the loss function in Equation~\eqref{eq:mdLGM_loss_decomp},
\begin{equation}
  f(\bd{\mu})=
  \left\lVert
    \bar{\bd{y}}^{\top}-\bd{\mu}^{\top}\bd{\Lambda}^{\top}
  \right\rVert^2
  =\left\lVert \bar{\bd{y}}-\bd{\Lambda}\bd{\mu}\right\rVert^2,
  \label{eq:mu_loss}
\end{equation}
is \(\bd{\mu}\), and its estimation can be carried out independently of the iterative procedure described above. Because Equation~\eqref{eq:mu_loss} corresponds to a standard linear regression problem in which \(\bd{\mu}\) is the parameter vector, \(\bd{\mu}\) can be estimated by the ordinary least squares estimator. All parameters $\bd{\theta}=
  \left(
    \bd{\mu}^{\top},
    \operatorname{vech}(\bd{\Phi})^{\top},
    \operatorname{diag}(\bd{\Psi})^{\top}
  \right)^{\top}
$
are thus estimated.

\section{Matrix Decomposition Latent Growth Model Tree}
SEM Tree partitions samples by recursive partitioning using external covariates and characterizes between-group differences through differences in the parameters of prespecified template models specified as SEMs. In SEM Tree, a tree is generated by repeatedly determining whether to split a node, which covariate should be used for the split, and at which cut-off value, based on amount of changes in the deviance ($-2 \times \text{log-likelihood}$) before and after splitting a node (or based on likelihood ratio test), obtained from ML estimation of the template model.

The essence of the MDLGM Tree proposed in the present study is that the template models are estimated by MDLGM rather than by ML estimation. In addition, node splitting is conducted using a criterion that differs from likelihood-based criterion. Because MDLGM Tree is expected to provide solutions similar to those obtained by ML while more efficiently avoiding improper solutions, it is expected to accurately detect subgroups (nodes), and to do so with lower computational cost, even when the true value(s) of variance parameter(s) in a subgroup (node) in the true classification structure is close to zero.

More formally, the algorithm of MDLGM Tree can be summarized as follows:

\begin{enumerate}[leftmargin=2em]
  \item Fit the template model, such as a linear LGM, to the parent node (whole data with \(N\) individuals), by MDLGM. Using the estimates, compute the model-implied mean and covariance structures, \(\bd{\mu}(\hat{\bd{\theta}})\) and \(\bd{\Sigma}(\hat{\bd{\theta}})\), and calculate the squared Mahalanobis distance between the data and the predicted values; \(D_{\mathrm{before}}=\sum_{i = 1}^{N}(\bm{y}_i-\boldsymbol{\mu}(\hat{\boldsymbol{\theta}}))^\top \bm{S}^{-1} (\bm{y}_i-\boldsymbol{\mu}(\hat{\boldsymbol{\theta}}))\), where $\bm{y}_i=(y_{i1},\dots,y_{iT})^\top$. 

  \item Split the data at a potential cut-off point of a covariate in the dataset into two child nodes, denoted by nodes $A$ and $B$. Fit the template model to each child node by MDLGM, then compute the sum of the distances, \(D_{\mathrm{after}}=\sum_{i = 1}^{N_A}(\bm{y}_{iA}-\boldsymbol{\mu}(\hat{\boldsymbol{\theta}}_A))^\top \bm{S}^{-1} (\bm{y}_{iA}-\boldsymbol{\mu}(\hat{\boldsymbol{\theta}}_A)) + \sum_{i = 1}^{N_B} (\bm{y}_{iB}-\boldsymbol{\mu}(\hat{\boldsymbol{\theta}}_B))^\top \bm{S}^{-1} (\bm{y}_{iB}-\boldsymbol{\mu}(\hat{\boldsymbol{\theta}}_B))\), where \(N_A+N_B=N\), and the subscripts $A$ and $B$ denote quantities associated with nodes $A$ and $B$, respectively.

  \item Repeat Step~2 for all potential cut-off points of all covariates. Determine the optimal covariate and its cut-off point that minimizes \(D_{\mathrm{after}}\).

  \item Compare the minimized value of \(D_{\mathrm{after}}\) with \(D_{\mathrm{before}}\), and split the parent node into two child nodes if the improvement in the distance is larger than a given level (\(cp\) value, such as whether the distance is reduced by at least \(1\%\) of \(D_{\mathrm{before}}\)). If no sufficiently large improvement is observed, the split is not conducted.

  \item Repeat Steps~1--4 for the all child nodes, treating them as parent nodes, until no further improvement in the distance is obtained.
\end{enumerate}

As in MDFA, MDLGM does not assume a probability distribution for the data. In light of this feature, in the present study, node splitting is basically conducted based on the Mahalanobis distance rather than on the log-likelihood.

\section{Simulation Study}
We conduct a simulation study to compare various aspects of performance between MDLGM Tree and SEM Tree. Specifically, we compare the methods in terms of the computational time required to estimate the tree, the accuracy of estimating the true number of nodes, the occurrence of improper solutions (i.e., whether variance parameters estimates are negative or close to zero, or whether the covariance matrix of the growth factors is singular), the classification accuracy of individuals into nodes, the bias and RMSE of the estimates of the parameters $\bd{\theta}$ in the template model (LGM) within each subgroup (node), and the correlations among the estimates of $\bd{\theta}$ obtained from the different tree estimation methods.

\subsection{Method}
\subsubsection{Treee estimation method}
In general, several modifications are possible in tree estimation with respect to the estimation algorithm and splitting criterion. Taking this into account, we included two settings for MDLGM Tree and four settings for SEM Tree, respectively, in the current simulation. For MDLGM Tree, in addition to the algorithm based on the Mahalanobis distance described in the previous section (MDLGM Tree1), we also examined another MDLGM Tree in which splits are determined based on the $cp$ criterion which is often used in SEM Tree, that is, the improvement in the deviance relative to that before splitting under the assumption of multivariate normality. Hereafter, this method is referred to as MDLGM Tree2. We developed custom functions for implementing these MDLGM Tree methods using the \texttt{MplusTrees} function included in the \texttt{MplusTrees} package in R \citep{Serang2025}.

The four settings for SEM Tree were as follows. The first setting fully corresponds to the algorithm described in the previous section; hereafter, this method is referred to as SEM Tree1. The second setting applies constrained ML estimation when implementing SEM Tree1, imposing constraints so that the estimated variance parameters included in the template model (LGM) are positive; this method is referred to as SEM Tree2. The third setting estimates the parameters in the template model (LGM) by Bayesian estimation using Gibbs sampling; specifically, $\bd{\theta}$ is estimated by the posterior mode, and the optimal split point is searched for based on the posterior mean of the deviance. This method is referred to as SEM Tree3. Finally, the fourth setting uses ML estimation but avoids a split if warnings are issued based on the \texttt{exclude.code} argument of the \texttt{semtree} function (\citealp{Brandmaier2025}), because it implies that an improper solution (e.g., the estimated covariance matrix of the growth factors becomes singular) occurs, even if a certain improvement in the log-likelihood based on $cp$ is observed. This method is referred to as SEM Tree4. Estimation for these four SEM Tree methods was conducted using custom functions based on the \texttt{MplusTrees} function in R and Mplus version 8 \citep{MuthenMuthen19982017}.

\subsubsection{Manipulated factors}
\textit{True and specified template models.}\quad We considered three combinations of the data-generating model (true model) and the specified template model. Specifically, these combinations were as follows: both the true and specified models were linear LGMs; both were quadratic LGMs; and the true model was a quadratic LGM whereas the specified model was a linear LGM.

\textit{True number of terminal nodes and data size.}\quad The true number of terminal nodes was fixed at 4, and the proportion of samples belonging to each (terminal) node was set to $.25$. The total sample size $N$ had two levels, $N=500$ and $N=2000$. Thus, the sample size in each (terminal) node was 125 when $N=500$, and 500 when $N=2000$. For the number of time points $T$, two conditions were considered, $T=5$ and $T=10$.

\textit{Variances of growth factors.}\quad Two levels were specified for the variances of the slope factors. When the true model was a linear LGM, based on the values of the growth factor variances in previous research \citep{Usami2019}, the variance of the intercept factor ($\phi_I^2$) was fixed at 1.0, and the variance of the slope factor ($\phi_S^2$) was manipulated to be either 0.1 or 1.0. The factor covariance ($\phi_{IS}$) was set so that the growth factor correlation was 0.3. The residual variance at each time point ($\psi_t^2$) was set so that the proportion of variance explained by the trajectory at each time point became 0.8. As a result, the minimum value of the specified residual variance was 0.250. These parameter values were assumed to be equal across the four terminal nodes.

When the true model was a quadratic LGM, following the same previous research, the variance of the intercept factor ($\phi_I^2$) was fixed at 1.0, the variance of the first-order slope factor ($\phi_{S1}^2$) was fixed at 0.1, and the variance of the second-order slope factor ($\phi_{S2}^2$) was manipulated to be either 0.01 or 0.03. The covariances among the growth factors were set so that the corresponding factor correlations were all 0.3. The residual variance at each time point ($\psi_t^2$) was set so that the proportion of variance explained by the trajectory at each time point became 0.8. These parameter values were also assumed to be equal across the four terminal nodes.

\textit{Degree of separation among terminal nodes.}\quad The degree of separation, which represents the (standardized) distance between the centers (means) of the terminal nodes, was also manipulated. When the true model was a linear LGM, the means of the intercept factor ($\mu_I$) and the slope factor ($\mu_S$) were set to 1.0 and 0.1, respectively. Using these factor means and the specified values of the factor (co)variances as described above, we first calculated the ideal trajectory values at each time point for the second and subsequent nodes so that the differences between the ideal trajectory values in each node and the model-implied means in the first node become $d$, $2d$, and $3d$ times the standard deviation, which was common across terminal nodes. Two values, $d=1$ and $d=2$, were specified for $d$, which represents the degree of separation in this sense. Thus, for example, when $d=1$, the ideal trajectory values in the second node were set so that its difference from the model-implied means in the first node was one standard deviation. Similarly, the ideal trajectory values in the third node were set so that its difference from the model-implied means in the first node was $2d=2$ standard deviations. We then set the means of the growth factors in each terminal node that optimally represented these specified ideal trajectory values in the sense of minimizing the sum of squared differences between model-implied means and ideal trajectory values across time points.

When the true model was a quadratic LGM, the mean of the intercept factor ($\mu_I$) in the first node was set to 1.0, the mean of the first-order slope factor ($\mu_{S1}$) was set to 0.1, and the mean of the second-order slope factor ($\mu_{S2}$) was set to 0.01. In the same manner as above, we first calculated the ideal trajectory values for the second and subsequent terminal nodes so that the differences between the ideal trajectory values in each node and the model-implied means in the first node become $d$, $2d$, and $3d$ times the standard deviation, under the setting of $d=1$ or $d=2$. We then set the means of the growth factors in each terminal node that optimally represented the specified ideal trajectory values in the sense of minimizing the sum of squared differences between model-implied means and ideal trajectory values across time points.

In this way, totally five factors were manipulated in each tree estimation method, yielding a total of $3$ (true and specified template model) $\times 2$ (sample size) $\times 2$ (the number of time points) $\times 2$ (variances of growth factors) $\times 2$ (degree of separation) $=48$ conditions. In each condition, data generation and estimation were repeated 100 times.

\subsubsection{Other settings and evaluation method}
\textit{Covariates.}\quad In addition to the data $\bd{y}$ to which the LGM was fitted, each generated data set included three external binary covariates, $z_1$, $z_2$, and $z_3$, used for splitting. Thus, the maximum depth of the estimated tree was 3. Among these covariates, $z_3$ was assumed to be unrelated to the true splitting and classification structure, and its values were generated independently from Be(0.5).

The covariates $(z_1,z_2)$ were generated as follows. The true covariate values were set to $(z^*_1,z^*_2)=(0,0)$ for samples belonging to the first terminal node, $(0,1)$ for those belonging to the second terminal node, $(1,0)$ for those belonging to the third terminal node, and $(1,1)$ for those belonging to the fourth terminal node. To represent imperfect correlations with \bd{y} and the influence of possible measurement errors in observed covariates, we assumed a situation in which $20\%$ of the samples belonging to each terminal node showed observed covariates that deviated from the true values. For example, for randomly selected $20\%$ of the samples belonging to the first node, the observed covariates $(z_1,z_2)$ were set to $(0,1)$, rather than $(0,0)$. Similarly, for $20\%$ of the samples belonging to the third node, the observed covariates were set to either $(0,1)$ or $(1,1)$, rather than $(1,0)$, with probability $1/2$ each.

\textit{Specification of the $cp$ value.}\quad A small preliminary simulation was conducted to search for an appropriate value of $cp$. Specifically, five data sets were generated under each of the conditions described above (except for the conditions in which the true model was a quadratic LGM whereas the specified model was a linear LGM), and the two methods MDLGM Tree1 and SEM Tree1 were applied to these data sets. We then examined in advance which value of $cp$, among $cp=0.01$, $0.005$, $0.003$, and $0.001$, yielded the highest accuracy in estimating the number of nodes for splits based on the Mahalanobis distance (MDLGM Tree1) and for splits based on the log-likelihood/deviance (SEM Tree1). As a result, for both methods, the highest estimation accuracy was obtained when $cp=0.003$ (see the Supplementary Materials S1 for the details). Therefore, in the current simulation, $cp$ was set to $0.003$ for all six estimation methods.

\textit{Initial values and multiple starts.}\quad Among six estimation methods (except for SEM Tree3 based on Bayes estimation), the initial values and convergence criteria  
were set to be the same. In MDLGM Tree1 and MDLGM Tree2,
to avoid possible local optima, estimates were obtained using three multiple starts, and the estimate with the smallest value of the loss function, $\lVert \bd{E} \rVert^2$, was adopted as the final estimate. The initial values were created by using the \texttt{lavInspect} function and \texttt{growth} function with the \texttt{do.fit} argument \texttt{FALSE} and \texttt{mimic} argument \texttt{Mplus} in the \texttt{lavaan} package \citep{Rosseel2025}, or by referring to the Mplus User's Guide \citep{MuthenMuthen19982017}. The convergence criteria was also set by referring to the Mplus User's Guide \citep{MuthenMuthen19982017}.

\textit{Specification of priors in SEM Tree3.}\quad When Bayesian estimation was conducted (SEM Tree3), prior distributions for the parameters were specified as follows, based on \cite{AsparouhovMuthen2021}. For the means of the growth factors, the minimum and maximum true values across all 48 conditions, calculated using the procedure described above, were 0.010 and 7.770, respectively. Therefore, a normal distribution with mean 0 and variance 100 was set for priors so that all of these true values were contained within the range of the mean $\pm 1$ standard deviation.

For the residual variances, the minimum and maximum values across all 48 conditions were 0.250 and 60.003, respectively. Therefore, an inverse-gamma distribution with shape parameter $\alpha=3$ and scale parameter $\beta=200$ was assumed so that all of these true values were contained within the range of the mean $\pm 1$ standard deviation. For the covariance matrix of the growth factors, an inverse-Wishart distribution was assumed. The scale matrix was specified as a diagonal matrix with diagonal elements equal to 20, and the degrees of freedom were set to 7 when the true model was a linear LGM and to 8 when the true model was a quadratic LGM. The rationale for these settings is as follows. Specifically, because the minimum and maximum values of the growth factor variances across all 48 conditions, obtained through the procedure described above, were 0.010 and 1.000, respectively, we specified the priors so that all of these true values were contained within the range of the mean $\pm 1$ standard deviation and so that the influence of the priors would not be overly strong. To this end, the marginal prior distribution for each growth factor variance was set to be an inverse-gamma distribution with shape parameter $\alpha=3$ and scale parameter $\beta=10$. Furthermore, the prior means of the covariances among the growth factors were set to 0. As a result, the inverse-Wishart distributions described above, which can express these prior specifications, were used. 

Following to the default of Mplus, in Gibbs sampling, two Markov chain Monte Carlo (MCMC) chains were used to estimate parameters. In all the chains, the number of iterations was set to 50,000, and the thin was set to 1. The convergence was evaluated by the Gelman-Rubin convergence criterion based on the potential scale reduction (PSR) factor \citep{GelmanRubin1992} and the threshold for the PSR value was set to 1.1. The first halves of the samples were discarded as being the part of burn-in phases, and parameters were estimated based on the second halves.

\textit{Evaluation of improper solutions.}\quad The occurrence of improper solutions was evaluated based on whether each estimated variance parameter was smaller than various thresholds, namely $10^{-2}$, $10^{-3}$, $10^{-4}$, and $10^{-5}$. In addition, following \cite{AsparouhovMuthen2022}, when the condition number calculated from the estimated covariance matrix of the growth factors, defined as the ratio of the smallest eigenvalue to the largest eigenvalue, was smaller than $10^{-10}$, the matrix was regarded as singular and the obtained solution was treated as improper.

\textit{Evaluation of classification accuracy and bias and RMSE of the estimate.}\quad Only when the number of terminal nodes ($=4$) was correctly estimated, we calculated, for each condition, the proportion of samples that were correctly assigned to their true nodes among those whose covariate values were not contaminated, corresponding to 80\% of the samples in each terminal node. Similarly, only when the number of terminal nodes was correctly estimated under conditions in which the true model and the template model were identical, we calculated the bias and RMSE of the estimate of each parameter. In addition, under these conditions, we calculated the correlations of estimated parameters in the 4 terminal nodes among different estimation methods.

\textit{Computational environment.}\quad The simulation code was created and executed in R 4.4.3 \citep{RCoreTeam2025}, and the simulation was conducted on a Surface Laptop for Business 7th Edition. The processor of this PC was an Intel(R) Core(TM) Ultra 7 266V (2.20 GHz), with 16.0 GB of RAM. The system type was a 64-bit operating system with an x64-based processor.

\subsection{Results}
\subsubsection{Computational time} 
\begin{figure} \centering \includegraphics[width=0.9\textwidth]{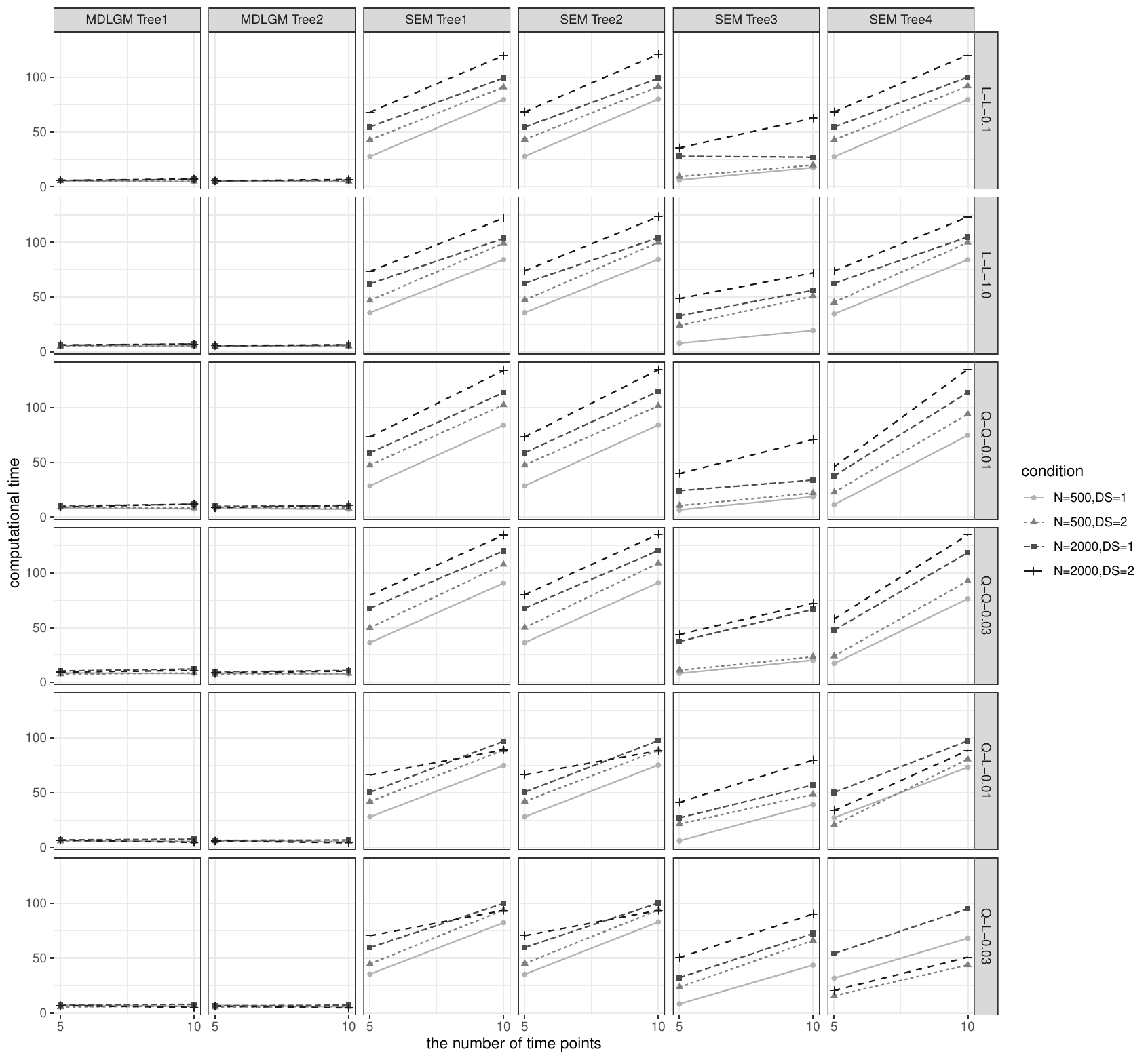} \caption{\\Average computational time (second) required to estimate a tree. SEM Tree1 denotes ML estimation, SEM Tree2 denotes constrained ML (CML) estimation, SEM Tree3 denotes Bayesian estimation, and SEM Tree4 denotes ML estimation with an algorithm that avoids node splitting based on any estimated model that produces a warning. MDLGM Tree1 and MDLGM Tree2 denote splitting based on the Mahalanobis distance and deviance, respectively. $DS$ denotes the degree of separation, and the three-element labels on the vertical axis (e.g., Q-L-0.01) indicate, in order, the true model (linear or quadratic), the template model (linear or quadratic), and the specified value of the slope factor variance.} \label{fig:comp_time} \end{figure}

Figure~\ref{fig:comp_time} shows the average computational time required to estimate a tree for each method across conditions. In the figure, $DS$ denotes the degree of separation ($d$), and the three-element labels on the vertical axis (e.g., Q-L-0.01) indicate, in order, the true model (linear or quadratic), the template model (linear or quadratic), and the specified value of the slope factor variance. 

The computational time of the MDLGM Tree was shorter under all conditions than that of ML-based SEM Tree methods (SEM Tree1, SEM Tree2, and SEM Tree4) and Bayesian SEM Tree method (SEM Tree3). In particular, when $T$ was large, the MDLGM Tree was more than 10 times faster on average. This suggests faster convergence in estimating the LGM (template model) in the MDLGM algorithm.  

In addition, whereas the computational time of the SEM Tree methods was relatively strongly affected by the number of time points, sample size and degree of separation, these factors had little apparent effect on the computational time of the MDLGM Tree methods. On the whole, in the SEM Tree methods, as the number of time points, sample size and degree of separation increased, the computational time increased. However, in some conditions, this was not the case. For example, under the conditions that the true model was a quadratic LGM whereas the specified model was a linear LGM, SEM Tree4 showed different tendencies from the other SEM Tree methods. Namely, the computational time of the SEM Tree4 became shorter when $DS = 2$. Since SEM Tree4 avoids node splitting based on any estimated model that produces a warning, the search space for SEM Tree4 may have been narrowed when $DS = 2$, where improper solutions were more likely to occur. This, in turn, may have resulted in shorter computational time. Likewise, when the model was misspecified, the computational times of SEM Tree1 and SEM Tree2 under the condition $T = 10$, $N = 2000$, and $DS = 2$ showed a slightly different trend from those of the other estimation methods. This may be attributable to the fact that, due to model misspecification, cases in which the true nodes were not appropriately recovered occurred more frequently.

\subsubsection{Estimation of the true number of nodes} 
\begin{figure} \centering \includegraphics[width=0.9\linewidth]{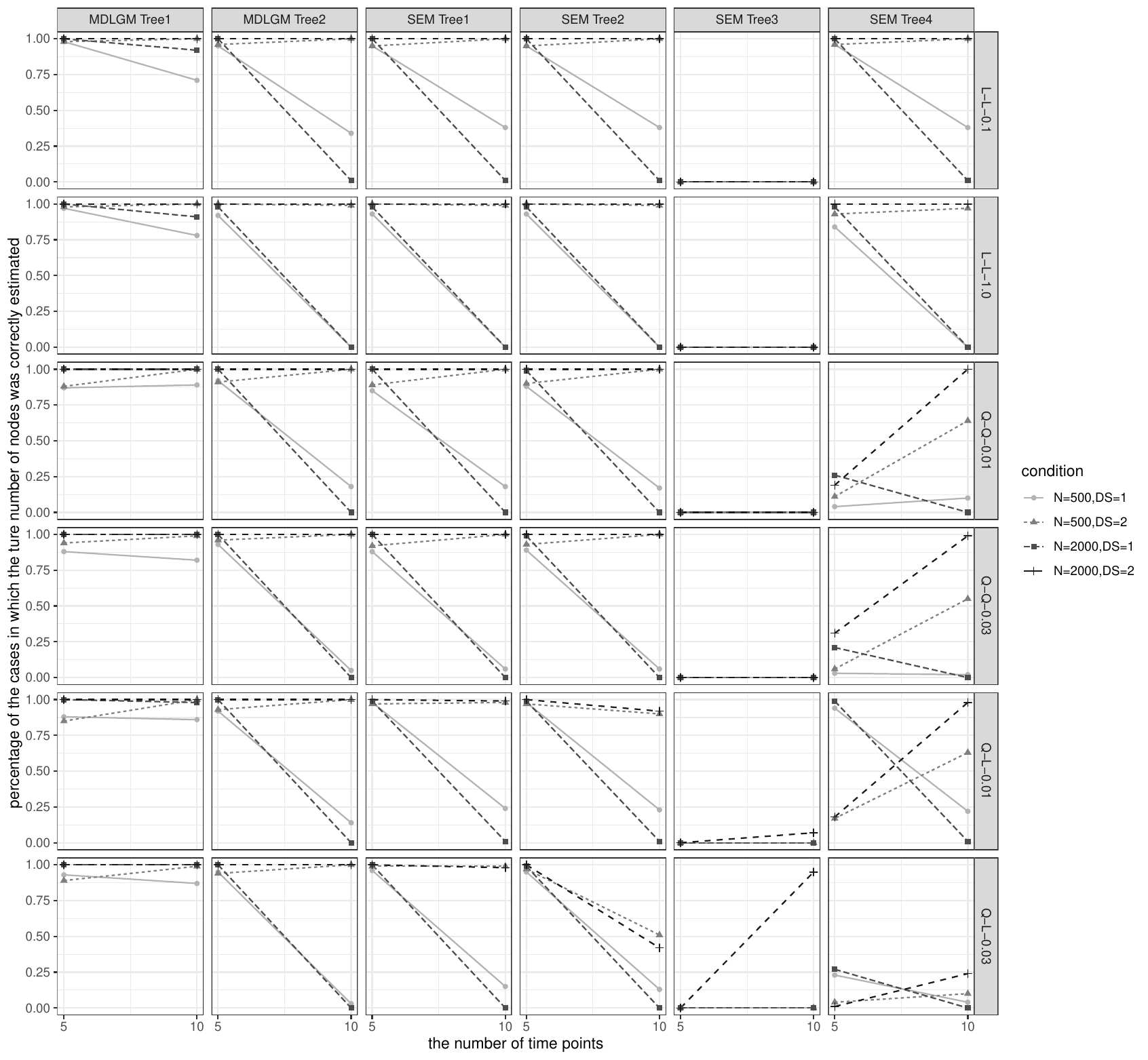} \caption{\\Proportion in which the true number of terminal nodes (= 4) was correctly estimated under each condition. SEM Tree1 denotes ML estimation, SEM Tree2 denotes constrained ML (CML) estimation, SEM Tree3 denotes Bayesian estimation, and SEM Tree4 denotes ML estimation with an algorithm that avoids node splitting based on any estimated model that produces a warning. MDLGM Tree1 and MDLGM Tree2 denote splitting based on the Mahalanobis distance and deviance, respectively. $DS$ denotes the degree of separation, and the three-element labels on the vertical axis (e.g., Q-L-0.01) indicate, in order, the true model (linear or quadratic), the template model (linear or quadratic), and the specified value of the slope factor variance.}
\label{fig:node_number_est} \end{figure}

Figure~\ref{fig:node_number_est} shows, for each estimation method and condition, the proportion in which the true number of terminal nodes (= 4) was correctly estimated. The Figure suggests that the proposed MDLGM Tree based on Mahalanobis distance (MDLGM Tree1) showed the proportion which remained high under all conditions.

Among the methods, the Bayesian SEM Tree (SEM Tree3) failed to correctly estimate the number of nodes under most conditions. One possible reason is the influence of the prior distribution, which may have led to an underestimation of the degree of improvement achieved by node splitting. This suggests that the performance of the SEM Tree3 can be highly sensitive to how the prior distributions are specified and how the splitting criterion is defined, and therefore its results should be interpreted with caution (see also the later subsection when default priors in Mplus were used). Only when $T = 10$, $DS = 2$, and the model was misspecified did SEM Tree3 show high accuracy.

In the MDLGM Tree and the SEM Tree methods that use log-likelihood-based splitting (MDLGM Tree2, SEM Tree1, SEM Tree2, and SEM Tree4), the accuracy of estimating the number of nodes tended to decrease when the number of time points increased to 10 under many conditions with $DS = 1$. This was likely because the differences between true nodes became less apparent at specific time points under those conditions because of the present procedure for the true values of parameters, in which those were specified so that the degree of separation between nodes at each time point would be as close as possible to the specified value ($DS$).

In addition, under the conditions that the true model was a quadratic LGM and $T=5$, SEM Tree4 shows lower estimation accuracy for the true number of nodes. As with the computational time discussed above, this may be because SEM Tree4 avoids node splitting based on any estimated model that produces a warning. Consequently, the search space for this method may have been narrowed under these conditions, where improper solutions were more likely to occur.

\subsubsection{Improper solutions} 
\begin{figure} \centering \includegraphics[width=0.9\linewidth]{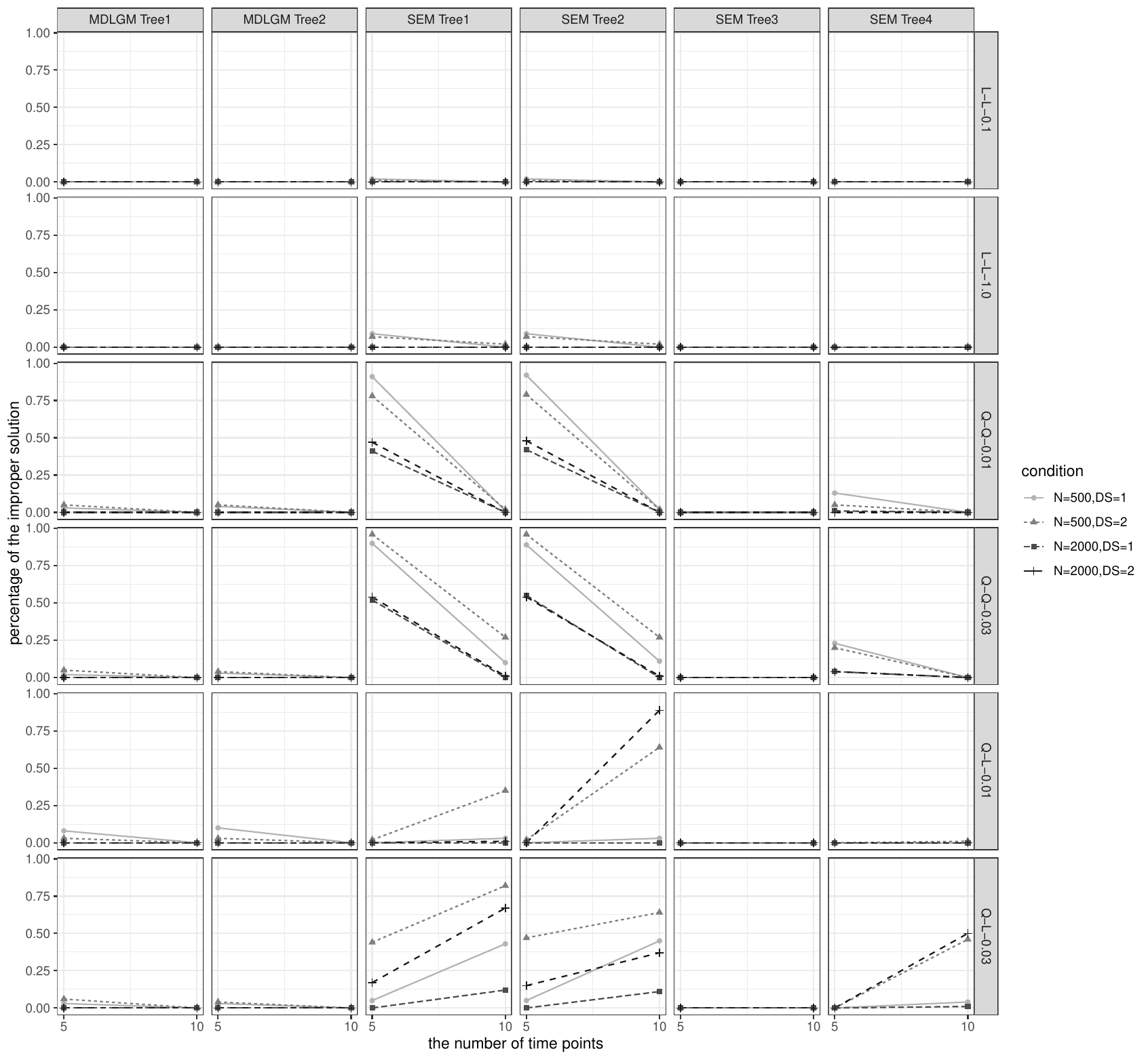} \caption{\\Proportion in which improper solutions occurred, using $10^{-5}$ as the threshold for variance parameters estimates. SEM Tree1 denotes ML estimation, SEM Tree2 denotes constrained ML (CML) estimation, SEM Tree3 denotes Bayesian estimation, and SEM Tree4 denotes ML estimation with an algorithm that avoids node splitting based on any estimated model that produces a warning. MDLGM Tree1 and MDLGM Tree2 denote splitting based on the Mahalanobis distance and deviance, respectively. $DS$ denotes the degree of separation, and the three-element labels on the vertical axis (e.g., Q-L-0.01) indicate, in order, the true model (linear or quadratic), the template model (linear or quadratic), and the specified value of the slope factor variance.} \label{fig:improper_soluion_5} \end{figure}

Figure~\ref{fig:improper_soluion_5} shows the proportion in which improper solutions occurred for each method and condition. For variance parameters, cases in which the estimate was smaller than $10^{-5}$ and thus stuck at the boundary were counted as improper solutions here. As expected given the nature of the estimation procedure, the MDLGM Tree methods produced almost no improper solutions. There are a few occurrence of improper solutions (at most about 5--10\%) under the condition that the true model is a quadratic LGM, mainly because estimated covariance matrix of growth factors became singular. The Bayesian SEM Tree (SEM Tree3), which utilizes prior distributions, showed no improper solutions. In contrast, as predicted by the previous research, the ML-based SEM Tree method (e.g., SEM Tree1) frequently showed improper solutions. In particular, under the condition that the true model was a quadratic LGM, there were the conditions in which the proportion exceeded 50\%. 

In the CML-based SEM Tree method (SEM Tree2), the improper solutions were observed just as frequently as in the SEM Tree1. Likewise, even for SEM Tree4, which avoids node splitting when an improper solution may occur, nonzero proportions were observed under some conditions. This suggests that, even when node splitting was not performed because of a warning and the number of nodes was consequently estimated to be one, an improper solution had already occurred during the fitting of the parent node.
Taken together, these results indicate that SEM Tree4 is not useful, either in terms of accurately estimating the true number of nodes or in terms of avoiding improper solutions.

In the SEM Tree methods using log-likelihood-based splitting, when both the true and specified models were quadratic LGMs, the occurrence of improper solutions decreased as the number of time points and sample size decreased. On the other hand, when the model was misspecified, the occurrence of improper solutions increased as the number of time points and the degree of separation increased, presumably reflecting the greater impact of model misspecification on the parameter estimates under these conditions.

These differences among the estimation methods remained essentially the same even when different thresholds ($10^{-2}$, $10^{-3}$, and $10^{-4}$) were used as the criterion for defining improper solutions based on the variance parameter estimates (see the Supplementary Materials S2 for details). However, when the threshold was set to $10^{-2}$, improper solutions were observed even with the MDLGM Trees. This result indicates that, although the variance parameter estimates in the MDLGM cannot become negative, they may nevertheless take very small positive values under certain conditions.

\subsubsection{Classification accuracy} 
\begin{figure} \centering \includegraphics[width=0.85\linewidth]{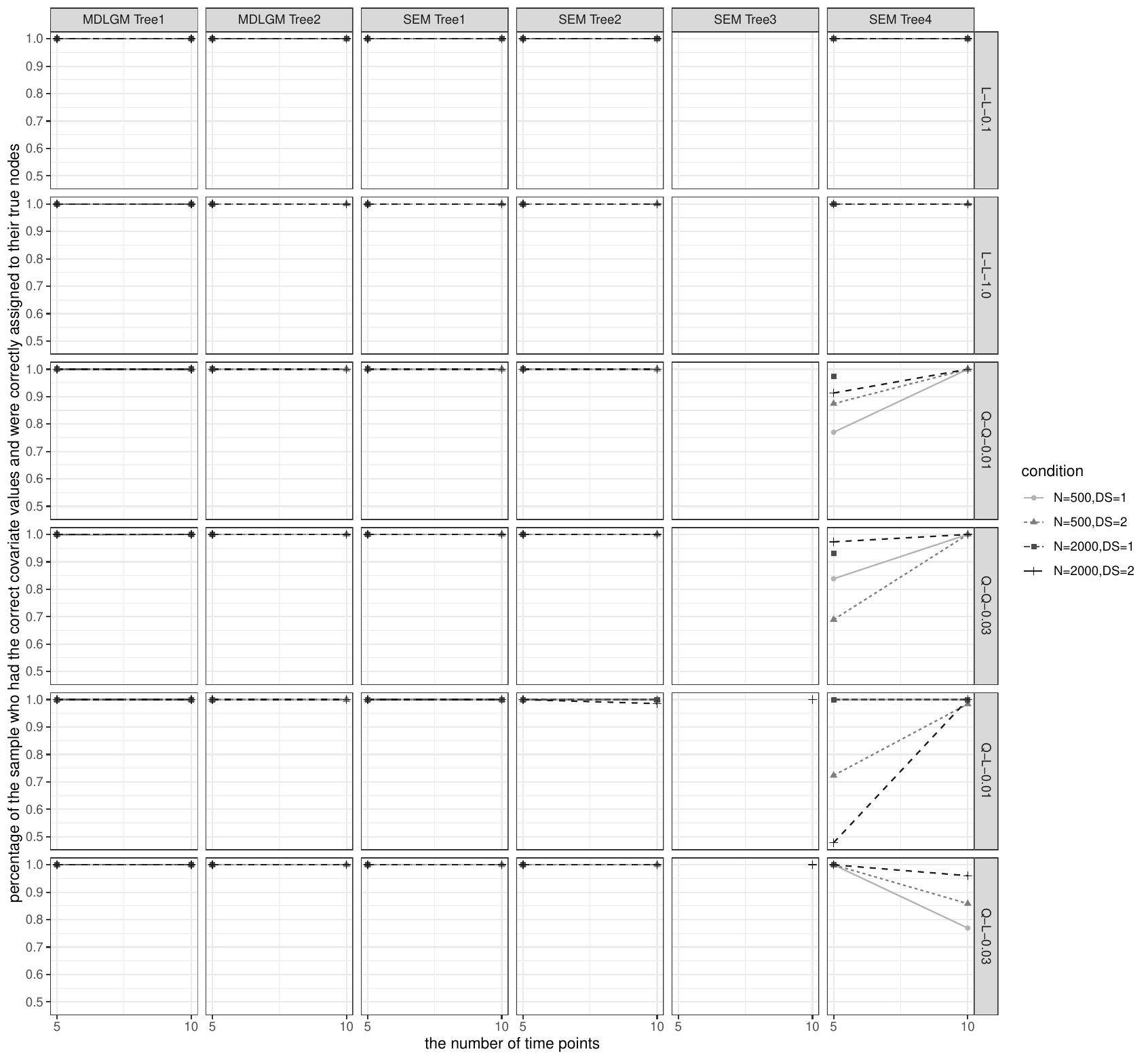} 
\caption{\\Average proportion of samples that were correctly classified to their true nodes among those whose covariate values were not contaminated over replications in which the number of nodes was correctly estimated. SEM Tree1 denotes ML estimation, SEM Tree2 denotes constrained ML (CML) estimation, SEM Tree3 denotes Bayesian estimation, and SEM Tree4 denotes ML estimation with an algorithm that avoids node splitting based on any estimated model that produces a warning. MDLGM Tree1 and MDLGM Tree2 denote splitting based on the Mahalanobis distance and deviance, respectively. $DS$ denotes the degree of separation, and the three-element labels on the vertical axis (e.g., Q-L-0.01) indicate, in order, the true model (linear or quadratic), the template model (linear or quadratic), and the specified value of the slope factor variance.}  \label{fig:node_membership_est} \end{figure}

Figure~\ref{fig:node_membership_est} shows, for each condition, the average proportion of individuals whose node membership was correctly classified, restricted to cases in which the number of nodes was correctly estimated and to individuals whose observed covariates were not contaminated. Here, results are not displayed for combinations of the estimation method and condition in which the number of nodes was never correctly estimated in 100 replications, such as Bayesian SEM Tree (SEM Tree3; see also the preceding result of estimating the number of nodes). 

The results indicate that, for each method, once the number of terminal nodes was correctly estimated, individuals were also generally classified into the correct nodes in the current simulation conditions. However, for SEM Tree4, the correct classification rate was not 1.0 and lower than those of the other methods under some conditions. SEM Tree4 avoids node splitting when a warning that may indicate an improper solution is issued. In the present case, it may have avoided node splitting based on $z_1$ and $z_2$ because of the occurrence of improper solutions, while node splitting based on $z_3$ was still performed, potentially leading to a decrease in the classification accuracy. The plausibility of this interpretation was confirmed through an additional simulation (see the Supplementary Materials S3 for the details).

\subsubsection{Bias and RMSE of the estimates} 
We calculated the bias and RMSE of the estimate of each parameter using only the data when the number of terminal nodes was correctly estimated under conditions in which the true model and the template model were identical. For this reason, the bias and RMSE of the SEM Tree3 could not be calculated.

Overall, in all methods including the MDLGM Tree, the biases of the parameter estimates were judged to be sufficiently small relative to the true values, and the RMSEs were generally comparable across methods. However, relatively large biases were observed for some variance parameters, likely because each node included individuals whose covariate values were contaminated. Comparisons across methods further showed that SEM Tree1 and SEM Tree2 exhibited larger biases than the other methods for some parameters because of the occurrence of improper solutions. SEM Tree4 also showed relatively large biases for some parameters, possibly because node splitting based on $z_3$ was performed. See the Supplementary Materials S4 for more detailed results. 

\subsubsection{Correlations among the estimates from different estimation methods} 
We calculated the correlations among the estimates from different estimation methods when the methods estimated the true number of nodes correctly under conditions in which the true and template model were identical. As a result, the correlations associated with the SEM Tree3 could not be calculated. 

On the whole, regardless of the type of parameter, the correlations among the estimation methods were very high, being close to 1. This indicates that all methods yielded similar parameter estimates, provided that they correctly identified the true number of terminal nodes. One noteworthy exception was that, when both the true and template models were quadratic LGMs and $T = 5$, the correlations involving SEM Tree4 were somewhat lower (although they were still at least approximately 0.85), likely because of the effect of node splitting based on $z_3$. See the Supplementary Materials S5 for more detailed results.

\subsubsection{Estimation accuracy of the true number of nodes when the Bayesian SEM Tree was conducted with non-informative prior distributions}
We conducted an additional simulation to investigate the effect of the priors on the accuracy of estimating the true number of nodes in the SEM Tree3. Here, we ran only SEM Tree3 using exactly the same settings as in the present simulation, with Mplus's default non-informative priors.

\begin{figure}
    \centering
    \includegraphics[width=0.5\linewidth]{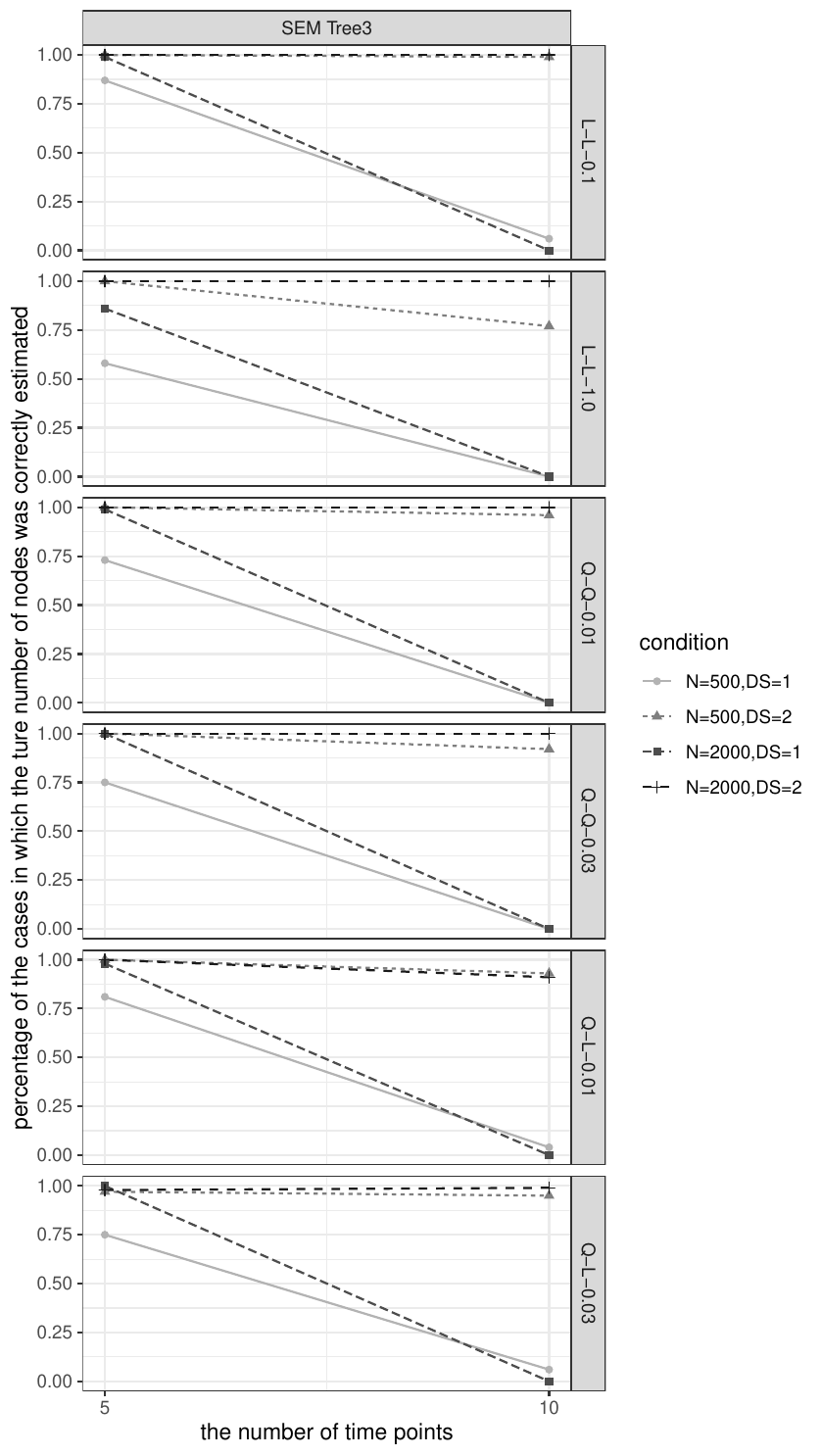}
    \caption{\\Proportion in which the true number of terminal nodes (= 4) was correctly estimated under each condition. SEM Tree3 denotes Bayesian estimation using the Mplus's default non-informative priors. $DS$ denotes the degree of separation, and the three-element labels on the vertical axis (e.g., Q-L-0.01) indicate, in order, the true model (linear or quadratic), the template model (linear or quadratic), and the specified value of the slope factor variance.}
    \label{fig:node_number_est_bayes}
\end{figure}

Figure~\ref{fig:node_number_est_bayes} shows, for each condition, the proportion of cases in which SEM Tree3 correctly estimated the true number of terminal nodes. It is clear that, when SEM Tree3 used Mplus's default non-informative priors, its accuracy in estimating the number of terminal nodes became comparable to that of the MDLGM Tree and the SEM Tree methods using log-likelihood-based splitting (MDLGM Tree2, SEM Tree1, SEM Tree2, and SEM Tree4; see also Figure~\ref{fig:node_number_est}). Moreover, the accuracy of SEM Tree3 was lower than that of MDLGM Tree1, particularly when $T$ was large ($T=10$). These results indicate that the performance of SEM Tree3 can be highly sensitive to how the prior distributions are specified and how the splitting criterion is defined.

\section{Summary and Future Work}
In this study, we developed MDLGM, which extends conventional MDFA to the context of LGMs, including factor means and factor (co)variances as parameters. We further proposed MDLGM Tree, which integrates MDLGM with decision-tree methods. Owing to its algorithm involving eigenvalue-decomposition, MDLGM Tree is expected to reduce the risk of improper solutions that may arise when estimating the LGM (template model) within each node. 

Consistent with this expectation, the simulation study showed that MDLGM Tree substantially reduced the risk of improper solutions and enabled faster tree estimation than SEM Tree. In particular, under conditions with a relatively large number of time points, around $T = 10$, MDLGM Tree was on average more than 10 times faster. It was also confirmed that, when proper solutions were obtained, MDLGM Tree yielded the solutions similar to those obtained by ML. In actual (SEM) Tree analyses, datasets often include more covariates than the three binary variables considered in this study, and these covariates are often continuous. Therefore, the computational advantages of MDLGM Tree are expected to become even more pronounced in practical applications. In addition, with regard to the Bayesian estimation in SEM Tree (SEM Tree3), it was suggested that the estimated tree structure may vary substantially depending not only on the choice of the splitting criterion but also on the specification of the priors. Moreover, in addition to the long computational time, Bayesian estimation may involve an increased risk of biased estimates when $N$ or $T$ is small. 

Taken together, MDLGM Tree is expected to be a useful model-based recursive partitioning method for future longitudinal data analyses, as it can be used to detect group differences in latent trajectories, identify external covariates that explain such differences, and classify individuals more accurately. Although MDLGM Tree would be useful as a stand-alone method, it may also be effectively used in conjunction with SEM Tree as a tool for sensitivity analysis. 

At the same time, it is also true that the current implementation of MDLGM Tree and the investigation procedure adopted in this study may have several limitations. In particular, future research is warranted from the following perspectives. First, as in the case of MDFA \citep{Terada2024}, further investigation of the theoretical properties of MDLGM-based estimators, such as statistical efficiency and consistency, is needed. 

Second, the present study was limited to relatively simple LGMs, such as linear and quadratic LGMs. Although these models have the advantage that the MDFA algorithm can be applied relatively straightforwardly to their estimation, LGMs can in principle be specified in various ways, including LGMs based on a free factor loadings approach in which some factor loadings are estimated (e.g., \citealp{Grimm2016}; \citealp{Usami2026a}). To extend MDLGM Tree to the classification method that accounts for such LGMs, it will be necessary to generalize the MDLGM algorithm proposed in this study so that it can be applied to SEMs more broadly. To achieve this goal, matrix decomposition-based estimation approaches that have recently been examined in contexts different from LGMs (e.g., \citealp{Usami2026b}) may provide a useful reference. 

Third, more detailed investigations, considering further simulation studies, are needed regarding how to set appropriate cutoff values ($cp$) and the number of multiple starts in MDLGM. This seems to be an especially important issue when constructing MDLGM Tree for the more general class of LGMs described above. 

Fourth, applying ensemble methods to MDLGM Tree, that is, developing MDLGM forests, is also an important direction for future research. Such an approach would be particularly useful in cases where researchers aim to evaluate the contribution of each covariate in explaining group differences or to predict individual trajectories more accurately. 

Fifth, it is also necessary to develop a general-purpose package that enables researchers to implement MDLGM Tree easily. Including the issue of forests noted above, such development is currently underway in the authors' research team. 

Sixth, the original MDFA and MDLGM primarily focus on point estimation of parameters, and their algorithms do not directly compute standard errors of the estimates. Bootstrap standard errors are  computationally costly, especially in the context of trees. In this regard, an important topic for future research is to examine the usefulness of methods for estimating standard errors using numerical differentiation and sandwich estimators, which have been discussed in contexts different from the MDLGM (\citealp{Usami2026b}). 

At present, SEM Tree has not been sufficiently applied in empirical research relative to its high potential, particularly because of its computational cost. Through the future intensive investigations and methodological improvements as discussed above, we hope that MDLGM Tree will be disseminated in a sound manner in psychological and behavioral science, and that the usefulness of longitudinal SEM approaches, including MDLGM Tree, will become more widely recognized among researchers working with longitudinal data, thereby further promoting their practical use.